\documentclass[times,sort&compress,3p]{elsarticle}
\journal{Journal of Multivariate Analysis}
\usepackage[labelfont=bf]{caption}

\usepackage{amsmath,amsfonts,amssymb,amsthm,booktabs,color,epsfig,graphicx,hyperref,url}

\theoremstyle{plain}
\newtheorem{theorem}{Theorem}

\newtheorem{lemma}{Lemma}

\theoremstyle{definition}

\usepackage{subfig}
\usepackage{xcolor}
\usepackage{amsfonts,dsfont} 
\usepackage{amsmath,amssymb,mathrsfs}
\usepackage{tikz}
\usetikzlibrary{positioning}
\usepackage{multirow} 
\usepackage{graphicx} 
\usepackage{hyperref, cleveref, autonum} 
\usepackage{rotating}

\newcommand{\bD}{{D}}

\newcommand{\bR}{{R}}

\newcommand{\bB}{{B}}

\newcommand{\bW}{{W}}

\newcommand{\bt}{{t}}

\newcommand{\btheta}{{\theta}}

\newcommand{\const}{\text{constant} \cdot}
\newcommand{\bzeta}{{\zeta}}
\newcommand{\bxi}{{\xi}}
\newcommand{\E}{\mathrm{E}}

\newcommand{\ml}[1]{\textcolor{black}{{#1}}}
\graphicspath{{./plots/}} 
\date{}
\begin{document}

\begin{frontmatter}

\title{Inference in Functional Linear Quantile Regression}

\author[1]{Meng Li\corref{mycorrespondingauthor}}
\author[2]{Kehui Wang}
\author[3]{Arnab Maity}
\author[3]{Ana-Maria Staicu}

\address[1]{Department of Statistics, Rice University, Houston, TX}
\address[2]{PPD Inc, Morrisville, NC}
\address[3]{Department of Statistics, North Carolina State University, Raleigh, NC}

\cortext[mycorrespondingauthor]{Corresponding author. Email address: \url{meng@rice.edu}}

\begin{abstract}
In this paper, we study statistical inference in functional quantile regression for scalar response and a functional covariate. Specifically, we consider a functional linear quantile regression model where the effect of the covariate on the quantile of the response is modeled through the inner product between the functional covariate and an unknown smooth regression parameter function that varies with the level of quantile. The objective is to test that the regression parameter is constant across several quantile levels of interest. The parameter function is estimated by combining ideas from functional principal component analysis and quantile regression. An adjusted Wald testing procedure is proposed for this hypothesis of interest, and its chi-square asymptotic null distribution is derived. The testing procedure is investigated numerically in simulations involving sparse and noisy functional covariates and in a capital bike share data application. The proposed approach is easy to implement and the {\tt R} code is published online at \url{https://github.com/xylimeng/fQR-testing}. 
\end{abstract}

\end{frontmatter}

\section{Introduction}
\label{sec:introduction} 
The advance in computation and technology generated an explosion of data that have functional characteristics. The need to analyze this type of data triggered a rapid growth of the functional data analysis (FDA) field; see~\cite{Ramsay+Silverman:05, Ferraty+Vieu:06} for two comprehensive treatments. Most research in functional data analysis has primarily focused on mean regression (see, for example, \cite{Yao+a:05,Jiang+Wang:10,Ger+:13,Iva+:15,Usset+:16}); only a few works accommodate higher-order moment effects~\citep{Staicu+:11,Li+Staicu+Bondell:15}. Quantile regression is appealing in many applications as it allows us to describe the entire conditional distribution of the response at various quantile levels. For example, in our capital bike share data application, it is of interest to study how the bike rental behavior of casual users in the previous day affects the upper quantiles of total bike rentals in the current day.

 Quantile regression models for scalar responses and functional covariates have been introduced in~\cite{Cardot+:05}. Functional quantile regression (fQR) models essentially extend the standard quantile regression framework to account for functional covariates: the effect of the covariate on a particular quantile of the response is modeled through the inner product between the functional covariate and an unknown smooth regression parameter function that varies with the level of quantile. Cardot et al. \cite{Cardot+:05} considered a smoothing splines-based approach to represent the functional covariates and derived its convergence rate; Kato \cite{Kato:12} studied principal component analysis (PCA)-based estimation and established a sharp convergence rate. In \cite{Crambes2013} and~\citep{Crambes2011} Crambes et al. discussed nonparametric quantile regression estimation and studied the theoretical properties of a support vector machine-based estimator, a method inspired from \cite{Li2007a}. Yao et al. \cite{Yao2017a} considered regularized partially fQR model by additionally incorporating high-dimensional scalar covariates. Shi et al. \cite{Shi2021} developed a procedure to test the adequacy of fQR based on functional PCA. Unlike these regularization and basis expansion-based methods, Ferraty et al. \cite{ferraty2005} and Chen and M\"{u}ller~\cite{chen+muller2012} estimated the conditional quantile function by inverting the corresponding conditional distribution function; they too studied consistency properties of the regression estimator. Nevertheless, hitherto there is no available work on statistical inference of the quantile regression estimator under fQR. Additionally, existing functional quantile regression research often assumes that the functional covariate is observed either completely on the domain or at very dense grids of points and typically with little or no error contamination.   
In this work, we are interested in formally assessing 
whether the effect of the true smooth signal of the covariate, varies across several quantile levels of interest of the response, when the smooth signal is observed at finite grids and possibly perturbed with error and a functional linear quantile regression model is assumed. This problem is important in its own right, yielding a more comprehensive description of the relationship between the covariate and the conditional distribution of the response. Furthermore, formally assessing such a hypothesis is critical when one wishes to improve the estimation accuracy of the conditional quantile of the response at some specified level. Specifically, suppose for several quantile levels around the specified level of interest, there is no evidence that the effect of the latent covariate on these quantile levels of the response differs. In that case, one can improve the accuracy in estimating the covariate effect on the response at the specified level of interest by borrowing information across these quantile levels. For example, in the case of standard quantile regression with vector predictors, there has been a rich literature on the so-called composite quantile regression to aggregate information across quantile levels~\citep{koenker1984, zou+yuan, zhao+xiao, liewen2014}.

In this paper, we assume a linear fQR model that relates the $\tau$th quantile level of the response to the covariate through the inner product between a bivariate regression coefficient function and the true covariate signal. In the case when the true signal is measured at the same time points across the study, one naive way to test the null hypothesis that the effect of the true covariate signal is constant across several quantile levels of interest, is to treat the discretely observed functional covariates as high dimensional covariate and apply standard testing procedures (Wald test) in linear quantile regression for vector covariates~\citep{koenker2005}. As expected, such an approach results in inflated type I error rates due to the high correlation between the repeated measurements corresponding to the same subject; the situation gets progressively worse when the covariate includes noise. Another alternative is to consider a single number summary of the covariate, such as average or median, and carry out this hypothesis testing by employing standard testing methods in quantile regression. Our numerical investigation of this direction shows that while the Type I error rates are preserved well, the power is substantially affected.

We propose to represent the latent smooth covariate and the quantile regression parameter function using the same orthogonal basis system; this reduces the inner product part of the linear fQR model to an infinite sum of products of basis coefficients of the smooth covariate and parameter function. There are various options of orthogonal basis types: we consider the data-driven basis that is formed by the leading eigenbasis functions of the covariance of the true covariate signal and use the percentage of variance explained criterion to determine a finite truncation for this basis. While using a finite basis system reduces the dimensionality of the problem, an important challenge is handling the variability of the basis coefficients of the smooth latent signal, called functional principal component (fPC) scores. We develop the asymptotic distributions of the quantile estimators based on the estimated fPC scores, when the functional covariate is sampled at a fine grid of points (dense design). Finally, we introduce an adjusted Wald test statistic and develop its asymptotic null distribution. The introduced testing procedure shows excellent numerical results even in situations when the functional covariate is sampled at few and irregular time points across the study (sparse design) and the measurements are contaminated by error.


{\color{black}{The theoretical study of the distribution of the quantile estimator based on the estimated fPC scores has important differences from the standard linear quantile regression with vector covariates. First,  the predictors, fPC scores, are unknown and require estimation, which in turn introduces uncertainty; by comparison the vector covariates are known in the standard quantile setting counterpart. We show that asymptotically the quantile estimators are still unbiased, but their variances are inflated. This implies that, in this reduced framework, a direct application of the Wald testing procedure for null hypotheses involving regression parameters is not appropriate. Second, dealing with estimated fPC scores in this situation is different from the measurement errors in predictors setting. For the latter, it is typically assumed that the measurement error and the true predictors are mutually independent or that the errors are independent across subjects~\citep{Wei+Carroll:09, Wang+Stefanski+Zhu:12,Wu+Ma+Yin:14}. However, in the functional data setting the resulting errors, due to the difference between the estimated fPC scores and the true scores, are dependent on the true predictors and are also dependent across subjects. As a result, the theoretical investigation requires more careful quantification in terms of the estimated scores and the use of quantile loss.}} 

This article makes three main contributions. First, we establish the asymptotic distribution of the coefficient estimator for both one single quantile level and multiple quantile levels for dense sampled functional covariates. 
To the best of our knowledge, this is the first work that studies inference of the quantile estimators; previous research in functional quantile regression focused on consistency and minimax rates (see~\cite{chen+muller2012,Kato:12}), and most literature on inference in FDA is limited to the context of functional linear regression \citep{Zhang+Chen:07, Horvath+:09, Kong+:16, Su+:17, Cao+:20}. Second, we propose an adjusted Wald test statistic to formally assess that the quantile regression parameter is constant across specified quantile levels and derive its asymptotic null distribution. Third, we consider cases where the functional covariate is observed sparsely and contaminated with noise and illustrate through detailed numerical investigation that the testing procedure continues to have excellent performance. Furthermore, we demonstrates the usage of the composite quantile regression and the corresponding advantage in terms of estimation and prediction accuracy, using a capital bike rental data set. 
Composite quantile regression is well known to improve the efficiency of the quantile estimators at a single quantile level, which becomes especially useful for extreme quantiles~\citep{kehuisinica}; nonetheless, more formal investigation of functional composite quantile regression is beyond the scope of this article.

The rest of the paper is organized as follows. Section~\ref{section:method} introduces the statistical framework, describes the null and alternative hypotheses, discusses a simpler approximation of the testing procedure, and presents the estimation approach. Section~\ref{sec:ch4.asym} develops the asymptotic normality of the proposed estimators, introduces the adjusted Wald test, and derives its null asymptotic distribution. Section~\ref{section:flqr.simulation} presents extensive simulation studies confirming the excellent performance of the proposed test procedure in various scenarios for both dense and sparse designs. Section~\ref{section:application} applies the proposed test to a bike rental data and illustrates the improvement of combined quantile regressions compared to a single level quantile regression after the proposed tests being used. Proofs of Theorem~\ref{th:asy.normal} and Theorem~\ref{th:wald.p0}, as well as some additional useful results, are given in Section~\ref{section:proof}.

\section{Methodology}
\label{section:method}

\subsection{Statistical framework} 
Suppose we observe data $\{Y_i, (t_{ij}, W_{ij})\}$ for $j \in \{1, \ldots, m_i\}$ and $i \in \{1, \ldots, n\}$, where $Y_i$ is a scalar response variable, $\{W_{i1}, \ldots, W_{im_i}\}$ is the evaluation of a latent and smooth process $X_i(\cdot)$ measured with noise at the finite grid of points $\{t_{i1}, \ldots, t_{im_i}\}$ for $t_{ij} \in \mathcal{T}$, and $\mathcal{T}$ is a bounded closed interval. It is assumed that the observed functional covariate is perturbed by white noise, i.e., $W_{ij} = X_i(t_{ij})+e_{ij}$, where $e_{ij}$ has mean 0 and variance $\sigma^2$. Furthermore, we assume that the true functional signal $X_i(\cdot) \in L^2(\mathcal{T})$ with $\mathcal{T} = [0,1]$, and $X_i(\cdot)$ are independent and identically distributed.  Our objective is to formally assess whether the smooth covariate signal $X_i(\cdot)$ has constant effect at specified quantile levels of the response.

Let $Q_{Y_i|X_i}(\tau)$ be the conditional $\tau$th quantile function of the response $Y_i$
given the true covariate signal $X_i(\cdot)$ where $\tau \in (0,1)$. We assume
the following linear fQR model:
\begin{equation}\label{model} 
Q_{Y_i|X_i}(\tau) = \beta_0(\tau) + \int_0^1 \beta(t, \tau)
X_i^c(t) dt,
\end{equation} 
where $\beta_0(\tau)$ is the quantile-level varying intercept function, and $\beta(t, \tau)$ is the bivariate regression coefficient function and the main object of interest. It is assumed that for a fixed quantile level $\tau$, $\beta(t, \tau)\in L^2[0,1]$ as a function of $t$. Here $X_i^c(t)$ is the de-meaned smooth covariate signal, defined as  $X_i^c(t) = X_i(t) - \E X_i(t)$. Model \eqref{model} is an extension of the standard linear quantile
regression model~\citep{koenker2005} to functional covariates. It was
first introduced by~\cite{Cardot+:05} and later considered by~\cite{chen+muller2012,Kato:12}. For simplicity, in the following it is assumed that the smooth covariate signal has zero mean, i.e., $\E X_i(t) =0$ for all $t\in [0,1]$.

Let $\mathcal{U} =  \{\tau_1, \ldots, \tau_L\}$ be a set with quantile levels of
interest where $\tau_1 <
\cdots < \tau_L$. Motivated by the reasons mentioned in Section~\ref{sec:introduction}, our goal is to test the null hypothesis:
\begin{equation}\label{H0:original}
H_0: \beta(\cdot, \tau_1) = \cdots = \beta(\cdot,\tau_L),
\end{equation}
against the alternative hypothesis $H_a: \beta(\cdot, \tau_{\ell}) \neq  \beta(\cdot,\tau_{\ell'}),~ \text{for some} ~ \ell \neq \ell'\in \{1,\ldots,L\}$. This null hypothesis involves infinite dimensional objects, which is very different from the common null hypotheses considered in quantile regression. 


One approach to simplify the null hypothesis is by using basis functions expansion. Specifically, let $\{ \phi_k(\cdot) \}_{k\geq 1}$ be an orthogonal basis in $L^2[0,1]$ such that $\int_0^1 \phi_k(t) \phi_{k'}(t) dt =0$ if $k\neq k'$ and $1$ if $k=k'$. We represent the unknown parameter function $\beta(\cdot, \tau)$ using this orthogonal basis $\beta(t, \tau) = \sum_{k \geq 1}
 \beta_k(\tau) \phi_k(t)$ where $\beta_k(\tau) = \int \beta(t, \tau)  \phi_k(t) dt $ are unknown parameter loadings varying with the quantile level $\tau$. It follows that the equality $\beta(\cdot, \tau_{\ell}) = \beta(\cdot, \tau_{\ell'})$ is equivalent to 
 $\beta_k(\tau_{\ell}) = \beta_{k} (\tau_{\ell'}),~k \geq 1$. Thus, the null hypothesis \eqref{H0:original} can be written as $H_0: \beta_k(\tau_1) = \beta_k(\tau_2) = \cdots = \beta_k(\tau_L)$ for $k \geq 1$. Furthermore, we represent the smooth covariate using the same basis function as $X_i(t) = \sum_{k \geq 1} \xi_{ ik} \phi_k(t)$ where $\xi_{ ik}  =\int X_i(t) \phi_k(t) \, dt$ are smooth covariate loadings. Then, the linear fQR model (\ref{model}) can be equivalently represented as $
Q_{Y_i|X_i}(\tau) = \beta_0(\tau) + \sum_{k = 1}^{\infty}\beta_k(\tau)\xi_{ik}$. In practice the infinite summation is typically truncated to some finite truncation $K$. As a result the fQR model can be approximated by 
\begin{equation}\label{model_finite} 
Q^K_{Y_i|X_i}(\tau) = \beta_0(\tau) + \sum_{k = 1}^{K}\beta_k(\tau)\xi_{ik},
\end{equation} 
and the null hypothesis to be tested can be approximated by a reduced version 
\begin{eqnarray}
\label{H0:finite} 
H_0^K: \begin{array}{cccc}
\begin{pmatrix} \beta_1(\tau_1)\\ \beta_2(\tau_1)\\ \vdots \\
\beta_{K}(\tau_{1})
\end{pmatrix} = &
\begin{pmatrix} \beta_1(\tau_2)\\ \beta_2{(\tau_2)}\\ \vdots \\
\beta_{K}(\tau_{2})
\end{pmatrix} = & \cdots & =
\begin{pmatrix} \beta_1(\tau_L)\\ \beta_2{(\tau_L)}\\ \vdots\\
\beta_{K}(\tau_L)
\end{pmatrix}
\end{array}.
\end{eqnarray}
Let ${\btheta}_\tau:=(\beta_0(\tau), \beta_1(\tau), \ldots, \beta_K(\tau))^T$ be the $(K+1)$-dimensional parameter vector and  
${\bzeta}:=( \btheta_{\tau_{1}}^{T}, \ldots, \btheta_{\tau_{L}}^{T} )^{T}$ the full quantile regression parameter vector of dimension $L(K+1)$. Then the reduced null hypothesis~\eqref{H0:finite}  can be equivalently re-written as $H_0^K: R \ \bzeta = 0$, where $R = R_1 \otimes R_2$ and
\[
\underset{(L-1)\times L} {R_1}=
\left[ {\begin{array}{cccccc}
	1 & -1 & 0 & \cdots & 0 & 0 \\      
	0 & 1 & -1 & \cdots & 0 & 0 \\
	\vdots & \vdots & \vdots & \ddots & \vdots& \vdots  \\    
	0 & 0 & 0 & \cdots & 1 & -1 \\    \end{array} } 
\right],
\qquad 	\underset{K\times (K+1)} {R_2} = [\textbf{0}_{K}, I_K].
\]
Here $\textbf{0}_{K}$ denotes the $K$-dimensional vector of zeros and $I_K$ is the $K\times K$ dimensional identity matrix.

If the loadings $\xi_{ik}$'s were known,  then model~\eqref{model_finite} is exactly the conventional quantile regression model. In such case, a standard Wald testing procedure for $H_0^K$ is typically formulated as $T_W = (R \widehat \bzeta)^T \ ( R \widehat \Gamma_{\widehat \bzeta} R^T)^{-1} \
R \widehat \bzeta$, {\color{black}{where $\widehat \bzeta$ is the quantile regression estimator of $\bzeta$ and $\widehat \Gamma_{\widehat \bzeta}$ is a consistent estimator of the covariance of ${\widehat \bzeta} $ conditional on the true loadings $\xi_{ik}$'s; see \cite[Chapter 3]{koenker2005} for a review of existing methods. }}However, in practice the loadings of the smooth covariate signal $\xi_{ik}$ are unknown, and a valid approach has to account for such uncertainty. 

Depending on the choice of the orthogonal basis, the approaches used to select the finite truncation $K$ and to develop the theoretical properties for the quantile regression estimators differ. Several choices have been commonly used in functional data analysis literature: Fourier basis functions~\citep{Staicu+:15}, Wavelet basis~\citep{Morris+Carroll:06} or orthogonal B-splines~\citep{Zhou+:08,Redd:12}. 
One important aspect to keep in mind when selecting the basis functions is how to handle the finite truncation $K$. In this paper we consider the orthogonal basis given by the eigenfunctions of the covariance of the smooth covariate signal $X_i(\cdot)$. Let $G(s,t): =\mathrm{Cov}\{X_i(s),X_i(t)\}$ be the covariance of $X_i(\cdot)$; Mercer's theorem gives the following spectral decomposition of the covariance $G(s,t)=\sum_{k=1}^{\infty}\lambda_{k}\phi_{k}(s)\phi_{k}(t),$ where $\{\phi_k(\cdot), \lambda_k\}_k$ are the pairs of eigenfunctions and corresponding eigenvalues. The eigenvalues $\lambda_k$'s are nondecreasing and nonnegative and the eigenfunctions $\phi_k(\cdot)$'s are mutually orthogonal functions in $L^2[0,1]$. Using the
Karhunen-Lo\`{e}ve expansion, the zero-mean smooth covariate $X_i(\cdot)$ can be represented as
$X_i(t)=\sum_{k=1}^{\infty}\xi_{ik}\phi_{k}(t),$
where $\xi_{ik}=\int_0^1 X_i(t)\phi_{k}(t)dt$ are commonly known as functional principal component (fPC) { scores} of $X_i(\cdot)$, satisfying that $\text{E}(\xi_{ik})
=0$, $\text{Var}(\xi_{ik}) = \lambda_k$ and uncorrelated over $k$. A popular way to select the finite truncation, or equivalently the number of leading eigenfunctions, is the percentage of variance explained; 
alternative options for selecting the finite truncation $K$ are considered in~\cite{Li+Wang+Carroll:13} and~\cite{lee2014model}. 

\subsection{Estimation procedure} \label{subsection:estimation}

We discuss estimation for the case when the functional covariate is observed on a fine grid of points, a setting known in the literature by the name of dense sampling design. Nevertheless, our procedure can be successfully applied to the case when the covariate is observed on an irregular sampling design with few points (sparse sampling design) and contaminated with noise, as illustrated later in the numerical investigation. When the sampling design is dense, and thus $m_i$ is very large for each $i$, a common approach in functional data analysis is ``smoothing first, then estimation"~\citep{Zhang+Chen:07}.
{
Specifically, we first reconstruct each trajectory $\widehat X_i(\cdot)$ from the data $(t_{ij}, W_{ij})|_{j=1}^{m_i}$ using penalized regression splines, while one can also use any other appropriate smoothing method such as the local polynomial kernel smoothing technique~\citep{Fan+Gijbels:96}. 
}Let $\bar X(\cdot)$ be the sample mean of these reconstructed trajectories and denote by $\widehat X_i^c(t) := \widehat X_i(t) -\bar X(t)$ the centered covariate. Furthermore, let $S(\cdot, \cdot)$ be the sample covariance of $\widehat X_i(t)$; the spectral decomposition of $S(\cdot, \cdot)$ yields the pairs of estimated eigenfunctions and eigenvalues $\{ \widehat \phi_k(\cdot), \widehat \lambda_k\}_k$. The theoretical properties of the estimated eigenfunctions $\widehat \phi_k(\cdot)$ have been well studied in the literature; see~\cite{Hal+Hos:06, Hall+:06, Zhang+Chen:07} among others. 
{
As eigenfunctions $\phi_k(\cdot)$ and $\widehat{\phi}_k(\cdot)$ are both defined up to a change in sign, we assume that the sign of $\widehat{\phi}_k(\cdot)$  is chosen such that $\int \phi_k(t) \widehat{\phi}_k(t) dt \geq 0$ throughout the paper.}
Finally the fPC scores $\xi_{ik}$ are estimated as $\widehat{\xi}_{ik} = \int \widehat{X}_i^c(t) \widehat{\phi}_k(t)dt$; in practice numerical integration is used to approximate the integral; see also~\cite{Li+:2010}. 

Using the estimated fPC scores $\widehat \xi_{ik}$'s, the quantile regression parameter of the approximated linear fQR model, ${\btheta}_{\tau}$, 
is estimated by 
\begin{equation}\label{eq:beta_n.zhat_new}
\widehat{\btheta}_{\tau} = \underset{(b_0, b_1, \ldots, b_K)^T \in \mathbb{R}^{K+1}}{\arg \min} \sum_{i = 1}^n \rho_\tau\left (y_i - b_0 - \sum_{k=1}^K \widehat \xi_{ik} b_k \right ),
\end{equation}
where $\rho_{\tau}(x)=x\{\tau-I(x<0) \}$ is the quantile loss function and $I(x<0)$ is the indicator function that equals $1$ if $x<0$ and $0$ otherwise. {\color{black}{Although throughout this article we focus on a homogeneous truncation level $K$ to ease presentation, the proposed method easily generalizes to the case in which $K$ varies with $\tau$.}} We next move on to studying the theoretical properties of the quantile regression estimator in (\ref{eq:beta_n.zhat_new}). 

\section{Theoretical properties} 
\label{sec:ch4.asym}

\subsection{Assumptions} 
Let $F_{i}(y)=P(Y_{i}<y|X_{i}(\cdot))$, and $f_i(\cdot)$ be the corresponding density function. We make the following assumptions:
\begin{enumerate}
	\item[A1.] 
	$\{Y_i, X_i(\cdot), e_i(\cdot)\}_{i = 1}^{n}$ are independent and identically distributed (i.i.d.) as $\{Y, X(\cdot), e(\cdot)\}$, and $X(\cdot)$ and $e(\cdot)$ are independent where $\mathrm{E} \{e(t)\} = 0$ and $\text{Cov}\{e(t), e(t')\} = \sigma^2 I(t = t')$  for any $t, t'$; 
	\item[A2.]
	The conditional distribution $F_i(\cdot)$ is twice continuously differentiable and the corresponding density function $f_i(\cdot)$ is uniformly bounded away from 0 and $\infty$ at points $Q_{Y_i|X_i}(\tau)$; 
	\item[A3.] The functional covariates $X(\cdot)$ satisfy that $\text{E}\{ X(t_1) X(t_2)X(t_3) X(t_4)\} < \infty$ uniformly for $(t_1, t_2, t_3, t_4) \in [0, 1]^4$;
	\item[A4.] There exists a finite number $p_0$ such that $\lambda_1 > \lambda_2 > \cdots > \lambda_{p_0} > 0$ and $\lambda_k = 0$ if $k > p_0$. 
\end{enumerate}

A2 and the i.i.d. assumption in A1 are standard in quantile regression with vector covariates; see \cite[][Ch. 4]{koenker2005}.  A1 assumes that the functional covariates $X_i(\cdot)$'s are observed with independent white noise $e_i(\cdot)$, making the model more realistic compared to error free assumptions made by~\cite{Kato:12}. The assumption A3 holds for Gaussian processes and is common in the FDA literature; for example, see~\cite{Hall+:06} and the discussion therein. 

Finally A4 requires that the functional covariate has a finite number of non-zero eigenvalues, \ml{making the approximate model~\eqref{model_finite} exact, with $K = p_0$}. This strong assumption has been employed previously in the literature ~\citep{Li+:2010,Li+Wang+Carroll:13}. \ml{In numerical studies, we found that A4 is not needed in order for the testing procedure to show excellent performance in terms of size and power; see, for example, the simulation study in Section~\ref{section:large.p.0} under the more general model~\eqref{model} when $p_0$ is divergent}. This seems to indicate that A4 is for theoretical convenience. 
One possible way to relax this assumption is to replace it with a condition on the number of principal components that are relevant in describing the dependence between the functional covariate and the response. \ml{Another possibility is to remove it entirely and show that the functional quantile regression $Q^K_{Y_i|X_i}(\tau)$ approximates the original model with negligible error. Nonetheless, our attempts to prove the main results by relaxing A4 in these directions have not been productive, partly due on one hand to the complication in the interweave of the quantile loss function and infinitely dimensional functional data and on the other hand to the focus on hypothesis testing, as opposed to estimation}. 
Specifically, A4 is critical to ensure a root-$n$ rate for the estimated coefficient functions formulated in Theorem~\ref{th:asy.normal} and subsequently to derive the test's null distribution. {\color{black}{As noted in the preceding section, even under A4, inference on fQR based on the estimated fPC scores differs from the standard multivariate quantile regression with vector covariates in the key aspect that estimation of the fPC scores induces a specific type of measurement error. Unlike the existing measurement error in covariates literature relying on certain independence assumptions~\citep{Wei+Carroll:09, Wang+Stefanski+Zhu:12,Wu+Ma+Yin:14}, measurement errors in the estimated fPC scores are dependent on the true predictors and are also dependent across subjects. This consideration requires a more careful quantification in terms of the estimated scores and the use of quantile loss.}} 
In this article, we focus on addressing the challenge posed by measurement error in quantile regression with intricate dependence induced by functional covariates, and leave developments to relax A4 to future research.


The following assumptions are commonly used when describing a dense sampling design ~\citep{Zhang+Chen:07, Li+:2010}. For convenient mathematical derivations, we assume that there are the same number of observations per subject, i.e., $m_i=m$ for all $i$. 
\begin{itemize}
	\item[B1.] The time points $t_{ij} \overset{i.i.d.}{\sim} g(\cdot)$ for $i \in \{1, \ldots, n\}$ and $j \in \{1, \ldots, m\},$ where the density $g(\cdot)$ has bounded support [0,1] and is continuously differentiable.
	\item[B2.] $m \geq C n^{c_m}$ where $c_m > 5/4$ and $C$ is some constant.  
\end{itemize}

For our theoretical development, we require the following condition for the kernel bandwidth $h_X$ that is used in smoothing the functional covariates. 
\begin{itemize}
	\item[C1.] $h_X = O(n^{-c_{m}/5})$.
\end{itemize}

\subsection{Asymptotic distribution} 
The following theorem gives the asymptotic distribution of the quantile estimator. Kato \cite{Kato:12} gave the minimax rate of the coefficient function estimation when there is no measurement error on the discrete functional covariates. The author assumed that the number of eigenvalues is infinite instead of finite as in our assumption A4. 
{\color{black}{Our established root-$n$ rate crucially depends on A4, which facilities downstream inference. One would need to properly scale the estimator using a slower rate and derive the asymptotic distribution, for both $\widehat{\theta}_{\tau}$ and test statistics constructed via $\widehat{\theta}_{\tau}$, should A4 be relaxed.}} We denote $\bD_0$ as the diagonal matrix whose diagonal entries are $(1, \lambda_1, \cdots, \lambda_{p_0})$ and $\bD_1(\tau) = \text{E}[f_i\{Q_{Y_i|X_i}(\tau)\} \bxi_i \bxi_i^T]$ which is positive definite, where $\bxi_i =(1, \xi_{i1}, \ldots, \xi_{ip_0})^T$. Similarly, we denote $\widehat{\xi}_{i} =(1, \widehat{\xi}_{i1}, \ldots, \widehat{\xi}_{ip_0})^T$. {\ml{When $K = p_0$, the hypothesis $H_0^K$ in~\eqref{H0:finite} is equivalent to $H_0$ in~\eqref{H0:original}, and the truncation model in (\ref{model_finite}) does not incur approximation error as the residual $\sum_{k > K} \beta_k(\tau) \xi_{ik}$ degenerates to zero owing to its zero variance.}}

\begin{theorem}
	\label{th:asy.normal}
	
	Denote by $\widehat{\theta}_{\tau}$ the quantile regression estimator defined by~ (\ref{eq:beta_n.zhat_new}) for $K=p_0$, where $\tau \in (0, 1)$. Under Conditions A1--A4, B1--B2, C1, we have 	
	\begin{eqnarray}
	\label{nine}
	\sqrt{n} (\widehat{\btheta}_{\tau}-\btheta_{\tau} )\overset{d}{\rightarrow} N\left\{0,\tau(1-\tau)\bD_{1}^{-1}(\tau)\bD_{0}\bD_{1}^{-1}(\tau) + {\Theta}_{\tau} {\Sigma}_0 {\Theta}_{\tau} \right\}, 
	\end{eqnarray}
	where ${\Theta}_{\tau} = 1_{(p_0 + 1)\times (p_0 + 1)} \otimes \theta^T_{\tau}$ and the matrix $\Sigma_0$ is defined in Section~\ref{section:proof} which does not depend on $\tau$.  {\color{black}{Moreover, $\widehat \bzeta = ( \widehat{\btheta}_{\tau_{1}}^{T}, \ldots, \widehat{\btheta}_{\tau_{L}}^{T} )^{T}$ is asymptotically multivariate normal centered at ${\bzeta}=( \btheta_{\tau_{1}}^{T}, \ldots, \btheta_{\tau_{L}}^{T} )^{T}$, and for $ 1\leq \ell \neq \ell'\leq L $ the asymptotic covariance matrix for $\widehat{\btheta}_{\tau_{\ell}}$ and $\widehat{\btheta}_{\tau_{\ell'}}$ is given by}}
	\begin{eqnarray}
	Acov \left\{    \sqrt{n} (\widehat{\btheta}_{\tau_{\ell}}-\btheta_{\tau_{\ell}}), \sqrt{n}(\widehat{\btheta}_{\tau_{\ell'}}-\btheta_{\tau_{\ell'}}) \right  \}
	=
	\label{eq:acov}
	\{\min(\tau_{\ell},\tau_{\ell'})-\tau_{\ell}\tau_{\ell'} \} \bD_{1}^{-1}(\tau_{\ell})\bD_{0}\bD_{1}^{-1}(\tau_{\ell'}) + {\Theta}_{\tau_{\ell}} {\Sigma}_0 {\Theta}_{\tau_{\ell'}}.\label{ten}
	\end{eqnarray}
\end{theorem}

Remark that the asymptotic covariances in both (\ref{nine}) and (\ref{ten}) contain two components: a Huber~\citep{huber1967} sandwich term that is typical in quantile regression theory and a ``variance inflation" term. Specifically, if the true scores $\bxi_{i} $'s were observed, then the asymptotic variance of $\widehat{\btheta}_{\tau}$ would be $\tau(1-\tau)\bD_{1}^{-1}(\tau)\bD_{0}\bD_{1}^{-1}(\tau)$, 
and the asymptotic covariance matrix for $\widehat{\btheta}_{\tau_{\ell}}, \widehat{\btheta}_{\tau_{\ell'}}$ would be $\{\min(\tau_{\ell},\tau_{\ell'})-\tau_{\ell}\tau_{\ell'} \} \bD_{1}^{-1}(\tau_{\ell})\bD_{0}\bD_{1}^{-1}(\tau_{\ell'})$; see ~\cite{pollard1991,koenker2005}. The variance inflation terms, ${\Theta}_{\tau} {\Sigma}_0 {\Theta}_{\tau}$ in (\ref{nine}) and ${\Theta}_{\tau_{\ell}} {\Sigma}_0 {\Theta}_{\tau_{\ell'}}$ in  (\ref{ten}), quantify the effect of uncertainty in estimating the fPC scores on the quantile regression estimators. Thus, when the covariates are functional data, the asymptotic distribution of $\widehat{\btheta}_{\tau}$ is unbiased but the variance is inflated where the variance inflation terms depend on the true parameter value $\btheta_{\tau}$.

The proof of Theorem~\ref{th:asy.normal} is detailed in Section~\ref{section:proof}. 
The reasoning follows two main steps: 1) approximate the estimated fPC scores $\widehat{\bxi}_i$'s by linear combinations of random vectors of the true fPC scores ${\bxi}_i$; and 2) show that the approximation error in the predictors is negligible to the quantile loss function. {\color{black}{Step 1 crucially relies on the dense design assumption B2. This allows to employ various bounds on both the estimated eigenfunctions and the difference $\widehat{X}_i(\cdot) - X_i(\cdot)$, which in turn enables us to derive a fine-grained characterization of the estimated scores (Lemma~\ref{lemma:z.hat.z}); see the supplementary materials for more detail.}}


\subsection{Adjusted Wald test} \label{section:wald-type test}

Using the asymptotic properties of the quantile regression estimators, we are now ready to develop a Wald type testing procedure for assessing the general null hypothesis (\ref{H0:original}) or its finite reduced version (\ref{H0:finite}) represented in vector form by $H_0^K: R \ \bzeta = 0$. Recall that ${\bzeta}=( \btheta_{\tau_{1}}^{T}, \ldots, \btheta_{\tau_{L}}^{T} )^{T}$ denotes the full quantile regression parameter, and $\widehat \bzeta = ( \widehat{\btheta}_{\tau_{1}}^{T}, \ldots, \widehat{\btheta}_{\tau_{L}}^{T} )^{T}$ is its estimator. 

\newcommand{\SigmaAdj}{\Gamma^a} 
\newcommand{\SigmaAdjHat}{\widehat{\Gamma}^a} 
We define a modified version of Wald test, called the \emph{adjusted Wald test}, by ignoring the variance inflation terms in the above asymptotic covariances. {\color{black}{Let $\Sigma(\tau_{\ell}, \tau_{\ell'}) = \sigma(\tau_{\ell}, \tau_{\ell'}) \bD_{1}^{-1}(\tau_{\ell})\bD_{0}\bD_{1}^{-1}(\tau_{\ell'})$ with $\sigma(\tau_{\ell}, \tau_{\ell'})$ set to $\tau_{\ell}(1-\tau_{\ell})$ if $\ell = \ell'$, and $\{\min(\tau_{\ell},\tau_{\ell'})-\tau_{\ell}\tau_{\ell'} \}$ otherwise. Then  
the asymptotic covariance matrix of $\widehat \bzeta$ without the inflation terms, denoted as $\SigmaAdj_{\widehat \bzeta}$ in which the superscript $a$ indicates the adjustment by ignoring the inflation terms, i.e., 
\newcommand{\notationAlpha}[1]{\Sigma(\tau_{#1}, \tau_{#1})}
\newcommand{\notationBeta}[2]{\Sigma(\tau_{#1}, \tau_{#2})}
\begin{equation} \underset{L(K + 1) \times L(K + 1)} {\SigmaAdj_{\widehat \bzeta}} = \begin{bmatrix}
  \notationAlpha{1} & \notationBeta{1}{2} & \cdots & \notationBeta{1}{L} \\
  \notationBeta{2}{1} & \notationAlpha{2} & \cdots & \notationBeta{2}{L} \\
  \vdots          & \vdots          & \ddots & \vdots          \\
  \notationBeta{L}{1} & \notationBeta{L}{2} & \cdots & \notationAlpha{L}
\end{bmatrix}. \label{eq:covariance}
\end{equation}
Let $\SigmaAdjHat_{\widehat \bzeta}$ be a consistent estimator of $ \SigmaAdj_{\widehat \bzeta}$ constructed similarly to (\ref{eq:covariance}) but with a consistent estimator $\widehat{\Sigma}(\tau_{\ell}, \tau_{\ell'})$ of ${\Sigma}(\tau_{\ell}, \tau_{\ell'})$. }}The adjusted Wald test is given by
\begin{eqnarray}\label{Waldtest}
T_n = n (R \widehat \bzeta)^T \ ( R \SigmaAdjHat_{\widehat \bzeta} R^T)^{-1} \
R \widehat \bzeta.
\end{eqnarray}
This test is not a proper Wald test as the covariance matrix used is not the valid covariance of ${\widehat \bzeta}$. The following result studies the asymptotic null distribution of $T_n$ assuming $K = p_0$.

\begin{theorem}
	\label{th:wald.p0}
	Assume the regularity conditions A1--A4, B1--B2 and C1 hold. If the null hypothesis is true, $\bR{\bzeta} = 0$, then the asymptotic distribution of $T_{n}$ is $\chi_{K}^{2}$.
\end{theorem}
The proof of this result relies on the observation that if ${\Gamma}_{\widehat \bzeta}$ is the proper covariance of $\widehat \bzeta$
as described by Theorem \ref{th:asy.normal}, then $\bR (\SigmaAdj_{\widehat \bzeta} - {\Gamma}_{\widehat \bzeta}) \bR^T = 0$. {\color{black}{Intuitively, this is because the inflation terms in (\ref{nine}) and (\ref{ten}) possess a sandwich structure with a constant matrix enclosed by $\Theta_{\tau}$, which is zeroed out if left multiplied by $R$ under the null hypothesis that $\bR{\bzeta} = 0$.}} Thus, although the estimation of the fPC scores yields inflated covariance of the regression estimator, its effect on testing the null hypothesis (\ref{H0:original}) is negligible. {\color{black}{Nevertheless, if one is interested in testing a different type of null hypothesis for $\bzeta$, such as nonlinear functionals, then this variance inflated term has to be taken into account for a proper testing procedure.}} 

We construct $\widehat{\Sigma}(\tau_{\ell}, \tau_{\ell'})$ for $1 \leq \ell, \ell' \leq L$ by a plug-in estimator that uses $\widehat{\bD}_0 = \sum_{i = 1}^n \widehat{\bxi}_{i}\widehat{\bxi}_{i}^{T}/n$ and {\color{black}{$\widehat{\bD}_1(\tau) = \sum_{i = 1}^n \widehat{f}_{i}({\bxi}_{i}^T {\theta}_{\tau})\widehat{\bxi}_{i}\widehat{\bxi}_{i}^{T}/n$ to estimate $\bD_0$ and $\bD_1(\tau)$}}, respectively. The consistency of these estimators can be proved by law of large numbers-based arguments together with Lemma~\ref{lemma:z.hat.z} that discusses the closeness between $\widehat{\bxi}_i$ and $\bxi_{i} $. {\color{black}{For the estimation of $f_i({\bxi}_{i}^T {\theta}_{\tau})$ in $D_1(\tau)$, we use the difference quotient method proposed by \cite{hendricks1992hierarchical} and substitute the estimates $\widehat{\xi}_i$ and $\widehat{\theta}_{\tau}$.}} Theorem~\ref{th:wald.p0} implies that, for testing the null hypothesis of equal functional covariate effect across various quantile levels, the common Wald test based on the estimated fPC scores provides a valid testing procedure. The adjusted Wald test, that disregards the variance component due to the estimation uncertainty of the fPC scores, has a chi-square asymptotic null distribution.

If the number of principal components $p_0$ is replaced \ml{by its consistent estimator $K_n$}, then the null distribution of the test statistic $T_n$ is approximately $\chi_{K_n}^{2}$ for large $n$. {\color{black}{In other words, the difference in the respective cumulative distribution functions, $P(T_n \leq t) - P(\chi_{K_n}^{2} \leq t)$, goes to zero for any $t \in \mathbb{R}$; it implies that the critical value of $T_n$ is asymptotically the same as that of $\chi_{K_n}^2$}}. \ml{Functional data analysis literature provides a rich menu of possibilities for selecting $p_0$, such as the percentage of variance explained (PVE) criterion and a Bayesian information criterion (BIC) proposed by~\cite{Li+Wang+Carroll:13}. The BIC method is proved to be consistent for both sparse and dense functional data. The PVE criterion is defined as $$K_n = \min\left\{p:  \sum_{i = 1}^p \widehat{\lambda}_i / \sum_{i = 1}^{q} \widehat{\lambda}_i \geq \text{PVE} \right\},$$ where $q$ is the number of estimated eigenvalues and PVE some user-defined threshold that approaches one. The widely used PVE approach also leads to consistent estimators of $p_0$ given that the number of estimated eigenvalues is greater than $p_0$ and eigenvalues are estimated consistently, which may well be true in many applications and particularly as suggested by our extensive simulation studies.
We shall use the method of PVE in the remaining sections, and we have found that it leads to accurate estimate of $p_0$ in finite sample performance. }

{
We would like to point out that the asymptotic power of the adjusted Wald test is obtainable using a non-central chi-square distribution. However, the expression is complicated without involving stronger assumptions on $X_i(\cdot)$'s, since the equation $\bR (\SigmaAdj_{\widehat \bzeta} - {\Gamma}_{\widehat \bzeta}) \bR^T = 0$ does not generally hold under the alternative thus a Wald-type test requires an estimate of the matrix $\Sigma_0$.  
}



\section{Simulation} 
\label{section:flqr.simulation}

\subsection{Settings}
\label{section:setting} 
The simulated data is of the form $\left\{Y_i, (t_{ij}, W_{ij})|_{j = 1}^{m_i} \right\}$ for $i \in \{1, \ldots, n\}$, where $Y_i$ is the scalar response and $W_{ij} =X_i(t_{ij}) +e_{ij}$ is the functional covariate contaminated with measurement error $e_{ij}$, $t_{ij}\in [0,1]$, and $X_i(\cdot)$ is the true functional covariate.  We generate the data from the following heteroscedastic model:
$Y_i = \int X_i(t) t dt + \{1 + \gamma \int X_i(t) t^2
dt\} \epsilon$ with $\epsilon \sim N(0, 1)$. This leads to a quantile regression model of the form (\ref{model}) 
with $\beta_0(\tau) = \Phi^{-1}(\tau)$, and $\beta(t, \tau) = t
+ \gamma t^2 \Phi^{-1}(\tau)$. {\color{black}{Note that the functional coefficient $\beta(t, \tau)$ is nonlinear in $t$ when $\gamma \neq 0$. Here the scalar $\gamma$ controls the heteroscedasticity and determines how the 
coefficient function $\beta(\cdot, \tau) (\tau \in \mathcal{U})$ varies across $\tau$. Specifically, if $\gamma =0$ then the effect of $X_i(\cdot)$ is constant across different quantile levels of $Y_i|X_i(\cdot)$, while if $\gamma\neq 0$ then the effect of $X_i(\cdot)$ varies across different quantile levels of $Y_i|X_i(\cdot)$.}}

\begin{figure}
	\centering 
	\begin{tabular}{cc}
		\includegraphics[width = 0.5\textwidth]{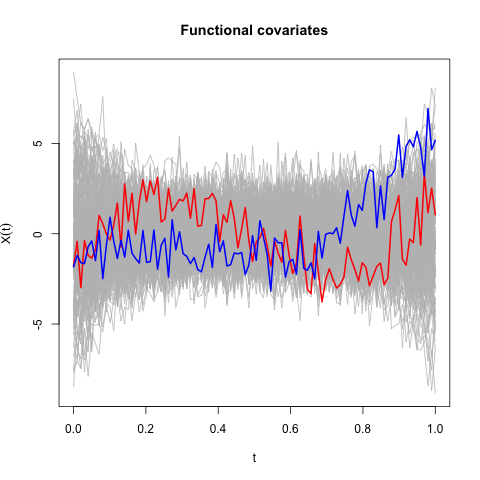} & 
		\includegraphics[width = 0.5\textwidth]{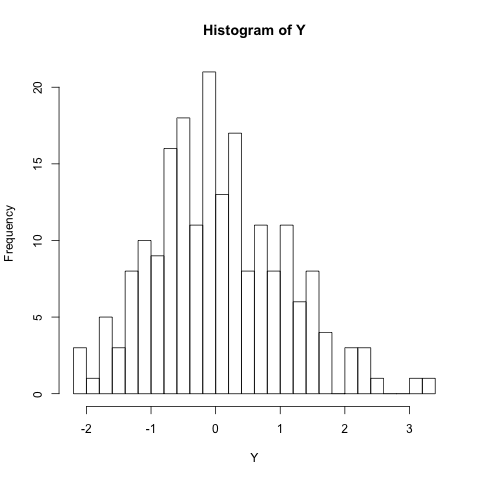} 
		\end{tabular} 
		\caption{Simulated data when $n = 200$ and $\gamma = 1$. The left panel plots the functional covariates and two randomly selected curves are highlighted in blue and red; the right panel is the histogram of the response. }
		\label{fig:sim.data} 
\end{figure}

The true functional covariate $X_i(\cdot)$ is generated from a Gaussian process with zero mean and covariance function $\textrm{cov}\{X_i(s), X_i(t)\} = \sum_{k\geq 1} \lambda_k \phi_k(s) \phi_k(t)$, where $\lambda_k = (1/2)^{k - 1}$ for $k = 1, 2, 3$ and $\lambda_k = 0$ for $k \geq 4$, and $\{\phi_k(\cdot)\}_k$ are the orthonormal Legendre polynomials on $[0,1]$: $\phi_1(t) = \sqrt{3}(2t - 1), \phi_2(t) = \sqrt{5}(6t^2 - 6t + 1), \phi_3(t) = \sqrt{7}(20t^3 - 30t^2 + 12t - 1)$. It is assumed that the measurement error $e_{ij}\sim N(0, \sigma^2)$. Fig.~\ref{fig:sim.data} plots simulated data when $n = 200, \gamma = 1$, and $\sigma = 1$. 

The objective is to test the null hypothesis $H_0: \beta(\cdot, \tau_{\ell})=\beta(\cdot,\tau_{\ell'})$ for $\tau_{\ell}, \tau_{\ell'}\in \mathcal{U}$, that the effect of the true functional covariate on the conditional distribution of the response is the same for all the quantile levels in a given set $\mathcal{U}$.  
When $\gamma = 0$, the coefficient function
$\beta(\cdot, \tau)$ is independent of $\tau$, which means that null hypothesis is true; when $\gamma\neq 0$ then $\beta(\cdot, \tau)$ is varies with $\tau$ and thus the null hypothesis is false. We consider two sets of quantile levels: $\mathcal{U}_1 =
\{0.1, 0.2, 0.3, 0.4\}$ for one-sided quantile levels, and $\mathcal{U}_2 = \{0.1, 0.2,
0.6, 0.7\}$ for two-sided quantile levels. 

We implement the proposed adjusted Wald test using a number of fPC selected via the PVE criterion with PVE=95\%. We use the {\tt R} package {\it refund}~\citep{refund:15} to estimate the fPC scores, where the individual trajectory is reconstructed using penalized regression splines via the function {\it gam} and the smoothing parameter is selected using the restricted maximum likelihood approach. 
We investigate the performance of the proposed test for low and high level of measurement error in the functional covariate ($\sigma = 0.05$ and $\sigma = 1$ respectively), for varying sample sizes $n$ from $100$ to $5000$. For the functional covariates, we consider a dense design in Section~\ref{section:sim.dense}, a sparse design in Section~\ref{section:sim.sparse}, and a setting when $p_0$ diverges in Section~\ref{section:large.p.0}.

\subsection{Dense design}
\label{section:sim.dense} 

We first consider a dense design for the functional covariates: the grid of points for each $i$ is an equispaced grid of $m_i=100$ timepoints in $[0,1]$. We are not aware of any testing procedures for testing the null hypothesis of constant effect at various quantile levels, when the covariate is functional; however we can exploit this particular setting and pretend the covariates are vectors and thus use or directly extend existing testing procedures from quantile regression. In particular, we consider three alternative approaches: (1) treat the observed functional covariate as vector and use the common Wald test for vector covariates in quantile regression (NaiveQR); (2) summarize observed functional covariates via a single number summary of the functional covariate in conjunction with the Wald test (SSQR); and (3) treat the observed functional covariate as a vector, reduce the dimensionality using principal component analysis and then apply the Wald test using the vector of principal component scores (pcaQR). For the pcaQR approach, the number of principal components are selected via PVE and using a level PVE=95\%. The Wald test for vector covariates in these three approaches is described in~\citep[Chapter 3.2.3]{koenker2005}. 

Table~\ref{table:type1_dense} summarizes the empirical Type I error rates of the adjusted Wald test when testing $H_0$ at one-sided quantile levels ($\mathcal{U}_1$) as well as two-sided quantile levels ($\mathcal{U}_2$), when the functional covariate is observed with large ($\sigma=1$) measurement error. The results are presented for three significance levels $\alpha=0.01$, $\alpha=0.05$ and $\alpha=0.10$; they indicate irrespective of quantile levels set or magnitude of the measurement error the Type I error rates are slightly inflated for moderate sample sizes. Nevertheless the empirical Type I error rates converge to the nominal level. The empirical Type I error rates for the alternative approaches are presented in Table~\ref{table:type1.others}. As expected the NaiveQR approach has very poor performance. The NaiveQR approach does hypothesis testing when the covariates are highly correlated; this leads to numerical instability due to singularity of the design matrix. Therefore NaiveQR produces many missing values (reported as ``--") in the table, and yields inflated empirical Type I error rates for any significance level. Results for $\sigma = 0.05$ are similar and omitted here. 

\begin{table}[t!] 
	\centering
	\caption{Type I error of the adjusted Wald-type test at significant level $\alpha \in \{0.01, 0.05, 0.10\}$ under dense design. We test $H_0$ at two sets of quantile levels:  $\mathcal{U}_1 =
		\{0.1, 0.2, 0.3, 0.4\}$ and $\mathcal{U}_2 = \{0.1, 0.2, 0.6, 0.7\}$. Results are based on $5000$ simulations.}
	\label{table:type1_dense}
	\hrule
	\begin{tabular}{cccllccccll}
		Scenario & $n$  & $0.01$ & 0.05 & 0.10 & Scenario & $n$  & $0.01$ & 0.05 & 0.10 \\ 
		\cmidrule(lr){1-5} \cmidrule(lr){6-10}
		& 100 & 0.021 & 0.060 & 0.104   & &   100 &     0.030 & 0.076 & 0.123 \\ 
		$\sigma = 1$ &  500 & 0.014 & 0.057 & 0.107   & $\sigma = 1$ &   500 &    0.015 & 0.062 & 0.116 \\ 
		$ \tau \in \mathcal{U}_1$ &  1000 & 0.017 & 0.052 & 0.106   & $ \tau \in \mathcal{U}_2$ &  1000 &     0.015 & 0.059 & 0.112 \\ 
		& 2000 & 0.011 & 0.051 & 0.101   & & 2000 &     0.010 & 0.053 & 0.103 \\
		& 5000 & 0.010 & 0.054 & 0.105   & & 5000 &     0.012 & 0.056 & 0.103 \\ 
	\end{tabular}
	\hrule 
\end{table} 

The pcaQR approach gives relatively good performance when the magnitude of the error is small ($\sigma=0.05$): the empirical Type I error is close to the nominal level in results not reported here. However, Table~\ref{table:type1_dense} shows that as the error variance increases ($\sigma=1$), the empirical rejection probabilities are either excessively inflated when $n \in \{1000, 2000, 5000\}$, or there are too many missing values when $n \in \{100, 500\}$. The results are not surprising, because in the case of large error variance, a direct application of principal component analysis yields a large number of principal components. As a consequence, the application of the classical Wald test for vector covariate leads to numerical instability due to singularity of the design matrix, in a similar way to the NaiveQR approach. The performance of SSQR approach is very good for all the scenarios considered and across various sample sizes: the empirical Type I error rates are close to the nominal levels. This is expected, as in the case when $H_0$ holds, the functional covariate effect is through its mean, and this effect is invariant over quantile levels.

\begin{table}
\centering
\caption{Type I error of alternative approaches at significant level $\alpha \in \{0.01, 0.05, 0.10\}$ under dense design. We test $H_0$ at two sets of quantile levels:  $\mathcal{U}_1 =
		\{0.1, 0.2, 0.3, 0.4\}$ and $\mathcal{U}_2 = \{0.1, 0.2, 0.6, 0.7\}$. Results are based on $5000$ simulations.
When one method returns error (due to singularity of the design matrix) in more than 20\% replications, we report it as ``--". }
\label{table:type1.others} 
\hrule 
\begin{tabular}{ccccccccccc}
& &  \multicolumn{3}{c}{NaiveQR} & \multicolumn{3}{c}{SSQR} & \multicolumn{3}{c}{pcaQR} \\ 
Scenario & $n$ & 0.01 & 0.05 & 0.10 & 0.01 & 0.05 & 0.10 & 0.01 & 0.05 & 0.10 \\ 
\cmidrule(lr){3-5} \cmidrule(lr){6-8} \cmidrule(lr){9-11}
&  100 &     -- & -- & -- & 0.008 & 0.033 & 0.071 & -- & -- & -- \\ 
$\sigma = 1$ &  500 &     -- & -- & -- & 0.008 & 0.036 & 0.080 & -- & -- & -- \\ 
$ \tau \in \mathcal{U}_1$ &  1000 &     -- & -- & -- & 0.010 & 0.049 & 0.092 & 0.996 & 0.999 & 1.000 \\ 
&  2000 &     1.000 & 1.000 & 1.000 & 0.009 & 0.048 & 0.097 & 1.000 & 1.000 & 1.000 \\ 
&  5000 &     1.000 & 1.000 & 1.000 & 0.008 & 0.053 & 0.099 & 0.999 & 1.000 & 1.000 \\ 
\\
&  100 &     -- & -- & -- & 0.009 & 0.040 & 0.077 & -- & -- & -- \\ 
$\sigma = 1$ &  500 &     -- & -- & -- & 0.009 & 0.050 & 0.096 & -- & -- & -- \\ 
$ \tau \in \mathcal{U}_2$ &  1000 &     1.000 & 1.000 & 1.000 & 0.009 & 0.046 & 0.095 & 1.000 & 1.000 & 1.000 \\ 
&  2000 &     1.000 & 1.000 & 1.000 & 0.010 & 0.048 & 0.099 & 1.000 & 1.000 & 1.000 \\ 
&  5000 &     1.000 & 1.000 & 1.000 & 0.011 & 0.051 & 0.100 & 1.000 & 1.000 & 1.000 \\ 
\end{tabular}
\hrule
\end{table}

\begin{figure}
\begin{tabular}{cccc}
\begin{sideways} \rule[0pt]{0.4in}{0pt} $\sigma = 1$ and $\tau \in \mathcal{U}_1$ \end{sideways}
& \includegraphics[trim = 20 30 20 50, clip, width = 0.3\textwidth]{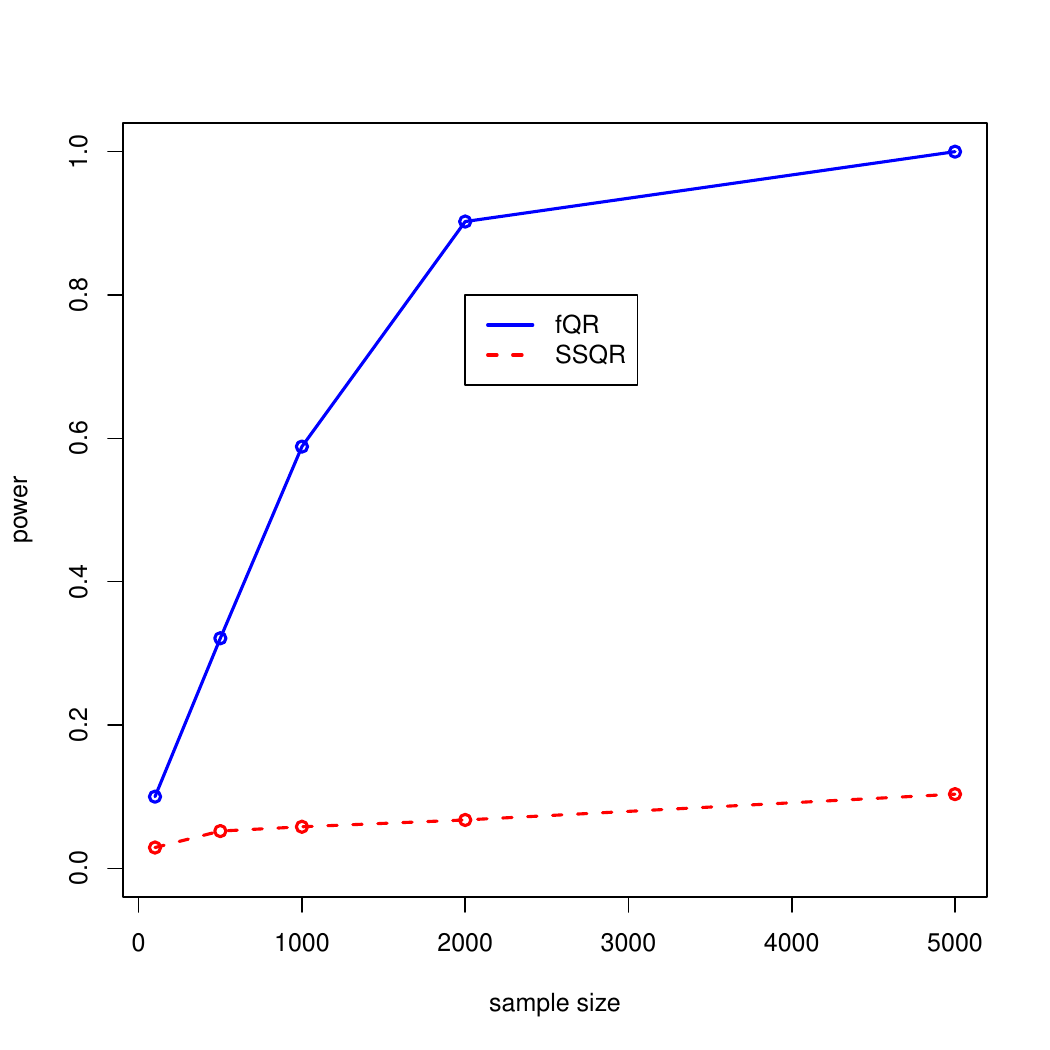}
& \includegraphics[trim = 20 30 20 50, clip, width = 0.3\textwidth]{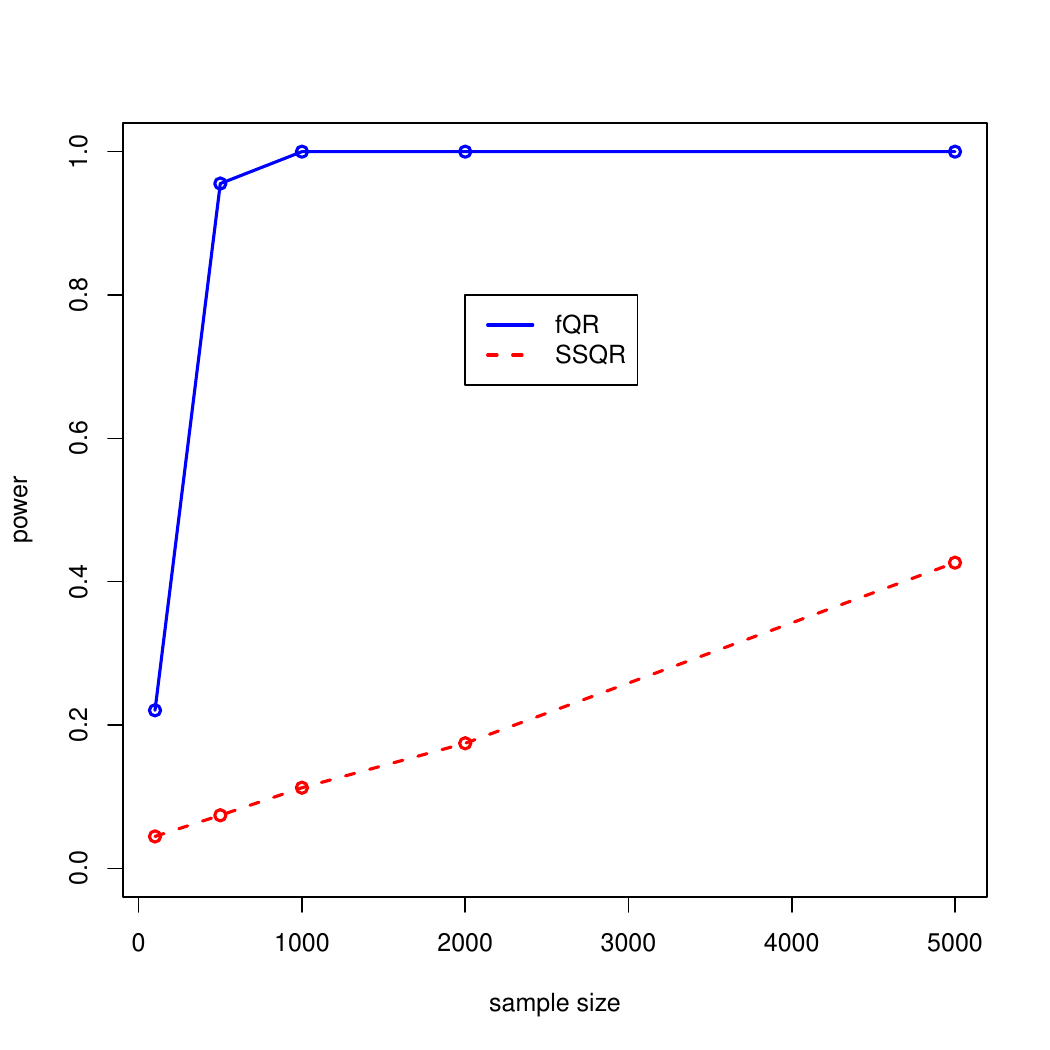}
& \includegraphics[trim = 20 30 20 50, clip, width = 0.3\textwidth]{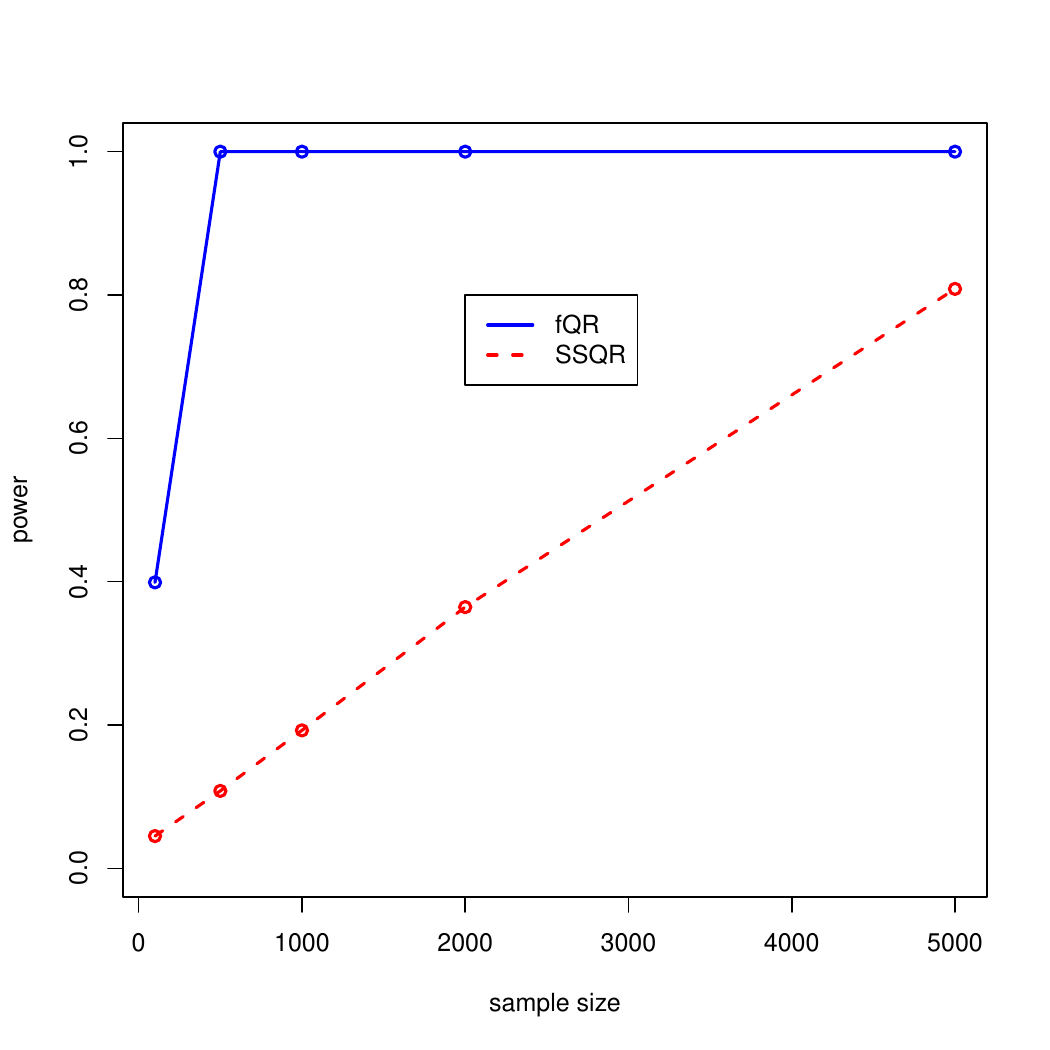} \\
\begin{sideways} \rule[0pt]{0.4in}{0pt} $\sigma = 1$ and $\tau \in \mathcal{U}_2$ \end{sideways}
& \includegraphics[trim = 20 30 20 50, clip, width = 0.3\textwidth]{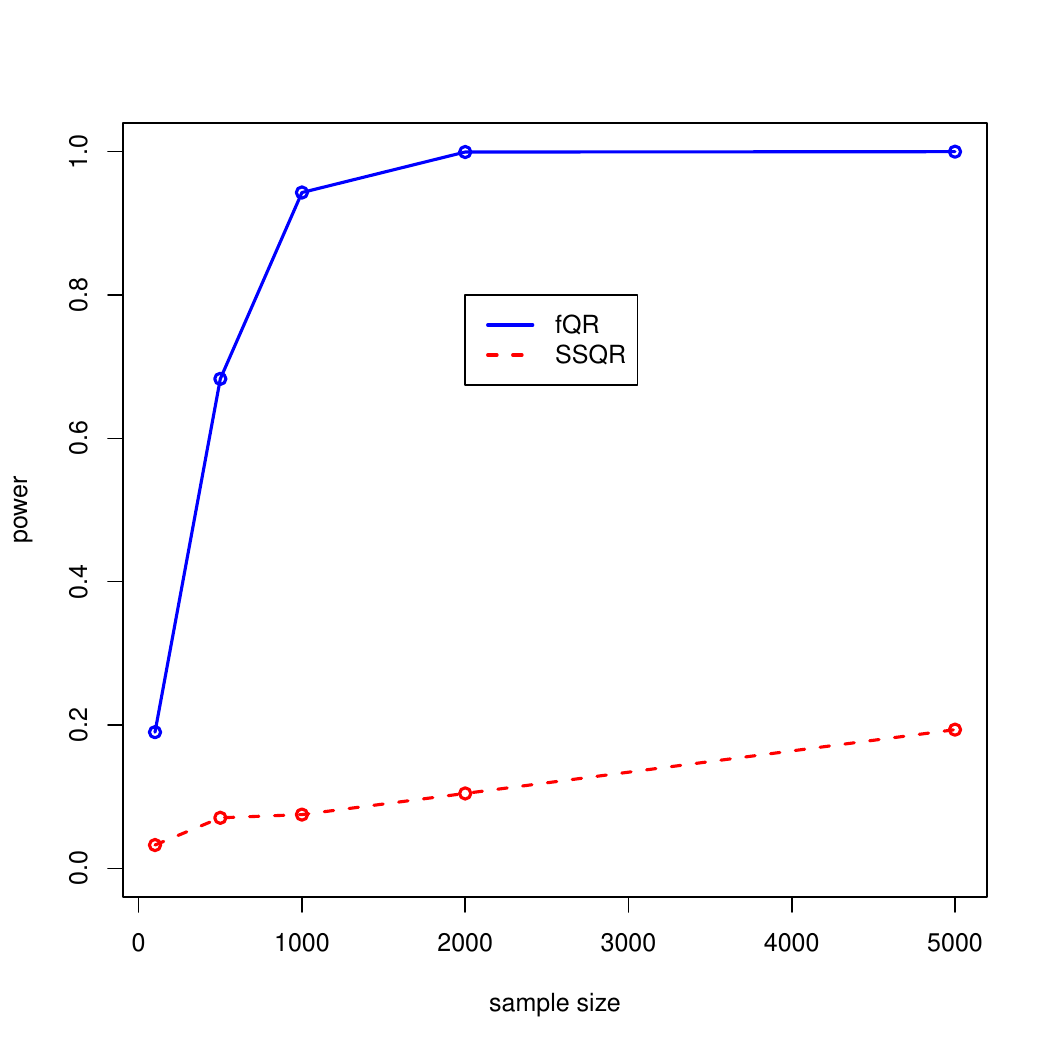}
& \includegraphics[trim = 20 30 20 50, clip, width = 0.3\textwidth]{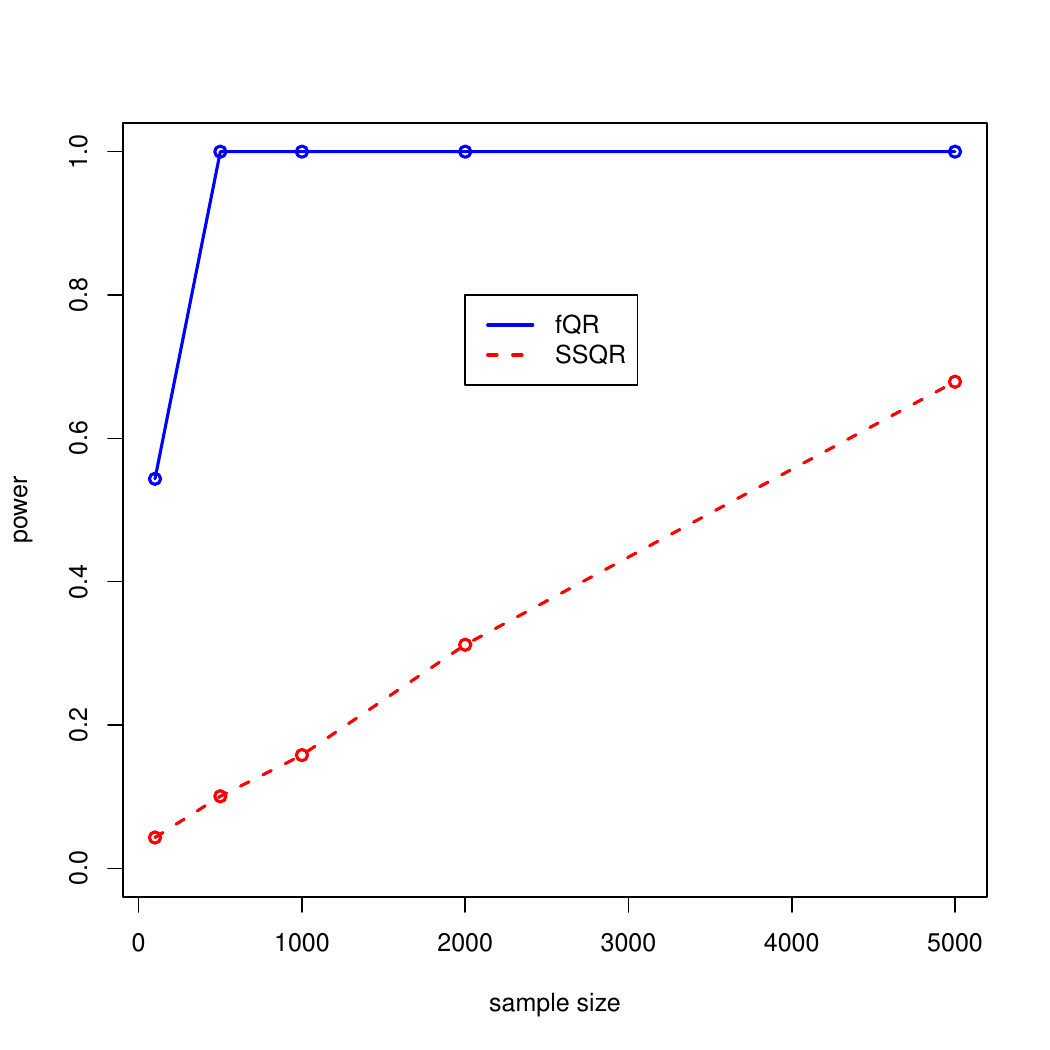}
& \includegraphics[trim = 20 30 20 50, clip, width = 0.3\textwidth]{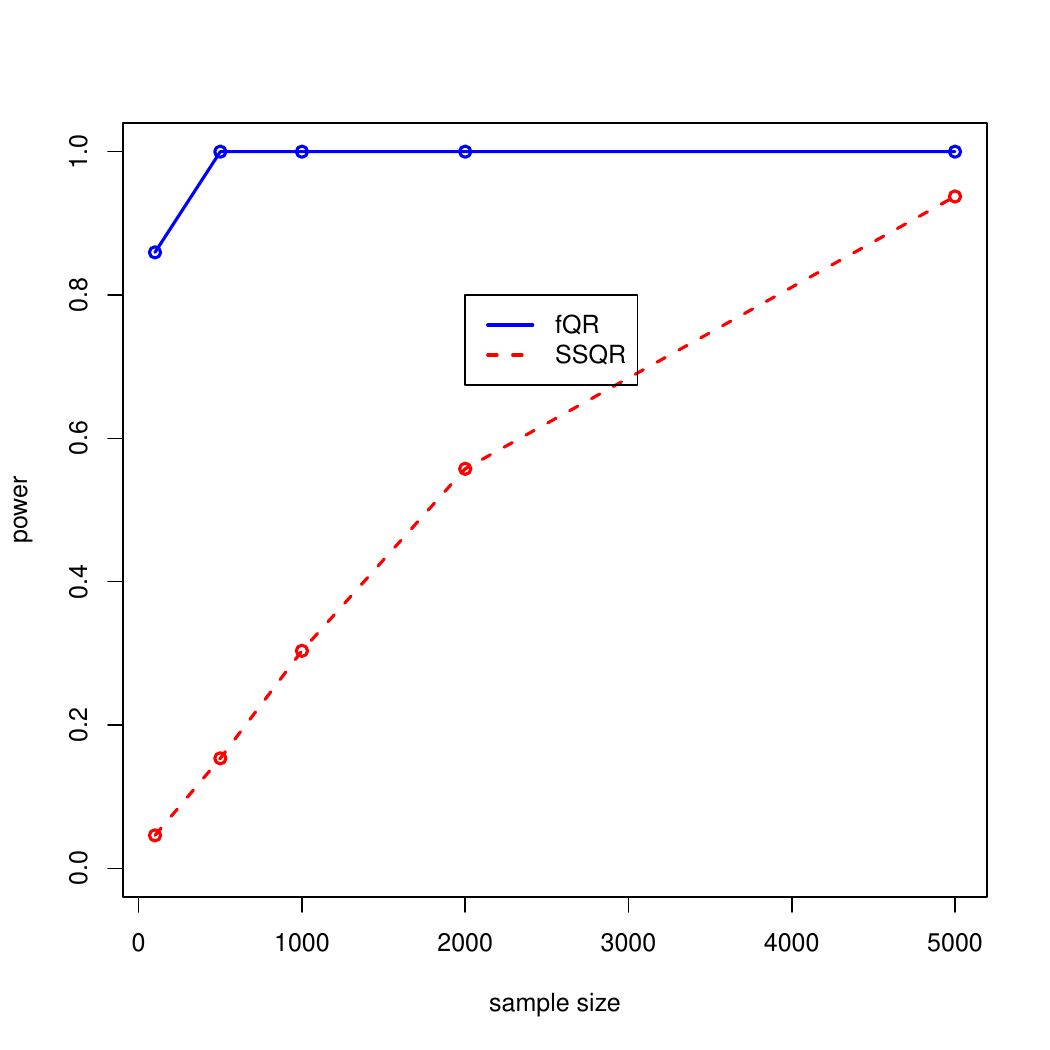} \\
& $\gamma = 0.5$ & $\gamma = 1$ & $\gamma = 1.5$ 
\end{tabular} 
\caption{Power curves of the adjusted Wald test and SSQR under dense design. We test $H_0$ at two sets of quantile levels: $\mathcal{U}_1 = \{0.1, 0.2, 0.3, 0.4\}$ and $\mathcal{U}_2 = \{0.1, 0.2, 0.6, 0.7\}$. The $x$-axis is the sample size $n \in \{100, 500, 1000, 2000, 5000\}$. Results are based on 2000 simulations. } 
\label{figure:power}
\end{figure}

Next we evaluate the performance in terms of empirical rejection probabilities when the null hypothesis is not true. We only focus on the proposed adjusted Wald testing and SSQR procedures, as they have the correct size. Fig.~\ref{figure:power} illustrates the power curves based on 2000 simulations for large noise with $\sigma=1$; the results are similar in the case of low noise ($\sigma=0.05$) and for brevity are not included. The adjusted Wald procedure is much more powerful than SSQR irrespective of the departure from the null hypothesis as reflected by the coefficient $\gamma$. For example, when $\gamma=1$ the probability to correctly reject $H_0$ using the adjusted Wald is about 100\% when the sample size is $500$ or more, whereas the counterpart obtained with SSQR is less than 70\% even when the sample size increases to 5000. These results are not surprising, as SSQR summarizes the entire functional covariate through a single scalar, while the proposed adjusted Wald test employs the full functional covariate. 


\subsection{Sparse design} \label{section:sim.sparse} 

Next, we study the performance of the adjusted Wald testing procedure when the functional covariate is observed sparsely and with measurement error. We set an overall grid of 101 equispaced points in $[0,1]$ and consider two settings: a `moderately sparse' sampling design with $m_i=50$ randomly generated time points to form $t_{i1}, \ldots, t_{im_i}$ for each $i$, and a `highly sparse' design with $m_i=10$. Other aspects of the data generating process follow the dense design described in the previous section. We use the adjusted Wald test which relies on sparse fPCA techniques, that estimate the fPC scores $\xi_{ik}$'s using conditional expectation proposed by~\cite{Yao+a:05}. 
When the sampling design of the functional covariate is sparse, there are no obvious reasonable alternative approaches to compare. Thus in this section we only discuss the performance of the proposed Wald-type procedure.

\begin{figure}[h!]
\begin{tabular}{cccc}
\begin{sideways} \rule[0pt]{0.4in}{0pt} $\sigma = 1$ and $\tau \in \mathcal{U}_1$ \end{sideways}
& \includegraphics[trim = 20 30 20 50, clip, width = 0.3\textwidth]{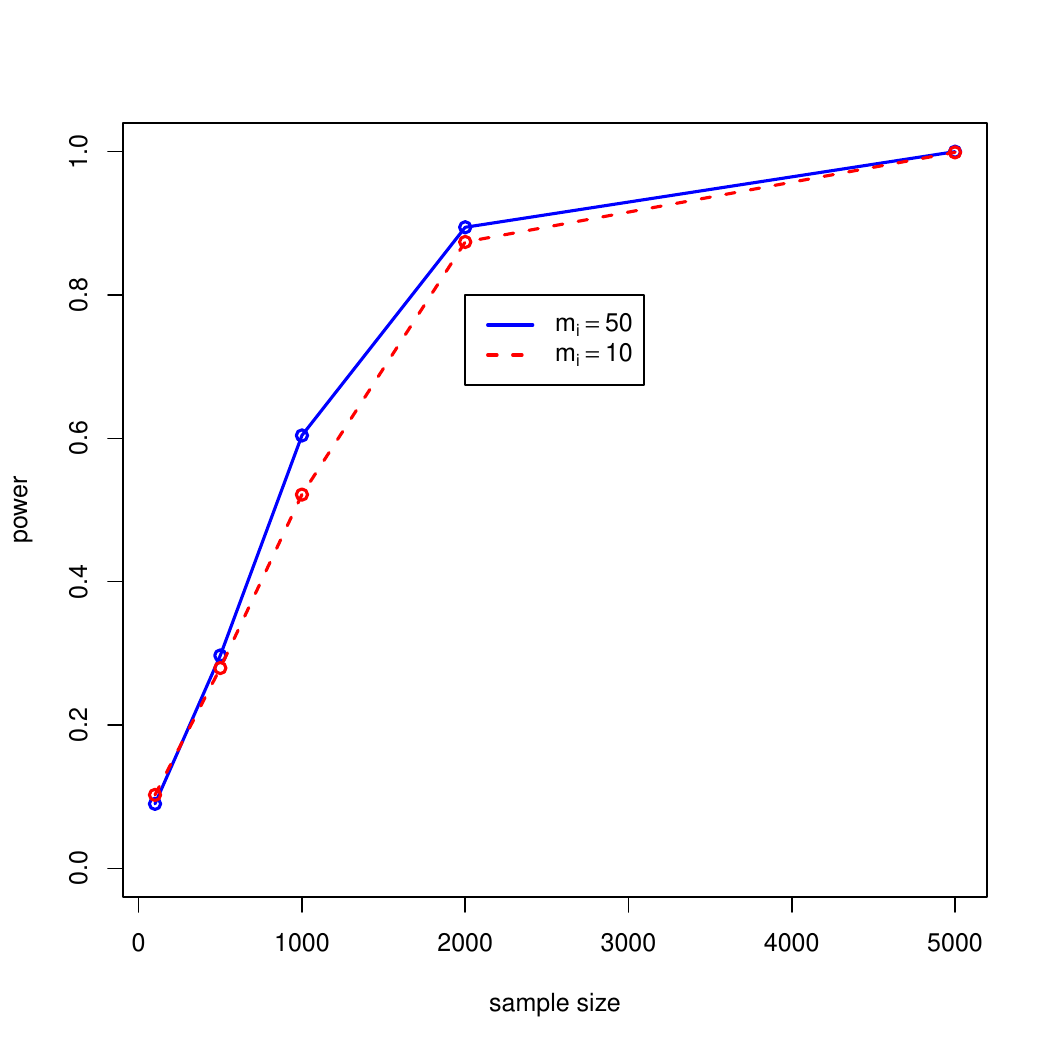}
& \includegraphics[trim = 20 30 20 50, clip, width = 0.3\textwidth]{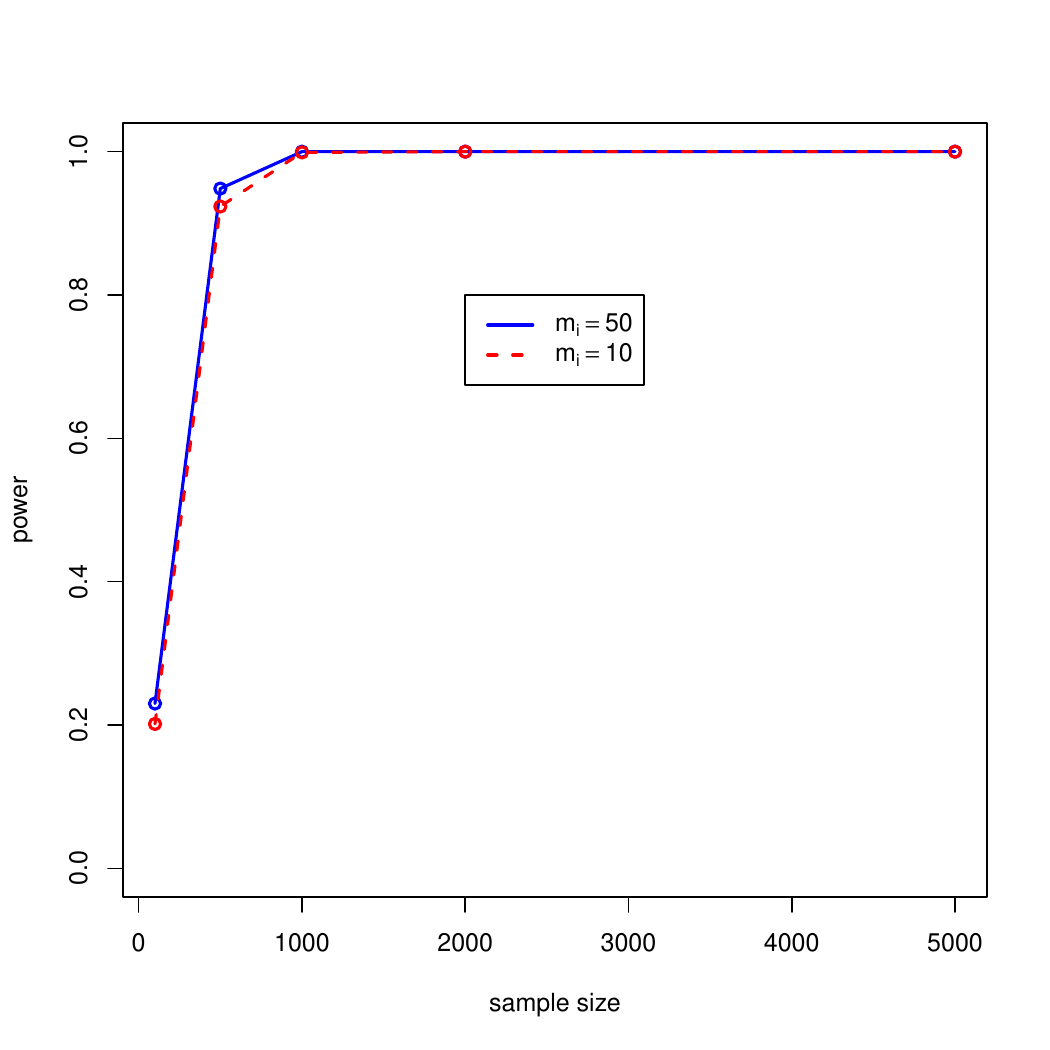}
& \includegraphics[trim = 20 30 20 50, clip, width = 0.3\textwidth]{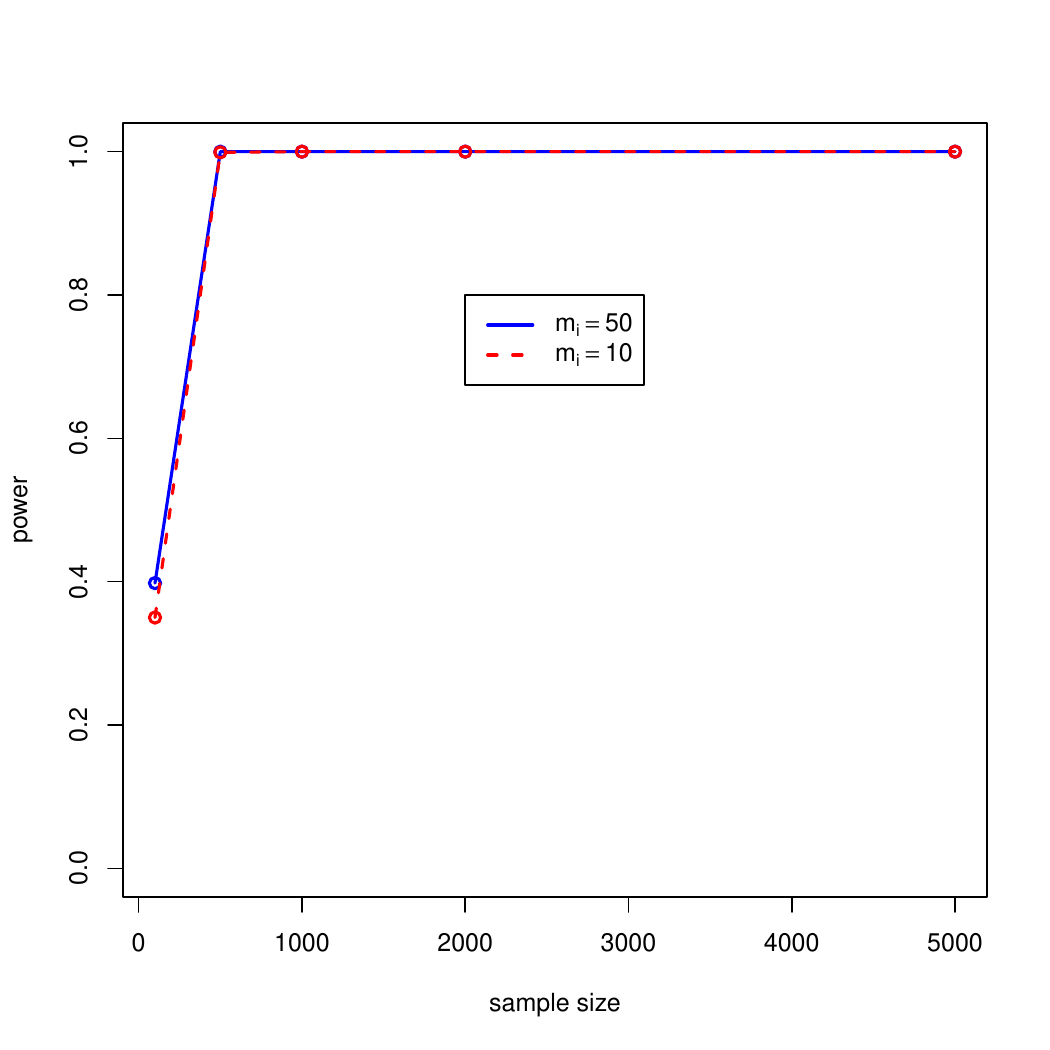} \\
\begin{sideways} \rule[0pt]{0.4in}{0pt} $\sigma = 1$ and $\tau \in \mathcal{U}_2$ \end{sideways}
& \includegraphics[trim = 20 30 20 50, clip, width = 0.3\textwidth]{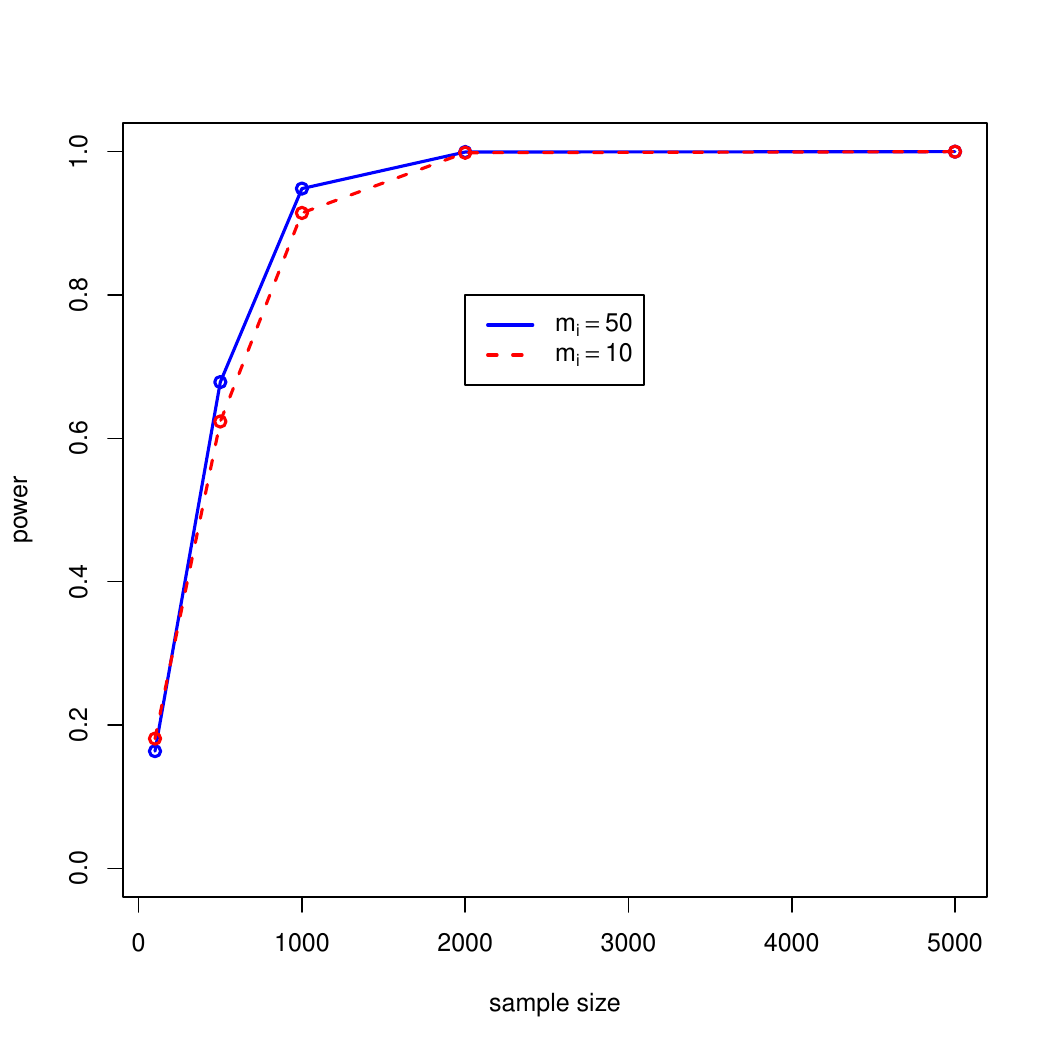}
& \includegraphics[trim = 20 30 20 50, clip, width = 0.3\textwidth]{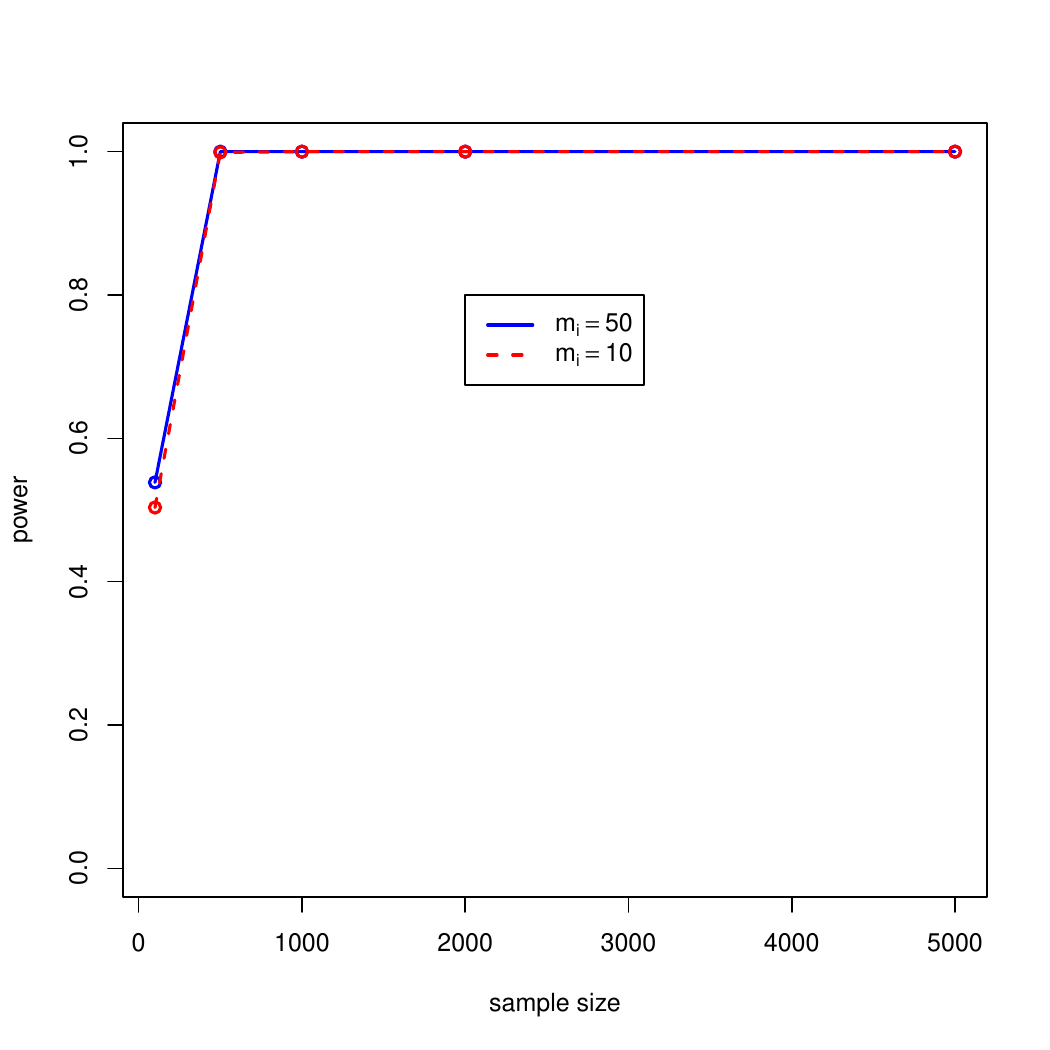}
& \includegraphics[trim = 20 30 20 50, clip, width = 0.3\textwidth]{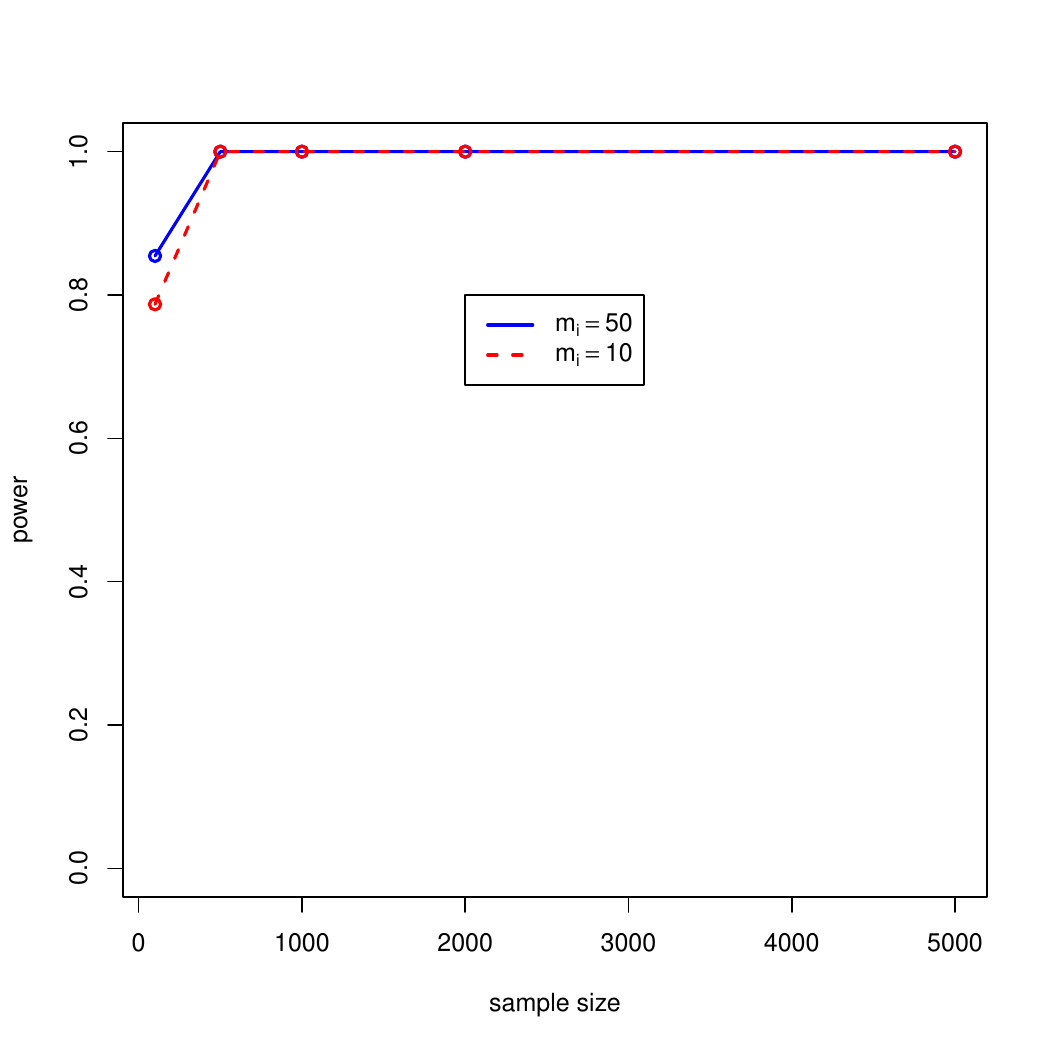} \\
& $\gamma = 0.5$ & $\gamma = 1$ & $\gamma = 1.5$ 
\end{tabular} 
\caption{Power curves of the adjusted Wald test for moderately sparse design with $m_i = 50$ (blue) and highly sparse design with $m_i = 10$ (red). We test $H_0$ at two sets of quantile levels: $\mathcal{U}_1 = \{0.1, 0.2, 0.3, 0.4\}$ and $\mathcal{U}_2 = \{0.1, 0.2, 0.6, 0.7\}$. The $x$-axis is the sample size $n \in \{100, 500, 1000, 2000, 5000\}$. Results are based on 2000 simulations.} 
\label{figure:power.sparse}
\end{figure}

Table~\ref{table:sparse.type.I} shows the empirical Type I error when the noise level $\sigma=1$. They show excellent performance of the adjusted Wald test in maintaining the nominal levels for moderately large sample size ($n=1000$ or larger) under both moderately sparse and sparse sampling design of the functional covariate. Fig.~\ref{figure:power.sparse} shows the power of the adjusted Wald test for moderately sparse and highly sparse designs for $\sigma=1$. It indicates that the sparsity of the functional covariates slightly affects the proposed functional Wald-type procedure, as expected. Nevertheless the adjusted Wald test continues to display excellent performance. The results are similar for low level of measurement error and for brevity are omitted here.

\renewcommand{\arraystretch}{1}
\begin{table}
\centering
\caption{Type I error of the adjusted Wald test at significance level $\alpha \in \{0.01, 0.05, 0.10\}$ under sparse design. We test $H_0$ at two sets of quantile levels: $\mathcal{U}_1 =
\{0.1, 0.2, 0.3, 0.4\}$ and $\mathcal{U}_2 = \{0.1, 0.2, 0.6, 0.7\}$. The missing rate is 50\% for moderate sparsity and 90\% for high sparsity. Results are based on 5000 simulations. }
\label{table:sparse.type.I}
\hrule
\begin{tabular}{cccccccc}
& &  \multicolumn{3}{c}{missing rate = 50\%}  &  \multicolumn{3}{c}{missing rate = 90\%}  \\ 
Scenario  & $n$ & 0.01 & 0.05 & 0.10 & 0.01 & 0.05 & 0.10 \\ \cmidrule(lr){3-5} \cmidrule(lr){6-8} 
   & 100 & 0.021  & 0.063  & 0.104  & 0.024  & 0.075  & 0.119  \\ 
$\sigma = 1$ &   500 & 0.014  & 0.055  & 0.104  & 0.011  & 0.058  & 0.110  \\ 
$ \tau \in \mathcal{U}_1$ &    1000 & 0.014  & 0.055  & 0.106  & 0.013  & 0.052  & 0.101  \\ 
  &  2000 & 0.011  & 0.055  & 0.106  & 0.013  & 0.053  & 0.103  \\ 
  &  5000 & 0.011  & 0.052  & 0.100  & 0.010  & 0.048  & 0.100  \\ 
     \\
&   100 & 0.026  & 0.075  & 0.120  & 0.034  & 0.092  & 0.143  \\ 
$\sigma = 1$ &  500 & 0.016  & 0.058  & 0.110  & 0.021  & 0.069  & 0.119  \\ 
$ \tau \in \mathcal{U}_2$ &   1000 & 0.013  & 0.057  & 0.106  & 0.014  & 0.063  & 0.114  \\ 
&  2000 & 0.011  & 0.053  & 0.100  & 0.011  & 0.056  & 0.108  \\ 
&  5000 & 0.010  & 0.048  & 0.103  & 0.011  & 0.049  & 0.100  \\ 
\end{tabular}
\hrule 
\end{table} 

\subsection{Divergent $p_0$}
\label{section:large.p.0} 
In this section, we study the performance of the proposed adjusted Wald test when Assumption A4 is violated.  We follow the same settings in Section~\ref{section:setting} but the eigen values and eigenfunctions to generate the functional covariate are given by
$\lambda_k = (1/2)^{k - 1}$ for $k \in \{1, \ldots, \lfloor\sqrt{n} \rfloor\}$ and $\lambda_k = 0$ for $k > \lfloor\sqrt{n} \rfloor$ where $\lfloor \cdot \rfloor$ is the floor function; the eigen function $\phi_k$ is the $k$th function in the Fourier basis $\{\sqrt{2} \cos(2 \pi t), \sqrt{2} \sin(2 \pi t), \sqrt{2} \cos( 4 \pi t), \sqrt{2} \sin(4 \pi t), \ldots \}$. We set $\sigma = 1$ for the measurement error in the functional covariate. 

\begin{figure}[h!]
	\begin{tabular}{cccc}
		\begin{sideways} \rule[0pt]{0.7in}{0pt} $\tau \in \mathcal{U}_1$ \end{sideways}
		& \includegraphics[trim = 20 30 20 50, clip, width = 0.3\textwidth]{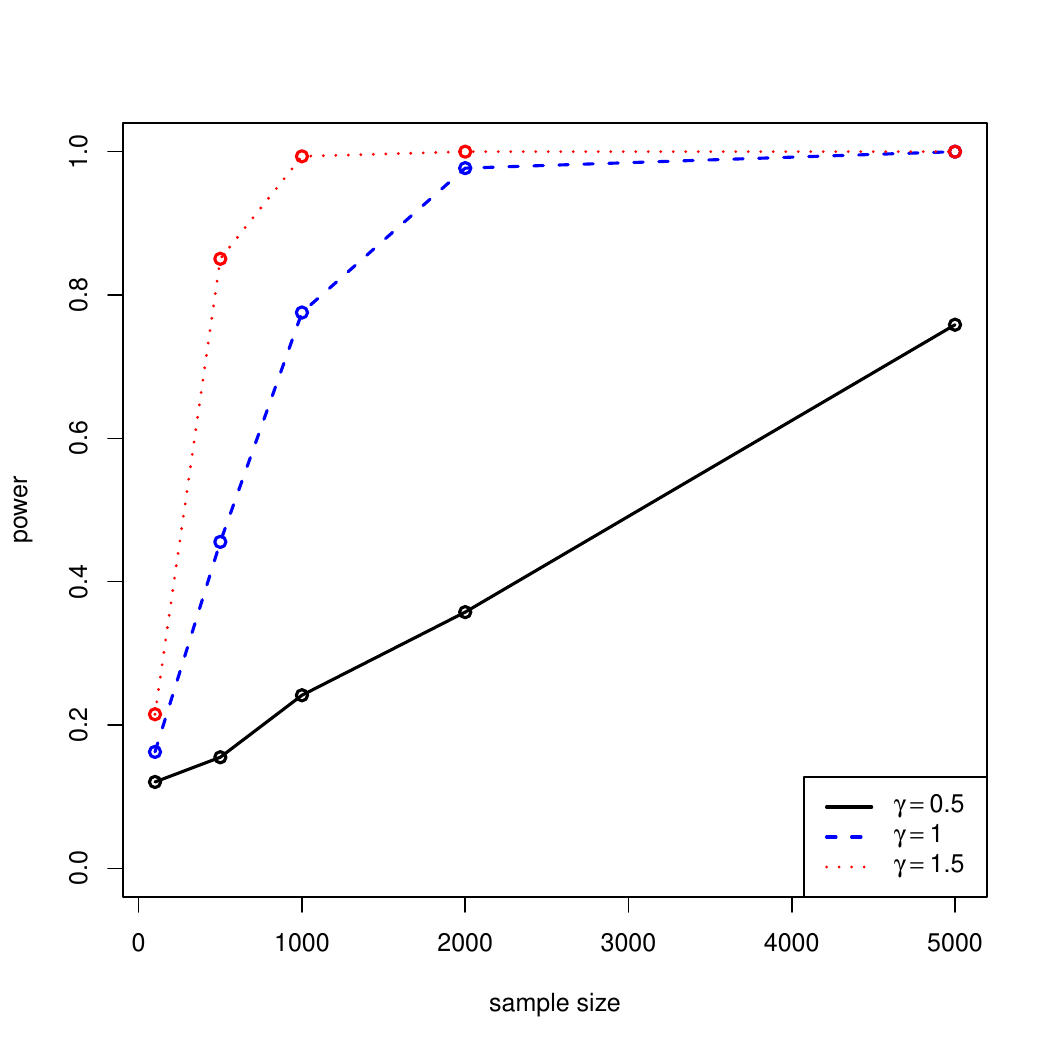}
		& \includegraphics[trim = 20 30 20 50, clip, width = 0.3\textwidth]{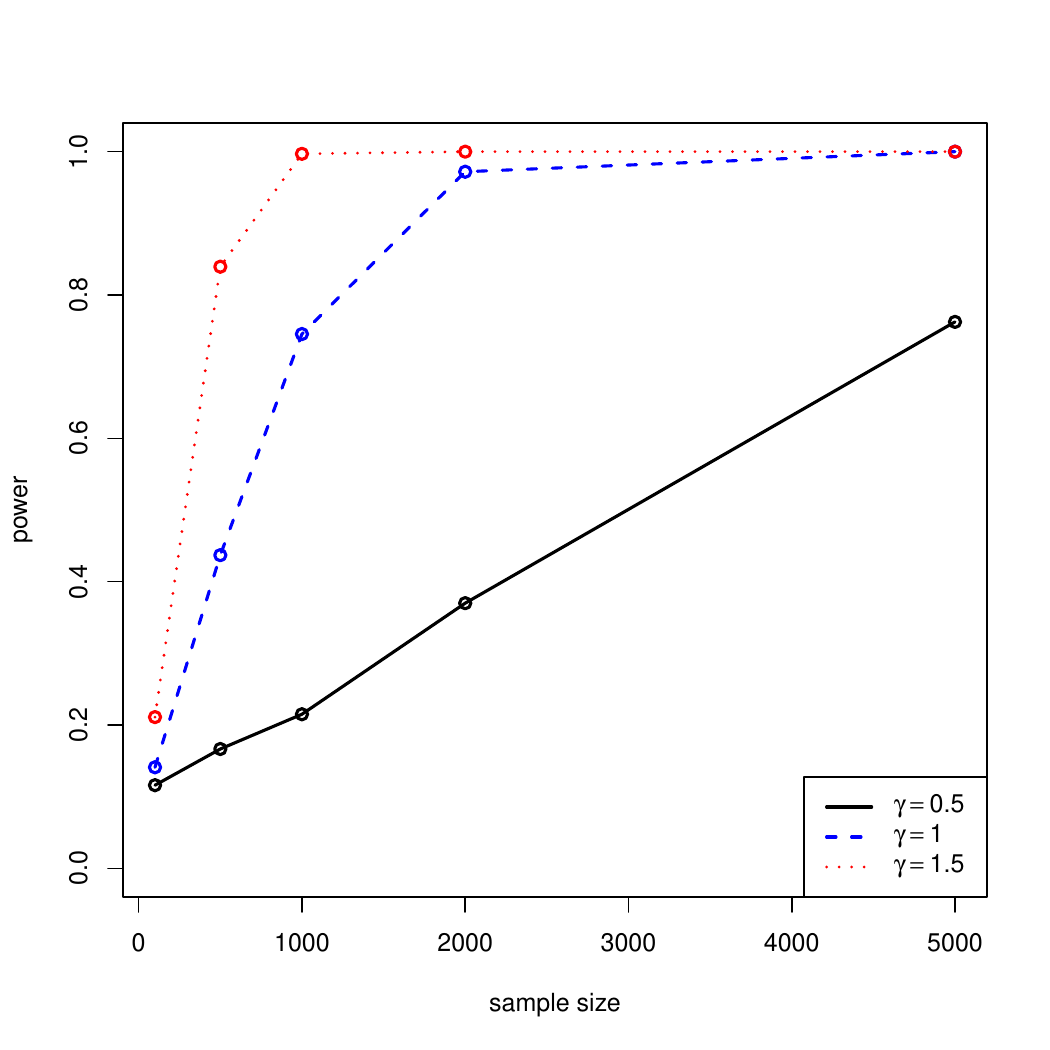}
		& \includegraphics[trim = 20 30 20 50, clip, width = 0.3\textwidth]{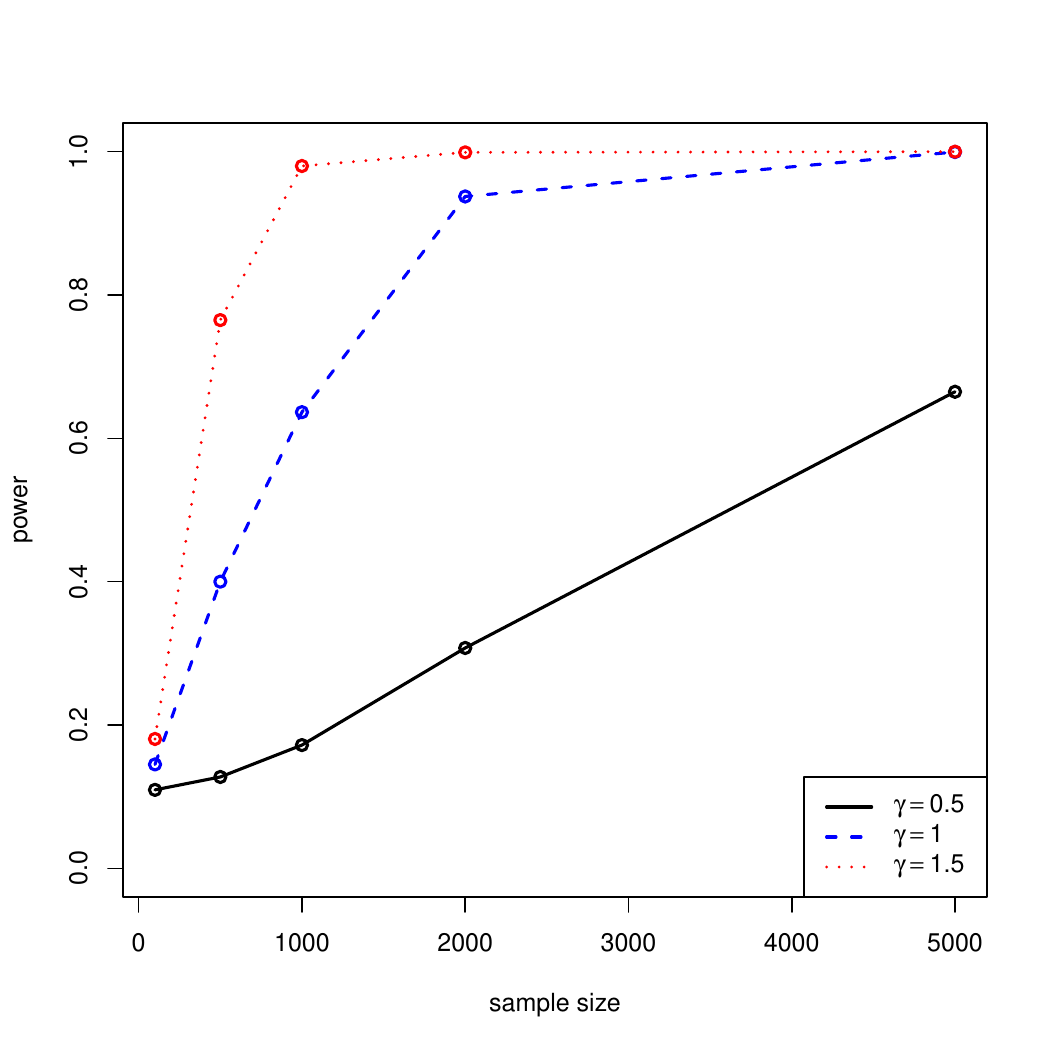} \\
		\begin{sideways} \rule[0pt]{0.7in}{0pt} $\tau \in \mathcal{U}_2$ \end{sideways}
		& \includegraphics[trim = 20 30 20 50, clip, width = 0.3\textwidth]{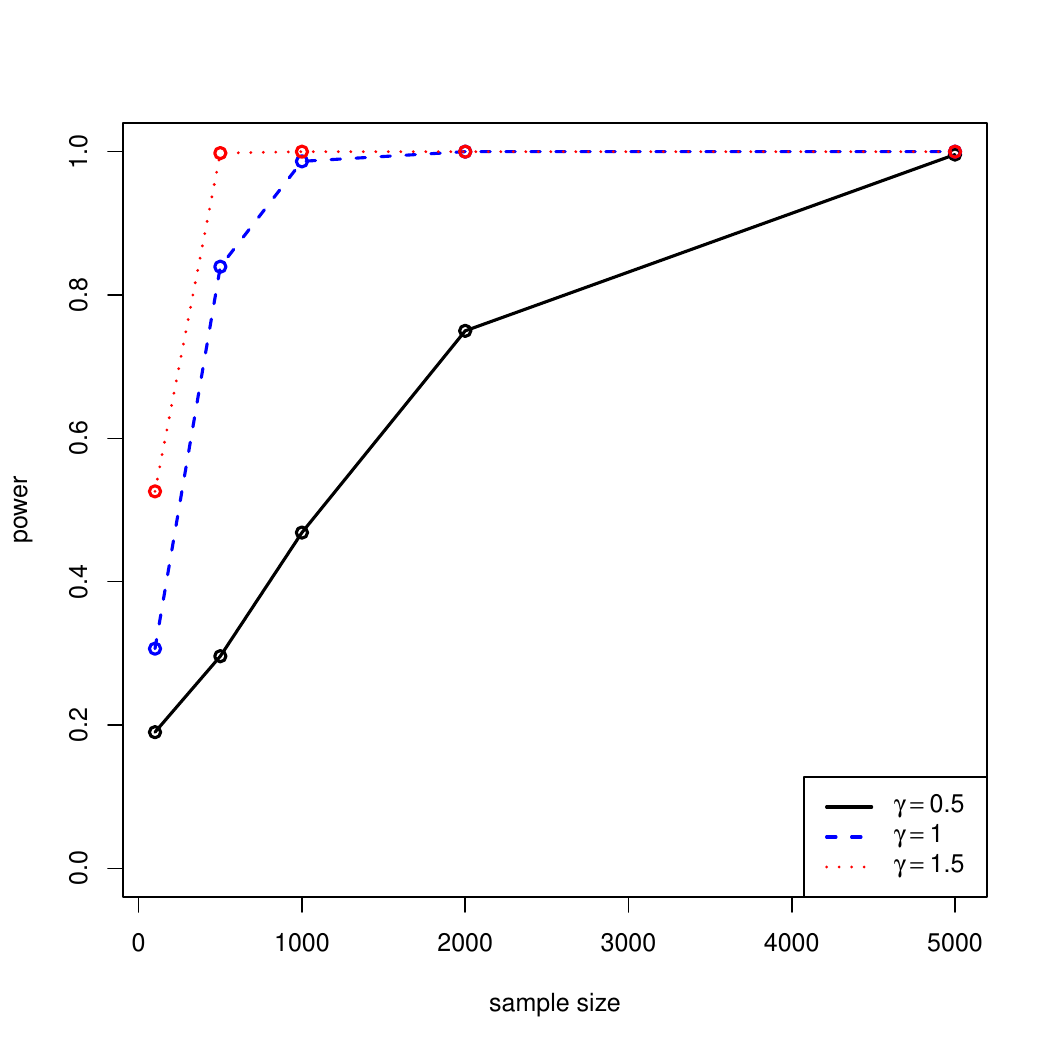}
		& \includegraphics[trim = 20 30 20 50, clip, width = 0.3\textwidth]{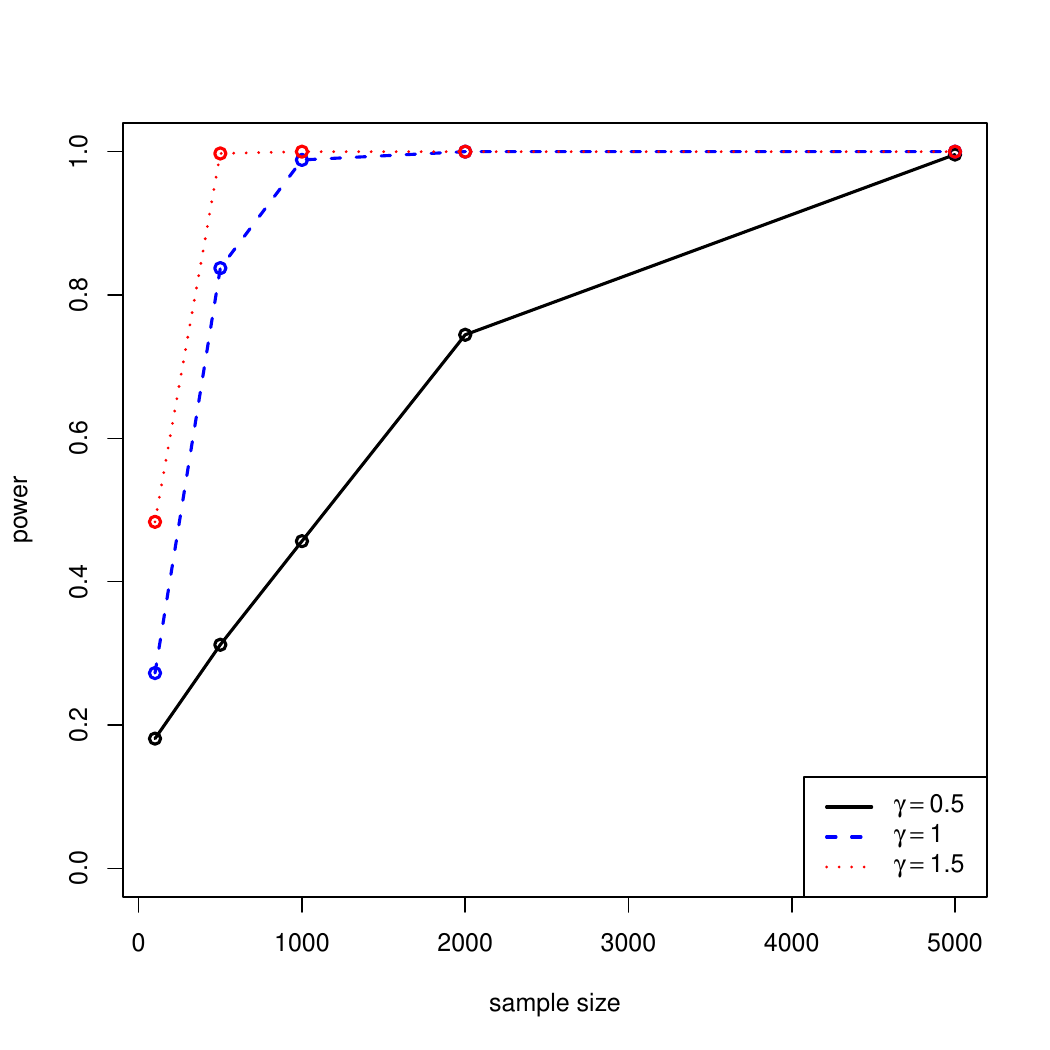}
		& \includegraphics[trim = 20 30 20 50, clip, width = 0.3\textwidth]{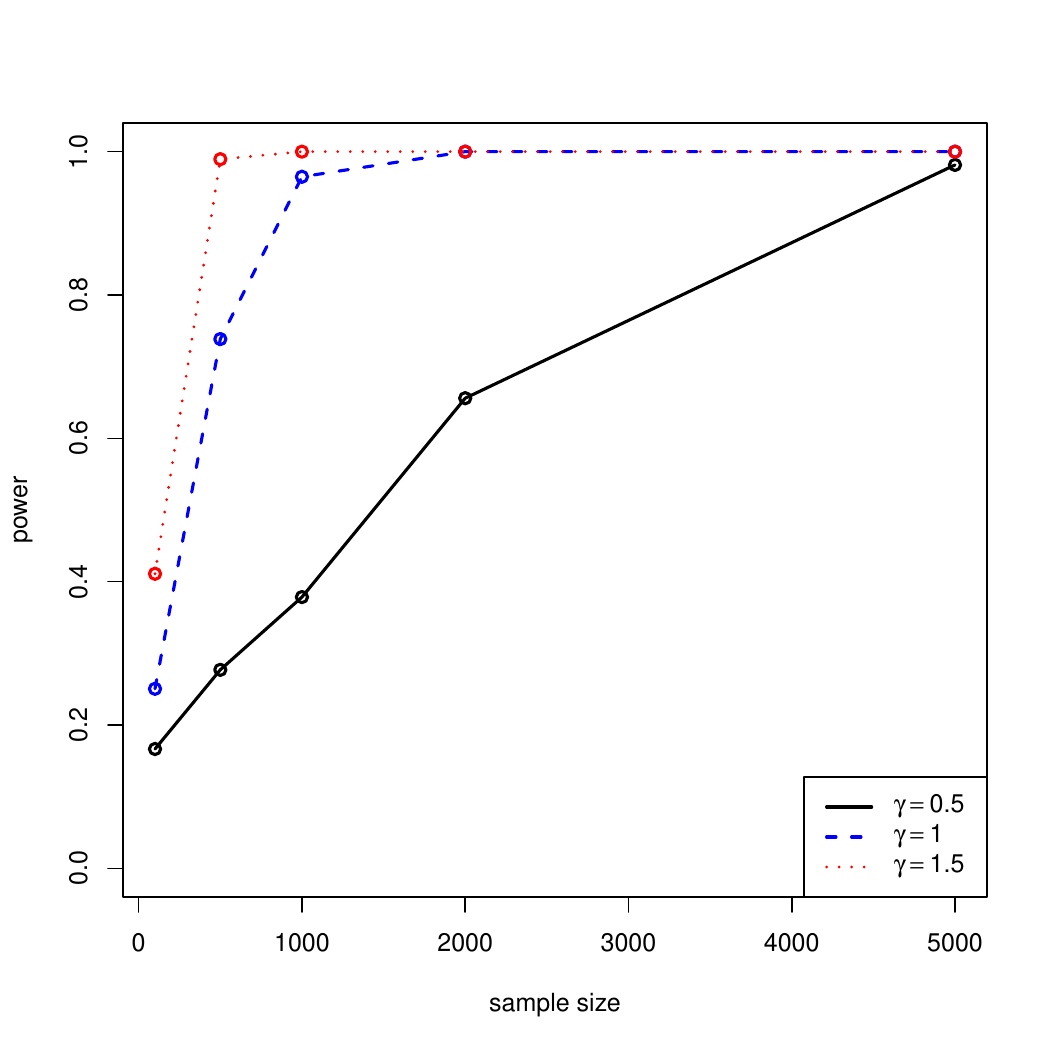} \\
		& No missing&  Missing rate = 50\% &  Missing rate = 90\% 
	\end{tabular} 
	\caption{Power curves of the adjusted Wald test when $p_0$ is divergent. We test $H_0$ at two sets of quantile levels: $\mathcal{U}_1 = \{0.1, 0.2, 0.3, 0.4\}$ and $\mathcal{U}_2 = \{0.1, 0.2, 0.6, 0.7\}$. In each plot, the $x$-axis is the sample size $n \in \{100, 500, 1000, 2000, 5000\}$, and a larger $\gamma$ corresponds to more deviation from the null hypothesis. Results are based on 2000 simulations.} 
	\label{figure:power.r2}
\end{figure}

Table~\ref{table:typeI.r2} presents the Type I error rates of the adjusted Wald test under various designs. We can see that even Assumption A4 is violated, the proposed test matches the nominal level when the sample size is large, for both dense and sparse designs. Fig.~\ref{figure:power.r2} plots the power curves when $\gamma \in \{0.5, 1, 1.5\}$, which indicates similar performance to the case where $p_0$ is a small constant. Therefore, it seems that the proposed Wald test continues to show desirable performance when Assumption A4 does not hold, at least under the simulation settings. A theoretical justification may be an interesting research topic.

\begin{table}[ht]
	\centering
	\caption{Type I error of the adjusted Wald test at significance level $\alpha \in \{0.01, 0.05, 0.10\}$ when $p_0$ is divergent under dense design (no missing) and sparse design. We test $H_0$ at two sets of quantile levels: $\mathcal{U}_1 =
\{0.1, 0.2, 0.3, 0.4\}$ and $\mathcal{U}_2 = \{0.1, 0.2, 0.6, 0.7\}$. The missing rate is 50\% for moderate sparsity and 90\% for high sparsity.  Results are based on 5000 simulations. 
	} 
	\label{table:typeI.r2} 
	\hrule
	\begin{tabular}{ccccccccccc}
		$\mathcal{U}$  & $n$ & \multicolumn{3}{c}{no missing} & \multicolumn{3}{c}{missing rate = 50\%}  &  \multicolumn{3}{c}{missing rate = 90\%}  \\ 
		& & 0.01 & 0.05 & 0.10 & 0.01 & 0.05 & 0.10 & 0.01 & 0.05 & 0.10\\ 
	\cmidrule(lr){3-5} \cmidrule(lr){6-8} \cmidrule(lr){9-11}
		\multirow{5}{*}{$\mathcal{U}_1$} 
		& 100 & 0.045 & 0.107 & 0.161 & 0.049 & 0.107 & 0.162 & 0.040 & 0.096 & 0.147 \\ 
		& 500 & 0.025 & 0.085 & 0.146 & 0.022 & 0.081 & 0.141 & 0.026 & 0.084 & 0.137 \\ 
		& 1000 & 0.015 & 0.067 & 0.123 & 0.019 & 0.072 & 0.132 & 0.013 & 0.062 & 0.120 \\ 
		& 2000 & 0.015 & 0.063 & 0.120 & 0.016 & 0.062 & 0.115 & 0.015 & 0.063 & 0.119 \\ 
		& 5000 & 0.010 & 0.052 & 0.107 & 0.010 & 0.059 & 0.117 & 0.012 & 0.055 & 0.101 \\ 
		\\
		\multirow{5}{*}{$\mathcal{U}_2$} 
		& 100 & 0.070 & 0.149 & 0.222 & 0.061 & 0.142 & 0.212 & 0.051 & 0.125 & 0.193 \\ 
		& 500 & 0.035 & 0.105 & 0.169 & 0.037 & 0.109 & 0.174 & 0.025 & 0.087 & 0.148 \\ 
		& 1000 & 0.025 & 0.084 & 0.148 & 0.022 & 0.079 & 0.140 & 0.024 & 0.077 & 0.132 \\ 
		& 2000 & 0.020 & 0.072 & 0.127 & 0.016 & 0.067 & 0.123 & 0.015 & 0.063 & 0.123 \\ 
		& 5000 & 0.013 & 0.056 & 0.111 & 0.014 & 0.058 & 0.104 & 0.012 & 0.057 & 0.110 \\
	\end{tabular}
	\hrule 
\end{table}

\section{Application}
\label{section:application}
\begin{figure}
\centering 
\begin{tabular}{cc}
	\includegraphics[trim = 20 20 0 0, clip = True, width = 0.42 \textwidth]{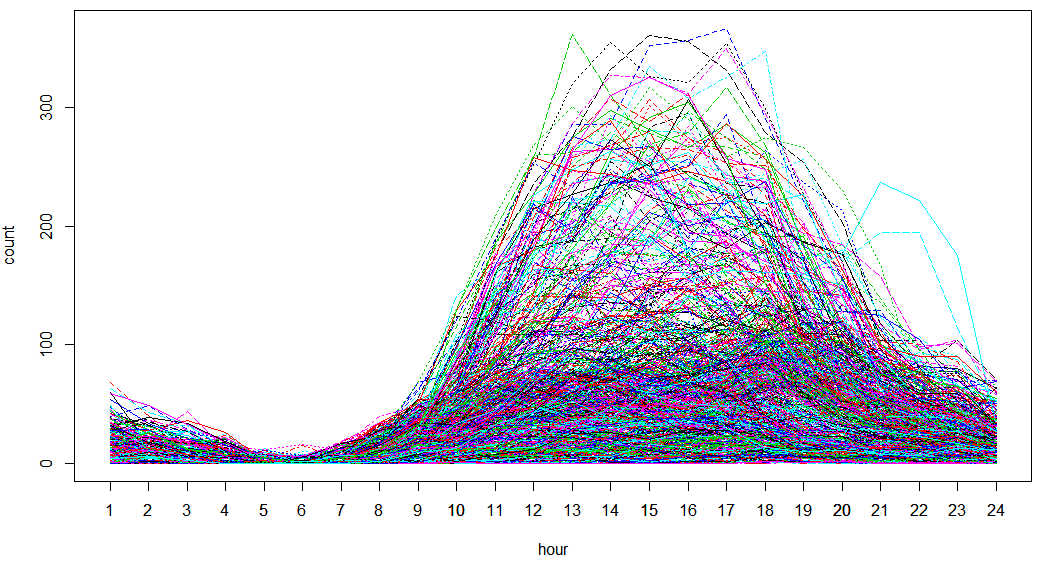} & 
	\includegraphics[trim = 0 40 0 52, clip = True, width = 0.5\textwidth]{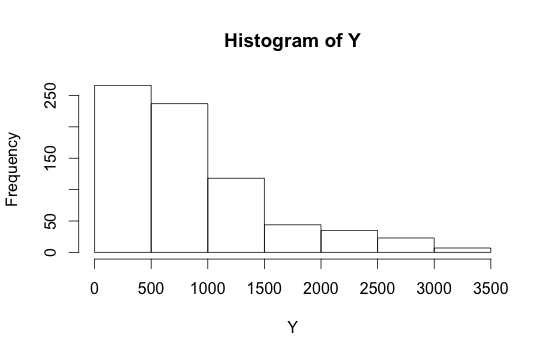} \\
	(a) Functional covariates $X_i(\cdot)$ & (b) Histogram of $y_i$
\end{tabular}
\caption{Bike rental data (casual users). The left panel plots hourly bike rentals for casual users on the previous day $X_i(t)$ for $t$ ranges from 0 to 24 hours, and the right panel plots the histogram of the total casual bike rentals on the current day $y_i$, where $i \in \{1, \ldots, 730\}$. } 
\label{figure:x} 
\end{figure}

In this section we consider the capital bike sharing study and discuss the application of the proposed testing procedure to formally assess whether the effect of the previous day casual bike rentals on the current day total bike rentals varies across several quantile levels. The bike data~\citep{Fan+Gam:14} is recorded by the Capital Bikeshare System (CBS), Washington D.C., USA, which is available at \url{http://capitalbikeshare.com/system-data}. 
As the new generation of bike rentals, bike sharing systems possess membership, rental and return automatically. With currently over 500 bike-share programs around the world~\citep{Larsen:13} and the fast growing trend, data analysis on these systems regarding the effects to public traffic and the environment has become popular. {\color{black}{The bike data includes hourly rented bikes for casual users that are collected during January 1st 2011 to December 31st 2012, for a total of 731 days. 

Our objective is to formally assess how the previous day casual bike rentals, $X_i(\cdot)$, affects the distribution 
of the current day total bike rentals counts, $Y_i$, where $i \in \{1, \ldots, 730\}$ denote the $i$th day starting from January 2nd 2011.}} A subsequent interest is to predict the $90\%$ quantile of the total casual bike rentals. Fig.~\ref{figure:x} plots the hourly profiles of casual bike rentals (left) and the histogram of the total casual bike rentals (right). 

We assume the functional quantile regression model (\ref{model}), $Q_{Y_i|X_i}(\tau) = \beta_0(\tau) + \int \beta(t, \tau) X_i^c(t) \, dt$, where $Y_i$ is the total bike casual bike rentals for the current day and $X_i(\cdot)$ is the true profile of the casual bike rentals recorded in the previous day. As described earlier $\beta_0(\cdot)$ is the quantile varying intercept function and $\beta(\cdot, \tau)$ is the slope parameter and quantifies the effect of the functional covariate at the $\tau$th quantile level of the distribution of the response. 

To address the first objective we consider a set of quantile levels and use the proposed testing procedure to test the null hypothesis
$$
H_0: \beta(\cdot, 0.20)=  \beta(\cdot, 0.40)=  \beta(\cdot, 0.60)= \beta(\cdot, 0.80).
$$

The number of fPC is selected using PVE = $99 \%$; this choice selects three fPC. We use the adjusted Wald test $T_n$ and its asymptotic null distribution; the resulting $p$-value is close to zero indicating overwhelming evidence that low and large number of bike rentals are affected differently by the hourly rentals on the previous day.

\begin{figure}
\centering 
\includegraphics[trim = 0 20 0 0, clip = True, width = 0.8\textwidth]{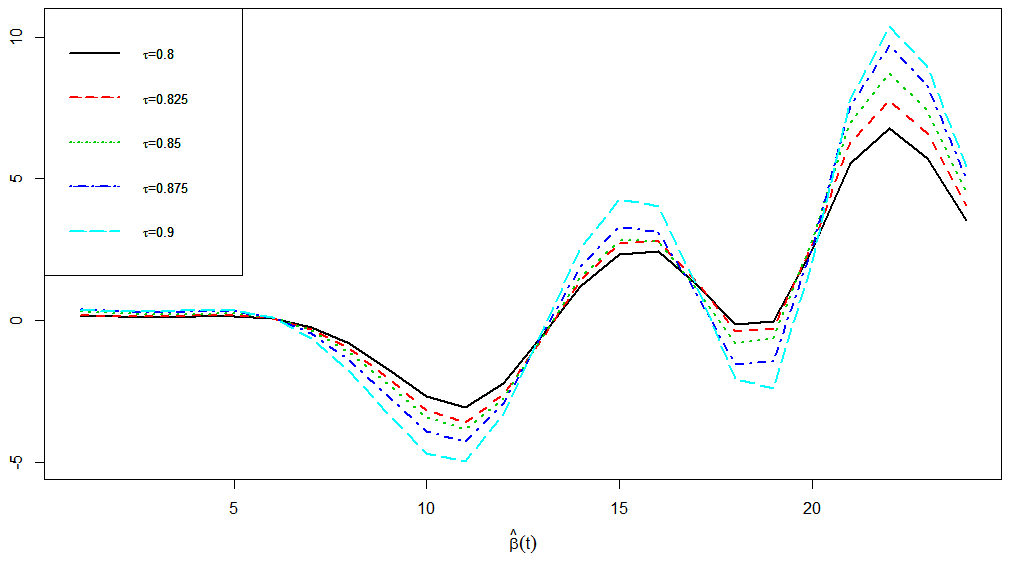}
\caption{Estimated ${\beta}(\cdot, \tau)$ by the proposed method at various quantile levels for the capital bike sharing study. The $x$-axis ranges from 0 to 24 hours. }
\label{figure:beta.hat} 
\end{figure}

Next we turn to the problem of predicting the $90\%$ quantile of the total bike rentals for the current day. When some quantile coefficients in a region of quantile levels are constant, we may improve the estimator's efficiency by borrowing information from neighboring quantiles to estimate the common coefficients, especially when the quantile level of interest is high. Here consider the quantile level set $\mathcal{U} = \{0.8, 0.825, 0.85, 0.875, 0.9\}$ around the $90\%$th quantile. We apply the proposed method to estimate the coefficient functions at various quantile levels $\mathcal{U}$ as shown in Fig.~\ref{figure:beta.hat}. The corresponding adjusted Wald test leads to a $p$-value = 0.466,  which suggests that the quantile coefficients are not significantly different across the quantile levels. We consider combined quantile regression at $\mathcal{U}$ by using the methods of quantile average estimator (QAE) and composite regression of quantiles (CRQ) with equal weights; see~\cite{koenker1984,kehuisinica} for more technical details. We denote the single quantile regression estimation at the 90th quantile by RQ.

\begin{figure}
	\centering
	\begin{tabular}{cc}
	\includegraphics[trim = 0 35 0 50, clip = True, scale = 0.45]{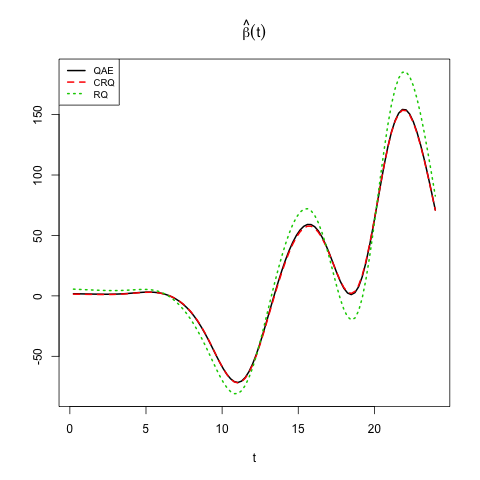} & \includegraphics[trim = 0 35 0 50, clip = True, scale = 0.45]{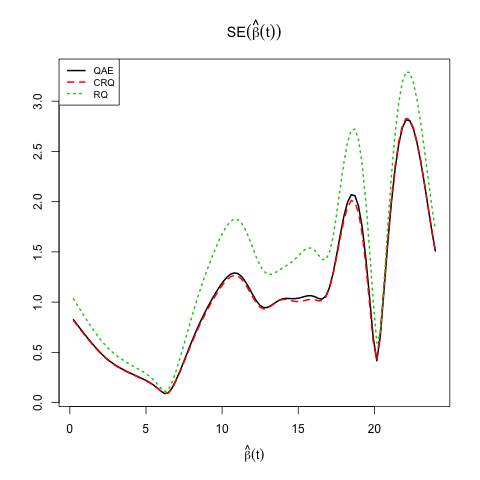} \\
	(a) Bootstrap mean & (b) Bootstrap standard error 
\end{tabular}
	\caption{Bootstrap means (left) and standard errors (right) when estimating $\beta(\cdot, 0.9)$ by QAE, CRQ, and RQ. QAE and CRQ reduce the standard error of RQ. The $x$-axis ranges from 0 to 24 hours.} 
	\label{fig:bs.data}
\end{figure}

We use $1000$ bootstrap samples to study the efficiency of the three estimators. Fig.~\ref{fig:bs.data} plots the bootstrap means and standard errors of the estimates of $\beta(\cdot, 0.9)$ by QAE, CRQ and RQ. The QAE and CRQ estimators have smaller standard errors uniformly for all $t$, indicating efficiency gain by combining information across quantile levels. We also observe that the number of fPC is either 3 or 4 in all bootstrap samples, suggesting that the assumption A4 is reasonable in this data application.

Furthermore, we conduct a cross-validation by randomly selecting $50\%$ of the data as the training data set and using the other half as the test data set. We use $1000$ replications and calculate the prediction error for each replication and each $\tau \in \mathcal{U}$ as follows: 
\begin{equation}
\label{eq:bike.pe}
  \mathrm{PE} = \sum_{i \in \text{ test sample}} \rho_{\tau}(y_i -
  \widehat{\bxi}_{i}^T \widehat{{\btheta}}_{\tau}),
\end{equation}
where the estimated coefficients $\widehat{{\btheta}}_{\tau}$ are based on the training data and the summation is over the test data. The RQ estimates are obtained separately at each $\tau \in \mathcal{U}$, while the QAE and CRQ estimates are shared across $\mathcal{U}$. 
The averaged prediction errors are reported in Table~\ref{table:cv}. We can see that the application of QAE and CRQ improves the prediction significantly for the $87.5\%$th and $90\%$th quantiles; differences among the three methods are not significant at the lower quantiles. This makes sense since data sparsity becomes more severe for more extreme quantile levels. Hence, incorporating lower quantile levels improves efficiency at higher levels, while it may not benefit the prediction performance at lower quantile levels by considering more extreme levels.

\begin{table}
\centering
\caption{Prediction errors from different methods averaged over 1000 cross-validations. The maximum standard error of each row is reported in the last column. QAE and CRQ that combine information at various quantile levels tend to yield smaller prediction errors than RQ at more extreme quantile levels. }
\label{table:cv} 
\hrule 
\begin{tabular}{ccccc}
 $\tau$ & QAE & CRQ & RQ & SE \\ 
0.8 & 154.163 & 153.073 & 152.396 & 0.277\\ 
  0.825 & 146.163 & 145.598 & 145.504 & 0.268\\ 
  0.85 & 137.028 & 136.758 & 137.071 & 0.259\\ 
  0.875 & 126.138 & 125.949 & 126.819 & 0.252\\ 
  0.9 & 112.774 & 112.842 & 113.823 & 0.238\\ 
\end{tabular} \hrule 
\end{table}

\section{Proofs} \label{section:proof}

In this section, we prove Theorem~\ref{th:asy.normal} and Theorem~\ref{th:wald.p0}, as well as auxiliary results needed in the proofs, including Lemma~\ref{lemma:z.hat.z}, Lemma~\ref{th:z1n.represent}, and Lemma~\ref{lemma:z2n}. We use $\|\cdot\|_{L^2}$ as the $L^2$-norm for a function and $\|\cdot\|$ as the Euclidean norm for a vector. 

\subsection{Proofs of Theorem~\ref{th:asy.normal} and Theorem~\ref{th:wald.p0}}
\begin{proof}[\textbf{\upshape Proof of Theorem~\ref{th:asy.normal}.}] 
The proof proceeds in three steps. In step 1, we approximate the estimated scores $\widehat{\bxi}_i$'s by linear combinations of $\bxi_i$'s. In step 2, we obtain the asymptotic distribution of $\widehat{\theta}_{\tau}$ at a single quantile level. In step 3, we extend the results in step 2 to multiple quantile levels. 

\noindent {Step 1 (Approximation of the estimated scores).} 
Most of the existing literature has been focused on establishing error bounds for estimated eigenvalues and eigenfunctions; see for example~\cite{Hal+Hos:06, Hal:09} and the discussion therein. The following lemma instead characterizes the accuracy in predicting the fPC scores.  

\begin{lemma}
	\label{lemma:z.hat.z}
	Under Assumptions A4, B1, B2 and C1, we have 
	\begin{equation}
	\label{eq:z.hat.z.expectation}
	\text{E} \|\widehat{\bxi}_i - \bxi_{i} \|^2  = o(n^{-1/2}). 
	\end{equation}
	\color{black}
	In addition, 
	\begin{equation}
	\label{eq:represent.difference}
	\underset{1 \leq i \leq n}{\max} \left| \widehat{\bxi}_i - \bxi_{i} - n^{-1/2} B \bxi_{i} \right| = O_p(n^{-1}),
	\end{equation}
	\color{black}
	where $\bB$ is a $(p_0 + 1)\times(p_0 + 1)$ dimensional matrix with the bottom right $p_0\times p_0$ block matrix equal to $\bB^+$ described next and the rest of the elements equal to zero. Here $\bB^+ = (b_{kk'})$ is a $p_0 \times p_0$ random matrix such that $b_{kk} = 0$ for $k \in \{1, \ldots, p_0\}$ and $b_{kk'} =  n^{-1/2}(\lambda_k - \lambda_{k'})^{-1}  \left(\sum_{i = 1}^n \xi_{ik} \xi_{ik'}\right)$ if $k \neq k'$.  
\end{lemma}

The result in~\eqref{eq:represent.difference} indicates that the leading term of $\widehat{\bxi}_i - \bxi_{i}$ is $n^{-1/2} \bB \bxi_{i}$, which is a linear combination of $\bxi_i$ with a random weight matrix $\bB$ that does not depend on $i$. 

\noindent {Step 2 (Quantile regression on estimated scores).} 
We focus on a single quantile level $\tau$ in this step. For any ${\delta} \in \mathbb{R}^{p_0 + 1}$, let 
\begin{equation}
\label{eq:Zn}
Z_n({\delta}) = \sum_{i = 1}^n \{ \rho_\tau(\widehat{u}_i - \widehat{\bxi}_i^T {\delta}/\sqrt{n}) - \rho_\tau(\widehat{u}_i) \},
\end{equation}
where $\widehat{u}_i = y_i - \widehat{\bxi}_i^T \btheta_{\tau}$. Then $Z_n({\delta})$ is a convex function which is minimized at $\widehat{{\delta}}_n =
\sqrt{n}(\widehat{\btheta}_{\tau}- \btheta_{\tau})$. Therefore, the asymptotic distribution of $\widehat{{\delta}}_n$ is determined by the limiting behavior of $Z_n({\delta})$. 
Let $\psi_\tau(t)=\tau - I(t < 0)$. According to the Knight's identity  \citep{knight}, we can decompose $Z_n({\delta})$ into two parts: $Z_n({\delta}) = Z_{1n}({\delta}) + Z_{2n}({\delta})$, where 
\begin{equation}
\label{eq:z1n}
Z_{1n}({\delta}) = -\frac{1}{\sqrt{n}} \sum_{i = 1}^n \widehat{\bxi}_i^T {\delta} \psi_\tau(\widehat{u}_i), \quad \quad 
Z_{2n}({\delta}) = \sum_{i = 1}^n \int_{0}^{ \widehat{\bxi}_i^T {\delta}/\sqrt{n}}\{ I(\widehat{u}_i \leq s) - I(\widehat{u}_i
\leq 0)\}ds = \sum_{i = 1}^n Z_{2ni}({\delta}).
\end{equation}
In order to show  \eqref{nine}, it is sufficient to prove that 
\begin{equation}
\label{eq:eq1.7.24}
Z_n({\delta}) \overset{d}{\rightarrow} -{\delta}^T W(\tau) + 
\frac{1}{2} {\delta}^T \bD_1 (\tau) {\delta},
\end{equation}
where $\bW(\tau) \sim N\left\{0, \tau(1 - \tau)\bD_0 + \bD_1(\tau) {\Sigma}_0(\tau) \bD_1(\tau)\right\}$, since one can apply the convexity lemma~\citep{pollard1991} to the quadratic form of $\delta$ in~\eqref{eq:eq1.7.24}. 

We next derive the limiting distributions of $Z_{1n}({\delta})$ and $Z_{2n}({\delta})$. For $Z_{1n}({\delta})$, similarly to its definition in~\eqref{eq:z1n}, we   
define $Z_{1n}^{*}({\delta})$ based on the true scores $\bxi_{i}$: 
\begin{equation}
\label{eq:z1n.star}
Z_{1n}^*({\delta}) = -\frac{1}{\sqrt{n}} \sum_{i = 1}^n \bxi_{i}^T {\delta} \psi_\tau({u}_i),
\end{equation}
where $u_ i = y_i - Q_{Y_i|X_i}(\tau) =  y_i - \bxi_{i}^T \btheta_{\tau} = y_i - \sum_{k = 0}^{p_0} \xi_{ik} \beta_k(\tau)$. 
By a direct application of the central limit theorem (CLT), we obtain that the asymptotic distribution of $Z_{1n}^*({\delta})$ is $N(0, \tau (1 - \tau) \delta^T D_0 \delta )$. However, when the predictors are estimated with errors, the difference $Z_{1n}({\delta}) - Z_{1n}^*({\delta})$ is non-negligible. Lemma \ref{th:z1n.represent} provides a representation of $Z_{1n}({\delta})$ by explicitly formulating this difference.
\begin{lemma}
	\label{th:z1n.represent}
	Under Assumptions A4, B1, B2 and C1, 
	\begin{equation}
	Z_{1n}({\delta}) = {\delta}^T \left[-\frac{1}{\sqrt{n}} \sum_{i = 1}^n \{ \bxi_{i} \psi_\tau({u}_i) - \bD_1(\tau) d_i \} \right] + o_p(1), 
	\end{equation}
	where ${d}_i = (0, d_{i1}, \ldots, d_{i p_0})^T$ and $d_{ik} = \sum_{r = 1, r \neq k}^{p_0} (\lambda_k - \lambda_r)^{-1} \xi_{ik} \xi_{ir} \beta_{r}(\tau), k \geq 1.$ 
\end{lemma}

Since $\bxi_{i} \psi_{\tau}(u_i) - \bD_1 (\tau) {d}_i$ are i.i.d., Lemma~\ref{th:z1n.represent} allows us to directly apply Linderberg's CLT to obtain the asymptotic distribution of $Z_{1n}({\delta})$. Note that $\text{E}\{\bxi_{i} \psi_{\tau}(u_i)\} = 0$ and $\text{Var} \{\bxi_{i} \psi_{\tau}(u_i)\} = \tau(1 - \tau)\bD_0.$ In addition, $\text{E} {d}_i = {0}$ because $\xi_{ik}$ and $\xi_{ir}$ are uncorrelated and have mean 0 (when $r \neq k$). Let the matrix ${\Sigma}(\tau)$ be the covariance matrix of $d_i$ whose first row and first column is all 0 and the $(k + 1, k' + 1)$th element ($k, k' = 1, \ldots, p_0$) is given by $\text{Cov}(d_{ik}, d_{ik'}) = \theta_{\tau}^T A^{k, k'} \theta_{\tau}$
for some $(p_0 + 1)$ by $(p_0 + 1)$ matrix $A^{k, k'}$.  The first row and first column of $A^{k, k'}$ are all 0, and simple calculation yields its bottom right block $A^{k, k', +} = (\sigma_{j, j'})$: 
\[
\sigma_{j, j'}  =\begin{cases}
0, & \text{ if } j  = k \text{ or }  j' = k',\\
\label{eq:covariance.inflat}
(\lambda_k - \lambda_j)^{-1}(\lambda_{k'} - \lambda_{j'})^{-1} \E(\xi_{1 k} \xi_{1 j} \xi_{1k'} \xi_{1 j'}), &  \text { otherwise. }
\end{cases}
\]
Let ${\Theta}_{\tau} = 1_{(p_0 + 1)\times (p_0 + 1)} \otimes \theta^T, $ and $\Sigma_0$ be a $(p_0 + 1)^2$ by $(p_0 + 1)^2$ matrix whose $(k + 1, k' + 1)$th block is $A^{k, k'}$ $(k, k' = 1, \ldots, p_0)$ and $(k + 1, k' + 1)$th block is $0_{(p_0 + 1) \times (p_0 + 1)}$ for $k = 0$ or $k' = 0$. Then $\Sigma(\tau)$ can be rewritten as $\Sigma(\tau) = \Theta_{\tau} \Sigma_0 \Theta_{\tau}^T$. 
Furthermore, we have 
\begin{equation}
\text{Cov}\{\bxi_{i} \psi_{\tau}(u_i), {d}_i\} = \text{E} \{\psi_{\tau}(u_i) \bxi_{i}^T {d}_i \}=
\E\{\bxi_{i}^T {d}_i \text{E} \psi_{\tau}(u_i) | \bxi_i \} = 0, 
 \end{equation}
 which leads to 
\begin{equation}
-\frac{1}{\sqrt{n}} \sum_{i = 1}^n \{ \bxi_{i} \psi_\tau({u}_i) - \bD_1(\tau) {d}_i\} \overset{d}{\rightarrow} N(0, \tau(1 - \tau)\bD_0 + \bD_1(\tau) \Sigma(\tau) \bD_1(\tau)). 
\end{equation} Hence, we have $Z_{1n}({\delta})  \overset{d}{\rightarrow} -{\delta}^T \bW(\tau)$ where $\bW(\tau)\sim N\left(0, \tau(1 - \tau)\bD_0 + \bD_1(\tau) {\Sigma}(\tau) \bD_1(\tau)\right)$. Consequently, the following result for $Z_{2n}({\delta})$ concludes the asymptotic distribution in~\eqref{eq:eq1.7.24}.


\begin{lemma}
	\label{lemma:z2n}
	Under Assumptions A4, B1, B2 and C1, we have 
	\begin{equation}
	Z_{2n}({\delta}) = \frac{1}{2}{\delta}^T \bD_1(\tau) {\delta} + o_p(1).
	\end{equation} 
\end{lemma}

\noindent {Step 3 (Asymptotic distributions across quantile levels).} When considering various quantile levels, the same arguments can be made via a convex optimization and the limiting distribution of the objective function. The asymptotic covariance in~\eqref{ten} is obtained by the covariance between 
$\bxi_{i} \psi_{\tau_{\ell}}({u}_i) + \bD_1(\tau_{\ell}) {d}_i(\tau_{\ell}) $ and $\bxi_{i} \psi_{\tau_{\ell'}}({u}_i) + \bD_1(\tau_{\ell'}) {d}_i(\tau_{\ell'}), $ following similar calculation as in~\eqref{eq:covariance.inflat}. 
\end{proof}

\begin{proof}[\textbf{\upshape Proof of Theorem~\ref{th:wald.p0}.}]
We just need to show that $\bR (\SigmaAdj_{\widehat \bzeta} - {\Gamma}_{\widehat \bzeta}) \bR^T = 0$. 
The $(\ell, \ell')$th block of the matrix $\SigmaAdj_{\widehat \bzeta} - {\Gamma}_{\widehat \bzeta}$ is $\Theta_{\tau_{\ell}} \Sigma_0 \Theta_{\tau_{\ell'}}$, where $1 \leq \ell, \ell' \leq L$. Therefore, we have $\SigmaAdj_{\widehat \bzeta} - {\Gamma}_{\widehat \bzeta} = A \Sigma_0 A^T$ where $A = (\Theta_{\tau_1}, \ldots, \Theta_{\tau_L})^T$ is a $(p_0+1)L \times (p_0+1)$ matrix. Noting that $\Theta_{\tau_{\ell}} = 1_{(p_0 + 1) \times (p_0 + 1)} \otimes \theta_{\ell}^T$ for $\ell \in \{ 1, \ldots, L\}$, we have $A^T = 1_{(p_0 + 1) \times (p_0 + 1)}  \otimes \bzeta^T $ and thus $A = 1_{(p_0 + 1) \times (p_0 + 1)}  \otimes \bzeta $. Therefore, when $R \bzeta = 0$, it follows that $R A = 1_{(p_0 + 1) \times (p_0 + 1)}  \otimes (R \bzeta) = 0$. This completes the proof.  
\end{proof} 

\subsection{Proofs of lemmas} \label{subsection:proof.lemmas}



\begin{proof}[\textbf{\upshape Proof of Lemma~\ref{lemma:z.hat.z}.}]
The bound in~\eqref{eq:z.hat.z.expectation} follows from standard bounds for the estimated eigenfunctions and covariance kernel in the FDA literature.  According to Theorem 1 in~\cite{Hal+Hos:06}, we have 
\begin{equation}
	\|\widehat{\phi}_k - \phi_k\|_{L^2} \leq 8^{1/2} s_k^{-1} |||\widehat{G} - G |||, 
\end{equation}
where $s_k = \min_{r \leq k} (\lambda_r - \lambda_{r + 1})$ and $|||\widehat{G} - G ||| = [\int_0^1\int_0^1 \{\widehat{G}(u,v) - G(u,v)\}^2 du dv]^{1/2}$. Therefore,
\begin{equation}
	|\widehat{z}_{ik} - \xi_{ik}| = \left|\int_0^1 X_i(t) \{\widehat{\phi}_k(t) - \phi_k(t)\}dt \right| \leq \|X_i\|_{L^2} \cdot \|\widehat{\phi}_k - \phi_k\|_{L^2} \leq \const \|X_i\|_{L^2} s_k^{-1} \cdot |||\widehat{G} - G |||, 
\end{equation}
which leads to $
\|\widehat{\bxi}_i - \bxi_{i} \| \leq \const \|X_i\|_{L^2} s_{p_0}^{-1}|||\widehat{G} - G |||. 
$
For any $c > 0$, invoking the bound $\text{E}|||\widehat{G} - G |||^c \leq \const n^{-c/2}$~\cite[Lemma 3.3]{Hal:09} leads to 
\begin{equation}
	\text{E} \|\widehat{\bxi}_i - \bxi_{i} \|^c \leq \const s_{p_0}^{-c} (\text{E}|||\widehat{G} - G |||^{2c})^{1/2} \leq \const s_{p_0}^{-c} n^{-c/2}. 
\end{equation}
Thus, for finite $p_0$, we have $\text{E} \|\widehat{\bxi}_i - \bxi_{i} \|^c = o(n^{-c/4})$; in particular, there holds $\sqrt{n} \text{E} \| \widehat{\bxi}^T_i  - \bxi^T_i\|^2 = o(1)$.

Next we prove the representation in~\eqref{eq:represent.difference}. Let $\widetilde{G}$ be the estimator of the kernel $G$ based on the fully observed covariate $X_i(\cdot)$, and recall that $\widehat{G}$ is the estimate based on the discretized  $W_{i j}$ with measurement error. Denote $\widetilde{\mathcal{Z}} = \sqrt{n}(\widetilde{G} - G)$ and $\widehat{\mathcal{Z}} = \sqrt{n}(\widehat{G} - G)$. We use the notation $\int \widehat{\mathcal{Z}} \phi_k \phi_{k'}$ to denote $\int_0^1 \int_0^1 \widehat{\mathcal{Z}}(u, v) \phi_k(u) \phi_{k'}(v) du dv$. 

Since $\{\phi_k: k \geq 1\}$ forms a basis of the $L^2$ space on [0, 1], we have $\widehat{\phi}_k = \sum_{k' = 1}^{\infty} a_{kk'} \phi_k', $ where $k \in \{1, \ldots, p_0\}$ and the generalized Fourier coefficients $a_{kk'} = \int_0^1 \widehat{\phi}_k(t) \phi_{k'}(t) dt$. Furthermore, we have the following expansion for $a_{kk'}$'s: 
\begin{equation}
	a_{kk}  = 1 + O_p(n^{-1}); \quad a_{kk'} = n^{-1/2}(\lambda_k - \lambda_{k'})^{-1} \int \widehat{\mathcal{Z}} \phi_k \phi_{k'} + O_p(n^{-1}) \;\text{if}\; k \neq k',
\end{equation} 
according to (2.6) and (2.7) in~\cite{Hal+Hos:06}. 
Therefore, for $k \in \{1, \ldots, p_0\}$, we have 
\begin{align}
	\int_0^1 X_i(t) \{\widehat{\phi}_k(t) - \phi_k(t)\} dt = \sum_{k' = 1}^{p_0} \{a_{kk'} - I(k' = k)\} \xi_{ik'} = \sum_{k' = 1, k' \neq k}^{p_0} n^{-1/2} (\lambda_k - \lambda_{k'})^{-1} \int \widehat{\mathcal{Z}} \phi_k \phi_{k'} \xi_{ik'} + O_p(n^{-1}). 
\end{align} 
A direct calculation gives that $$ \int \widetilde{\mathcal{Z}} \phi_k \phi_{k'} = n^{-1/2} \sum_{i = 1}^n \xi_{ik} \xi_{ik'} - n^{1/2} \bar{\xi}_k \bar{\xi}_{k'}$$ 
for $k, k' = 1, \ldots, p_0$ and $k \neq k'$, where $\bar{\xi}_k = n^{-1}\sum_{i = 1}^n \xi_{ik}$. Since $n^{1/2} \bar{\xi}_k \bar{\xi}_{k'} = n^{-1/2} \cdot (n^{1/2} \bar{\xi}_k) \cdot (n^{1/2} \bar{\xi}_{k'}) = n^{-1/2} \cdot O_p(1) \cdot O_p(1) = O_p(n^{-1/2})$, we have $\int \widetilde{\mathcal{Z}} \phi_k \phi_{k'} = n^{-1/2} \sum_{i = 1}^n \xi_{ik} \xi_{ik'} + O_p(n^{-1/2})$. The same approximation holds when using $\widehat{\mathcal{Z}}$ since $\widehat{\mathcal{Z}} - \widetilde{\mathcal{Z}}$ is uniformly $o_p(n^{-1/2})$ as shown by~\cite{Zhang+Chen:07}. Consequently, 
\begin{equation}
	\label{eq:eq1.7.25} 
	\int_0^1 X_i(t) \{\widehat{\phi}_k(t) - \phi_k(t)\} dt = \sum_{k' = 1, k' \neq k}^{p_0} n^{-1}(\lambda_k - \lambda_{k'})^{-1} \left(\sum_{i = 1}^n \xi_{ik} \xi_{ik'}\right) \xi_{ik'} + O_p(n^{-1}). 
\end{equation}
This approximation will not be affected if we use $\widehat{X}_i(\cdot)$ instead of the true curve $X_i(\cdot)$ because the difference $\widehat{X}_i(\cdot) - X_i(\cdot)$ is negligible uniformly for all $i$ (e.g., see Theorem 2 in~\cite{Zhang+Chen:07} or Lemma 1 in~\cite{Zhu+:14}). Let a $p_0$-dimension random matrix $\bB^+ = (b_{kk'})$ where $b_{kk'} = 0$ if $k = k'$ and $b_{kk'} =  n^{-1/2} (\lambda_k - \lambda_{k'})^{-1} \left(\sum_{i = 1}^n \xi_{ik} \xi_{ik'}\right)$ if $k \neq k'$. Let $\bB$ be a $(p_0 + 1)\times(p_0 + 1)$ zero matrix but the bottom right block is replaced by $\bB^+$, then the right hand side in~\eqref{eq:eq1.7.25} becomes $n^{-1/2} \bB \bxi_i + O_p(n^{-1})$. Consequently, we have 
\begin{equation}
	\widehat{\bxi}_{i} - \bxi_{i} = n^{-1/2} \bB \bxi_i + O_p(n^{-1}).
\end{equation}
{\color{black}{This completes the proof by noting that all the stochastic bounds starting from $O_p(n^{-1})$ in $a_{kk}$ do not depend on $i$.}}
\end{proof} 


\begin{proof}[\textbf{\upshape Proof of Lemma~\ref{th:z1n.represent}.}]
We first decompose the difference between $Z_{1n}({\delta})$ and $Z_{1n}^{*}({\delta})$ into three parts $S_1, S_2$ and $S_3$ as follows: 
\begin{align}
	Z_{1n}({\delta}) & - Z_{1n}^{*}({\delta}) =  -\frac{1}{\sqrt{n}} \sum_{i = 1}^n \widehat{\bxi}_i^T {\delta} \psi_\tau(\widehat{u}_i) + \frac{1}{\sqrt{n}} \sum_{i = 1}^n {z}_i^T {\delta} \psi_\tau({u}_i) \\ 
	& =  \left( -\frac{1}{\sqrt{n}} \sum_{i = 1}^n (\widehat{\bxi}_i^T - \bxi_{i}^T) {\delta} \{\psi_\tau(\widehat{u}_i) - \psi_\tau(u_i)\} \right) + \left(
	  -\frac{1}{\sqrt{n}} \sum_{i = 1}^n (\widehat{\bxi}_i^T - \bxi_{i}^T) {\delta} \psi_\tau(u_i) \right) + \left( -\frac{1}{\sqrt{n}} \sum_{i = 1}^n \bxi_{i}^T {\delta} \{\psi_\tau(\widehat{u}_i) - \psi_\tau(u_i)\} \right)\\
	& =:  \; S_1 + S_2 + S_3. 
\end{align}
The proof proceeds in three steps: $S_2 = o_p(1)$ ({Step i}), $S_1 = o_p(1)$ ({Step ii}), and $S_3 = n^{-1/2} \delta^T D_1(\tau) \sum_{i = 1}^n d_i + o_p(1)$ ({Step iii}). {Step i} and {Step ii} 
indicate that the first two terms $S_1$ and $S_2$ are negligible, and it is sufficient to show that $\text{E}(S_2^2) = o(1)$ and $\text{E} |S_1| = o(1)$ according to Chebyshev's inequality. The third term $S_3$ is challenging to analyze since the function of $\psi_{\tau}(\cdot)$ is not differentiable. In {Step iii}, we approximate the term $S_3$ mainly using the uniform approximation on $\psi_{\tau}(\cdot)$. 

\noindent{{Step i.}} First notice that $\text{E}\{\psi_{\tau}(u_i)|\bxi_{i},\widehat{\bxi}_i\}=0$ and $\text{E}\{\psi_{\tau}(u_i)^2|\bxi_{i},\widehat{\bxi}_i\}= \tau - \tau^2$. Therefore, we have $\text{E}(S_2)= 0$, and further 
\begin{align}
	\text{E}(S_{2}^2) = \text{E} \left\{\frac{1}{\sqrt{n}}\sum_{i=1}^n (\widehat{\bxi}_{i}^T{\delta}-\bxi_{i}^T{\delta})\psi_{\tau}(u_i)\right\}^2 =  \frac{1}{n}\sum_{i=1}^n \sum_{i'=1}^n \text{E}\left\{(\widehat{\bxi}_{i}^T-\bxi_i^T ){\delta}\psi_\tau(u_i)\cdot (\widehat{\bxi}_{{i'}}^T-\bxi_{i'}^T) {\delta}\psi_\tau(u_{i'}) \right\}.
\end{align}
For $i=i'$,
\begin{align}
	\text{E}\left\{(\widehat{\bxi}_{i}^T-\bxi_i^T ){\delta}\psi_\tau(u_i)\cdot (\widehat{\bxi}_{i'}^T-\bxi_{i'}^T) {\delta}\psi_\tau(u_{i'}) \right\} & = \text{E}\left\{(\widehat{\bxi}_i^T{\delta}-\bxi_i^T{\delta})\psi_\tau(u_i) \right\}^2\\
	&= \text{E}\left\{(\widehat{\bxi}_i^T{\delta}-\bxi_i^T{\delta})^2\text{E}\{\psi_\tau^2(u_i)|\bxi_i,\widehat{\bxi}_i \} \right\} =\tau(1-\tau)\E\left\{(\widehat{\bxi}_i^T{\delta}-\bxi_i^T{\delta})^2\right\}.
\end{align}
Since $\widehat{\bxi}_i$ are identically distributed for all $i$, we have $\E\left\{(\widehat{\bxi}_i^T{\delta}-\bxi_i^T{\delta})^2\right\} = \E\left\{(\widehat{\bxi}_1^T{\delta}-\bxi_1^T{\delta})^2\right\}.$
For $i\neq i'$, we have 
$
	\text{E}\left\{(\widehat{\bxi}_{i}^T-\bxi_i^T ){\delta}\psi_\tau(u_i)\cdot (\widehat{\bxi}_{i'}^T-\bxi_{i'}^T) {\delta}\psi_\tau(u_{i'}) \right\}=0
$ by noting that  
\begin{equation}
	\text{E}\{\psi_\tau(u_i)\psi_\tau(u_{i'})|\bxi_i,\widehat{\bxi}_i, \bxi_{i'},\widehat{\bxi}_{i'} \} = \text{E}\{\psi_\tau(u_i)|\bxi_i,\widehat{\bxi}_i, \bxi_{i'},\widehat{\bxi}_{i'} \}\cdot\text{E}\{\psi_\tau(u_{i'})|\bxi_i,\widehat{\bxi}_i, \bxi_{i'},\widehat{\bxi}_{i'} \}= 0.
\end{equation}
Therefore, $\text{E}(S_2^2)= \tau(1- \tau)\text{E}\left\{(\widehat{\bxi}_1^T{\delta}-\bxi_1^T{\delta})^2\right\} = O(\text{E} \|\widehat{\bxi}_i - \bxi_{i} \|^2) = o(1).$

\noindent{{Step ii.}}
For $S_1$, we first introduce the notation
\begin{equation}
	\label{eq:delta.def}
	\Delta_i  = \text{E}( \psi_{\tau}(\widehat{u}_i) | \bxi_{i}, \widehat{\bxi}_i) = \tau-F_i(\widehat{\bxi}_i^T \btheta_{\tau} )=F_i(\bxi_{i}^T \btheta_{\tau} )-F_i(\widehat{\bxi}_i^T \btheta_{\tau} ). 
\end{equation}
For each $i$, this random variable $\Delta_i$ satisfies that 
\begin{align}
	\label{eq:dummy1.11.18}
	\Delta_i & =  \text{E}( \psi_{\tau}(\widehat{u}_i) - \psi_{\tau}(u_i) | \bxi_{i}, \widehat{\bxi}_i), \\
	\label{eq:dummy2.11.18}
	|\Delta_i| & =  \text{E}( |\psi_{\tau}(\widehat{u}_i) - \psi_{\tau}(u_i)| | \bxi_{i}, \widehat{\bxi}_i).
\end{align}
The result given in~\eqref{eq:dummy1.11.18} is obtained by noting that $\psi_{\tau}(u_i)$ has mean 0 conditional on $\bxi_{i}$,  while~\eqref{eq:dummy2.11.18} holds because $|\psi_\tau(\widehat{u}_i) - \psi_\tau({u}_i)| = I\{\min(\widehat{\bxi}_i\btheta_{\tau},\bxi_{i}\btheta_{\tau})<y_i<\max(\widehat{\bxi}_i\btheta_{\tau} ,\bxi_{i}\btheta_{\tau}  )\}.$

By Taylor's theorem, for any $a, b \in \mathbb{R},$ we have
$$ F(a + b) - F(a) = f(a)b + b^2 \int_0^1 f'(a + tb)(1 - t) dt =: f(a)
b + \frac{ b^2}{2} R(a, b),$$
where $|R(a,b)| \leq C_0$. Therefore, 
\begin{equation}
	\Delta_i = -(\widehat{\bxi}_i^T \btheta_{\tau}  - \bxi_{i}^T \btheta_{\tau} ) f_i(\bxi_{i}^T \btheta_{\tau} ) + (\widehat{\bxi}_i^T \btheta_{\tau}  - \bxi_{i}^T \btheta_{\tau} )^2 R(\widehat{\bxi}_i^T, \bxi_{i}^T),
\end{equation}
where $|R(\widehat{\bxi}_i^T, \bxi_{i}^T)| \leq 2C_0$. We also have the bound 
\begin{equation}
	\text{E} \Delta_i^2 \leq \const \text{E} \|\widehat{\bxi}_i - \bxi_{i} \|^2  = o(n^{-1/2}).
\end{equation}
Therefore, $|S_1| 
\leq  \frac{1}{\sqrt{n}} \sum_{i = 1}^n |(\widehat{\bxi}_i^T {\delta} - \bxi_{i}^T {\delta})| \cdot  |\psi_\tau(\widehat{u}_i) - \psi_\tau({u}_i)|$ and consequently 
\begin{align}
	\text{E} |S_1| 
	& \leq \frac{1}{\sqrt{n}}\text{E} \{\sum_{i = 1}^n |(\widehat{\bxi}_i^T {\delta} - \bxi_{i}^T {\delta})| |\Delta_i| \} \\
	&= \sqrt{n} \text{E} |(\widehat{\bxi}_1^T {\delta} - \bxi_1^T {\delta}) \Delta_1| \leq \sqrt{n \text{E} \|\widehat{\bxi}_i - \bxi_{i} \|^2 \text{E} \Delta_1^2} = o(1). 
\end{align} 

\noindent{{Step iii.}}
	Define 
	\begin{equation}
		R_n(\bt) = \sum_{i = 1}^n \bxi_{i} \{\psi_{\tau}(u_i - \bxi_{i}^T \bt ) - \psi_{\tau}(u_i)\},
	\end{equation}
	for any vector such that $\|\bt\| \leq C$ for some constant $C$. Then the uniform approximation~\citep{He+Shao:96} indicates that 
	\begin{equation}
		\sup \| R_n(\bt) - \text{E} \{R_n(\bt)\} \| = O_p(n^{1/2} (\log n) \|\bt\|^{1/2}). 
	\end{equation} 
	On the other hand, 
	\begin{align}
		\text{E} \{R_n(\bt)\} & = \sum_{i = 1}^n \text{E} [\bxi_{i} \{F_i(\bxi_{i}^T \btheta_{\tau} ) - F_i(\bxi_{i}^T \btheta_{\tau}  - \bxi_{i}^T \bt)\}] = n \text{E} [\bxi_1 \{F_1(\bxi_1^T \btheta_{\tau} ) - F_1(\bxi_1^T \btheta_{\tau}  - \bxi_1^T \bt)\} ]\\
		& = - n \text{E} \bxi_1 \bxi_1^T f_1(\bxi_1^T \btheta_{\tau} ) \bt + O(n \text{E} \|\bxi_1\|^3 \| \bt\|^2) = -n \bD_1(\tau) \bt + O(n \|\bt\|^2). 
	\end{align}
	Therefore, 
	\begin{equation}
		\label{eq:dummy3.11.18}
		R_n(\bt) = -n \bD_1(\tau) \bt + O(n \|\bt\|^2) + O_p\left(n^{1/2} (\log n) \|\bt\|^{1/2}\right). 
	\end{equation}
	Note that $\widehat{u}_i = u_i + \bxi_{i}^T \btheta_{\tau}  - \widehat{\bxi}_i^T \btheta_{\tau} $ and $\widehat{\bxi}_i - \bxi_i = \bB \bxi_i$ up to a negligible term $O_p(n^{-1})$. Then  
	\begin{equation}
		-\frac{1}{\sqrt{n}} \sum_{i = 1}^n \bxi_{i}^T {\delta}\{ \psi_\tau(\widehat{u}_i) - \psi_\tau({u}_i)\} = -n^{-1/2} R_n(n^{-1/2} \bB \btheta_{\tau} ) + o_p(1),
	\end{equation} 
	where the term $o_p(1)$ is obtained by the same arguments used in Step ii via conditional expectation and Taylor theorem. Substituting $\bt = n^{-1/2} \bB \btheta_{\tau}$ into~\eqref{eq:dummy3.11.18} and noting that $\|n^{-1/2} \bB \btheta_{\tau}  \| = O_p(n^{-1/2})$, we obtain that $R_n(n^{-1/2} \bB \btheta_{\tau} ) =  -n^{1/2} \bD_1(\tau)  \bB \btheta_{\tau}  + O(1) + O_p(n^{1/4} \log n)$, leading to   
	\begin{equation}
		\label{eq:dummy4.11.18}
		S_3 = -\frac{1}{\sqrt{n}} \sum_{i = 1}^n \bxi_{i}^T {\delta}\{ \psi_\tau(\widehat{u}_i) - \psi_\tau({u}_i)\} = {\delta}^T \bD_1(\tau)  \bB \btheta_{\tau}  + o_p(1). 
	\end{equation}
	%
	According to the definition of $B$ in~\eqref{eq:represent.difference}, it is easy to verify that $B \theta_{\tau} = n^{-1/2} \sum_{i = 1}^n d_i$, where ${d}_i = (0, d_{i1}, \ldots, d_{i p_0}) $ and 
	$d_{ik} = \sum_{r = 1, r \neq k}^{p_0} (\lambda_k - \lambda_r)^{-1} \xi_{ik} \xi_{ir} \beta_r(\tau)$ for $k \geq 1.$ Therefore, it follows that $S_3 = n^{-1/2} \delta^T D_1(\tau) \sum_{i = 1}^n d_i + o_p(1)$, which concludes Step iii. This completes the proof.  
\end{proof}

\begin{proof}[\textbf{\upshape Proof of Lemma~\ref{lemma:z2n}.}] 
Recall that $Z_{2n} = \sum_{i = 1}^n Z_{2ni},$ where 
	\begin{equation}
		Z_{2ni}({\delta}) = \int_{0}^{\widehat{\bxi}_i^T {\delta}/\sqrt{n}}
		\{I(\widehat{u}_i \leq s) - I(\widehat{u}_i \leq 0)\} ds. 
	\end{equation}
	First, we have 
	\begin{equation}
		\text{E}[Z_{2ni}({\delta}) | \bxi_{i}, \widehat{\bxi}_i] =  \int_{0}^{\widehat{\bxi}_i^T
			{\delta}/\sqrt{n}} F_i(\widehat{\bxi}_i^T \btheta_{\tau}  + s) -  F_i(\widehat{\bxi}_i^T
		\btheta_{\tau} ) ds = \frac{1}{\sqrt{n}} \int_0^{\widehat{\bxi}_i^T {\delta}}  F_i(\widehat{\bxi}_i^T \btheta_{\tau}  + \frac{t}{\sqrt{n}}) -  F_i(\widehat{\bxi}_i^T
		\btheta_{\tau} ) dt. 
	\end{equation}
	Therefore, by Taylor's theorem, we have 
	$$  \text{E}[Z_{2ni}({\delta}) | \bxi_{i}, \widehat{\bxi}_i] = \frac{1}{\sqrt{n}} \int_0^{\widehat{\bxi}_i^T {\delta}} f_i(\widehat{\bxi}_i^T
	\btheta_{\tau} )\frac{t}{\sqrt{n}} + \frac{t^2}{2n} R(\widehat{\bxi}_i^T {\delta}, \frac{t}{\sqrt{n}}) dt = \frac{1}{2n} {\delta}^T \widehat{\bxi}_i f_i(\widehat{\bxi}_i^T
	\btheta_{\tau} ) \widehat{\bxi}_i^{T} {\delta} + R_{n,i}, 
	$$
	where $R_{ni}$ is the remainder satisfying that $|R_{ni}| \leq c n^{-3/2} |\widehat{\bxi}_i^T {\delta}|^3$.  Consequently, 
	$$  \text{E}[Z_{2ni}({\delta}) | \bxi_{i}, \widehat{\bxi}_i] =   \frac{1}{2} \cdot {\delta}^T \frac{1}{n} \widehat{\bxi}_i f_i(\widehat{\bxi}_i^T
	\btheta_{\tau} ) \widehat{\bxi}_i^{T} {\delta} + R_{ni}.  $$
	Therefore, the unconditional expectation of $Z_{2ni}({\delta})$ is 
	$$
	\text{E} Z_{2ni}({\delta})  = \text{E} \{ \text{E} [Z_{2ni}({\delta})|\bxi_{i}, \widehat{\bxi}_i]
	\} =  \frac{1}{2} \cdot {\delta}^T \text{E} \left( \frac{1}{n} \widehat{\bxi}_i f_i(\widehat{\bxi}_i^T
	\btheta_{\tau} ) \widehat{\bxi}_i^{T} \right) {\delta} + \text{E} R_n = \frac{1}{2} \cdot \frac{1}{n} {\delta}^T \bD_1(\tau)  {\delta} + \text{E}(R_{ni}),
	$$ 
	leading to 
	$$\text{E} Z_{2n}=\frac{1}{2} {\delta}^T \bD_1(\tau)  {\delta} + \sum_{i=1}^{n}\text{E} (R_{ni}).$$
	The second term $\sum_{i=1}^{n}\text{E} (R_{ni})$ is negligible because 
	\begin{eqnarray}
		\left|\sum_{i=1}^{n}\text{E} (R_{ni})\right|& \leq& \sum_{i=1}^n \text{E}|R_{ni}|
		\leq c n^{-3/2}\sum_{i=1}^n \text{E} |\widehat{\bxi}_{i}^{T}{\delta}|^{3} = c n^{-1/2} \text{E} |\widehat{\bxi}_{1}^{T}{\delta}|^{3}\\
		&\leq& O(n^{-1/2}) \cdot \left(\text{E}\|\widehat{\bxi}_{1}\|^3_2 \right)\|{\delta}\|_2^3 =  O(n^{-1/2}) \cdot O(1) = o(1),
	\end{eqnarray}
	where the last step is due to the fact that $\|\widehat{\bxi}_{1}\|_2=\|\widehat{X}_{1}\|_2\leq \|\widehat{X}_{1}-X_1\|_2 +\|X_{1}\|_2$. 
	
	We next will show that $\underset{i = 1, \ldots, n}{\max} \|\bxi_{i}\|/\sqrt{n}\overset{p}{\rightarrow} 0.$ Note that $\|\bxi_{i}\|^2 =  1+\xi_{i1}^2+\ldots+\xi_{ip_0}^2$ for $i \in \{1,\ldots,n\}$, and $\|\bxi_{i}\|^2$'s are i.i.d. with a finite second moment $ \text{E}\|\bxi_{i}\|^2 = 1+\lambda_1+\ldots+\lambda_{p_0} < \infty$. For any $\epsilon > 0$, we have
	\begin{align}
		P\left(\underset{i = 1, \ldots, n}{\max}\frac{\|\bxi_{i}\|}{\sqrt{n}} > \epsilon\right) & \leq \sum_{i = 1}^n P(\|\bxi_i\| > \sqrt{n} \epsilon) \leq \frac{1}{n \epsilon^2} \sum_{i = 1}^n \text E\{\|\bxi_i\|^2 I(\|\bxi_i\| > \sqrt{n} \epsilon)\} \\ & = \frac{1}{\epsilon^2}\text E\{\|\bxi_1\|^2 I(\|\bxi_1\| > \sqrt{n} \epsilon)\} \rightarrow 0, 
	\end{align} 
	according to the dominated convergence theorem. It implies that $\underset{i = 1, \ldots, n}{\max} \|\widehat{\bxi}_{i}\|/\sqrt{n} = o_p(1)$ since $\|\widehat{\bxi}_{i} - \bxi_i\| = o_p(1)$ uniformly for all $i$'s. 
	Consequently, $\text{Var}(Z_{2n}|{\bxi}_i\text{'s}, \widehat{\bxi}_i\text{'s}) \leq \underset{i = 1, \ldots, n}{\max} \|\widehat{\bxi}_{i}^T {\delta}\|/\sqrt{n} \cdot \text{E}(Z_{2ni}|{\bxi}_i\text{'s}, \widehat{\bxi}_i\text{'s}) = o_p(1)$, i.e., the conditional variance converges to 0 in probability. Therefore, following the martingale argument in the proof of Theorem 2 in~\cite{pollard1991}, we have $ Z_{2n}-\text{E}(Z_{2n})=o_{p}(1)$. This completes the proof.
\end{proof} 

\section*{Acknowledgments} 
We thank the Editor, the Associate Editor and two anonymous referees for constructive comments that helped to improve the paper. The authors would like to acknowledge the support of NSF DMS 2015569 (Li) and DMS 1454942 (Staicu) and NIH 5P01 CA142538-09 (Staicu). 	
	

\bibliographystyle{myjmva}
\bibliography{citation}

\begin{thebibliography}{52}
\expandafter\ifx\csname natexlab\endcsname\relax\def\natexlab#1{#1}\fi
\providecommand{\bibinfo}[2]{#2}
\ifx\xfnm\relax \def\xfnm[#1]{\unskip,\space#1}\fi
\bibitem[{Cao et~al.(2020)Cao, Wang and Wang}]{Cao+:20}
\bibinfo{author}{G.~Cao}, \bibinfo{author}{S.~Wang}, \bibinfo{author}{L.~Wang},
  \bibinfo{title}{Estimation and inference for functional linear regression
  models with partially varying regression coefficients},
  \bibinfo{journal}{Stat} \bibinfo{volume}{9} (\bibinfo{year}{2020})
  \bibinfo{pages}{e286}.
\bibitem[{Cardot et~al.(2005)Cardot, Crambes and Sarda}]{Cardot+:05}
\bibinfo{author}{H.~Cardot}, \bibinfo{author}{C.~Crambes},
  \bibinfo{author}{P.~Sarda}, \bibinfo{title}{Quantile regression when the
  covariates are functions}, \bibinfo{journal}{Nonparametric Statistics}
  \bibinfo{volume}{17} (\bibinfo{year}{2005}) \bibinfo{pages}{841--856}.
\bibitem[{Chen and M{\"u}ller(2012)}]{chen+muller2012}
\bibinfo{author}{K.~Chen}, \bibinfo{author}{H.-G. M{\"u}ller},
  \bibinfo{title}{Conditional quantile analysis when covariates are functions,
  with application to growth data}, \bibinfo{journal}{Journal of the Royal
  Statistical Society: Series B (Statistical Methodology)} \bibinfo{volume}{74}
  (\bibinfo{year}{2012}) \bibinfo{pages}{67--89}.
\bibitem[{Crambes et~al.(2011)Crambes, Gannoun and Henchiri}]{Crambes2011}
\bibinfo{author}{C.~Crambes}, \bibinfo{author}{A.~Gannoun},
  \bibinfo{author}{Y.~Henchiri}, \bibinfo{title}{{Weak consistency of the
  Support Vector Machine Quantile Regression approach when covariates are
  functions}}, \bibinfo{journal}{Statistics and Probability Letters}
  \bibinfo{volume}{81} (\bibinfo{year}{2011}) \bibinfo{pages}{1847--1858}.
\bibitem[{Crambes et~al.(2013)Crambes, Gannoun and Henchiri}]{Crambes2013}
\bibinfo{author}{C.~Crambes}, \bibinfo{author}{A.~Gannoun},
  \bibinfo{author}{Y.~Henchiri}, \bibinfo{title}{{Support vector machine
  quantile regression approach for functional data: Simulation and application
  studies}}, \bibinfo{journal}{Journal of Multivariate Analysis}
  \bibinfo{volume}{121} (\bibinfo{year}{2013}) \bibinfo{pages}{50--68}.
\bibitem[{Fan and Gijbels(1996)}]{Fan+Gijbels:96}
\bibinfo{author}{J.~Fan}, \bibinfo{author}{I.~Gijbels}, \bibinfo{title}{Local
  Polynomial Modelling and Its Applications}, Monographs on Statistics and
  Applied Probability, \bibinfo{publisher}{Chapman \& Hall},
  \bibinfo{year}{1996}.
\bibitem[{Fanaee-T and Gama(2014)}]{Fan+Gam:14}
\bibinfo{author}{H.~Fanaee-T}, \bibinfo{author}{J.~Gama}, \bibinfo{title}{Event
  labeling combining ensemble detectors and background knowledge},
  \bibinfo{journal}{Progress in Artificial Intelligence} \bibinfo{volume}{2}
  (\bibinfo{year}{2014}) \bibinfo{pages}{113--127}.
\bibitem[{Ferraty et~al.(2005)Ferraty, Rabhi and Vieu}]{ferraty2005}
\bibinfo{author}{F.~Ferraty}, \bibinfo{author}{A.~Rabhi},
  \bibinfo{author}{P.~Vieu}, \bibinfo{title}{Conditional quantiles for
  dependent functional data with application to the climatic \textit{{E}l
  {N}i{\~n}o} phenomenon}, \bibinfo{journal}{Sankhy{\=a}: The Indian Journal of
  Statistics} \bibinfo{volume}{67} (\bibinfo{year}{2005})
  \bibinfo{pages}{378--398}.
\bibitem[{Ferraty and Vieu(2006)}]{Ferraty+Vieu:06}
\bibinfo{author}{F.~Ferraty}, \bibinfo{author}{P.~Vieu},
  \bibinfo{title}{Nonparametric Functional Data Analysis},
  \bibinfo{publisher}{Springer}, \bibinfo{address}{New York},
  \bibinfo{year}{2006}.
\bibitem[{Gertheiss et~al.(2013)Gertheiss, Maity and Staicu}]{Ger+:13}
\bibinfo{author}{J.~Gertheiss}, \bibinfo{author}{A.~Maity},
  \bibinfo{author}{A.-M. Staicu}, \bibinfo{title}{Variable selection in
  generalized functional linear models}, \bibinfo{journal}{Stat}
  \bibinfo{volume}{2} (\bibinfo{year}{2013}) \bibinfo{pages}{86--101}.
\bibitem[{Hall and Hosseini-Nasab(2006)}]{Hal+Hos:06}
\bibinfo{author}{P.~Hall}, \bibinfo{author}{M.~Hosseini-Nasab},
  \bibinfo{title}{On properties of functional principal components analysis},
  \bibinfo{journal}{Journal of the Royal Statistical Society: Series B
  (Statistical Methodology)} \bibinfo{volume}{68} (\bibinfo{year}{2006})
  \bibinfo{pages}{109--126}.
\bibitem[{Hall and Hosseini-Nasab(2009)}]{Hal:09}
\bibinfo{author}{P.~Hall}, \bibinfo{author}{M.~Hosseini-Nasab},
  \bibinfo{title}{Theory for high-order bounds in functional principal
  components analysis}, \bibinfo{journal}{Mathematical Proceedings of the
  Cambridge Philosophical Society} \bibinfo{volume}{146} (\bibinfo{year}{2009})
  \bibinfo{pages}{225--256}.
\bibitem[{Hall et~al.(2006)Hall, M\"{u}ller and Wang}]{Hall+:06}
\bibinfo{author}{P.~Hall}, \bibinfo{author}{H.-G. M\"{u}ller},
  \bibinfo{author}{J.-L. Wang}, \bibinfo{title}{Properties of principal
  component methods for functional and longitudinal data analysis},
  \bibinfo{journal}{The Annals of Statistics} \bibinfo{volume}{34}
  (\bibinfo{year}{2006}) \bibinfo{pages}{1493--1517}.
\bibitem[{He and Shao(1996)}]{He+Shao:96}
\bibinfo{author}{X.~He}, \bibinfo{author}{Q.-M. Shao}, \bibinfo{title}{A
  general {B}ahadur representation of {$M$}-estimators and its application to
  linear regression with nonstochastic designs}, \bibinfo{journal}{The Annals
  of Statistics} \bibinfo{volume}{24} (\bibinfo{year}{1996})
  \bibinfo{pages}{2608--2630}.
\bibitem[{Hendricks and Koenker(1992)}]{hendricks1992hierarchical}
\bibinfo{author}{W.~Hendricks}, \bibinfo{author}{R.~Koenker},
  \bibinfo{title}{Hierarchical spline models for conditional quantiles and the
  demand for electricity}, \bibinfo{journal}{Journal of the American
  statistical Association} \bibinfo{volume}{87} (\bibinfo{year}{1992})
  \bibinfo{pages}{58--68}.
\bibitem[{Horvath et~al.(2009)Horvath, Kokoszka and Reimherr}]{Horvath+:09}
\bibinfo{author}{L.~Horvath}, \bibinfo{author}{P.~Kokoszka},
  \bibinfo{author}{M.~Reimherr}, \bibinfo{title}{Two sample inference in
  functional linear models}, \bibinfo{journal}{The Canadian Journal of
  Statistics} \bibinfo{volume}{37} (\bibinfo{year}{2009})
  \bibinfo{pages}{571--591}.
\bibitem[{Huang et~al.(2015)Huang, Scheipl, Goldsmith, Gellar, Harezlak,
  McLean, Swihart, Xiao, Crainiceanu and Reiss}]{refund:15}
\bibinfo{author}{L.~Huang}, \bibinfo{author}{F.~Scheipl},
  \bibinfo{author}{J.~Goldsmith}, \bibinfo{author}{J.~Gellar},
  \bibinfo{author}{J.~Harezlak}, \bibinfo{author}{M.~W. McLean},
  \bibinfo{author}{B.~Swihart}, \bibinfo{author}{L.~Xiao},
  \bibinfo{author}{C.~Crainiceanu}, \bibinfo{author}{P.~Reiss},
  \bibinfo{title}{refund: Regression with Functional Data},
  \bibinfo{year}{2015}. \bibinfo{note}{R package version 0.1-13}.
\bibitem[{Huber(1967)}]{huber1967}
\bibinfo{author}{P.~J. Huber}, \bibinfo{title}{The behavior of maximum
  likelihood estimates under nonstandard conditions}, in:
  \bibinfo{booktitle}{Proceedings of the Fifth Berkeley Symposium on
  Mathematical Statistics and Probability, {V}ol. {I}: {S}tatistics},
  \bibinfo{publisher}{University of California Press: Berkeley, CA, USA},
  \bibinfo{year}{1967}, pp. \bibinfo{pages}{221--233}.
\bibitem[{Ivanescu et~al.(2015)Ivanescu, Staicu, Scheipl and Greven}]{Iva+:15}
\bibinfo{author}{A.~E. Ivanescu}, \bibinfo{author}{A.-M. Staicu},
  \bibinfo{author}{F.~Scheipl}, \bibinfo{author}{S.~Greven},
  \bibinfo{title}{Penalized function-on-function regression},
  \bibinfo{journal}{Computational Statistics} \bibinfo{volume}{30}
  (\bibinfo{year}{2015}) \bibinfo{pages}{539--568}.
\bibitem[{Jiang and Wang(2010)}]{Jiang+Wang:10}
\bibinfo{author}{C.-R. Jiang}, \bibinfo{author}{J.-L. Wang},
  \bibinfo{title}{{Covariate adjusted functional principal components analysis
  for longitudinal data.}}, \bibinfo{journal}{The Annals of Statistics}
  \bibinfo{volume}{38} (\bibinfo{year}{2010}) \bibinfo{pages}{1194--1226}.
\bibitem[{Jiang et~al.(2014)Jiang, Bondell and Wang}]{liewen2014}
\bibinfo{author}{L.~Jiang}, \bibinfo{author}{H.~D. Bondell},
  \bibinfo{author}{H.~Wang}, \bibinfo{title}{Interquantile shrinkage and
  variable selection in quantile regression}, \bibinfo{journal}{Computational
  Statistics \& Data Analysis} \bibinfo{volume}{69} (\bibinfo{year}{2014})
  \bibinfo{pages}{208--219}.
\bibitem[{Kato(2012)}]{Kato:12}
\bibinfo{author}{K.~Kato}, \bibinfo{title}{Estimation in functional linear
  quantile regression}, \bibinfo{journal}{The Annals of Statistics}
  \bibinfo{volume}{40} (\bibinfo{year}{2012}) \bibinfo{pages}{3108--3136}.
\bibitem[{Knight(1998)}]{knight}
\bibinfo{author}{K.~Knight}, \bibinfo{title}{Limiting distributions for {$L_1$}
  regression estimators under general conditions}, \bibinfo{journal}{The Annals
  of Statistics} \bibinfo{volume}{26} (\bibinfo{year}{1998})
  \bibinfo{pages}{755--770}.
\bibitem[{Koenker(1984)}]{koenker1984}
\bibinfo{author}{R.~Koenker}, \bibinfo{title}{A note on {L}-estimates for
  linear models}, \bibinfo{journal}{Statistics \& Probability Letters}
  \bibinfo{volume}{2} (\bibinfo{year}{1984}) \bibinfo{pages}{323--325}.
\bibitem[{Koenker(2005)}]{koenker2005}
\bibinfo{author}{R.~Koenker}, \bibinfo{title}{Quantile Regression},
  volume~\bibinfo{volume}{38}, \bibinfo{publisher}{Cambridge university press},
  \bibinfo{year}{2005}.
\bibitem[{Kong et~al.(2016)Kong, Staicu and Maity}]{Kong+:16}
\bibinfo{author}{D.~Kong}, \bibinfo{author}{A.-M. Staicu},
  \bibinfo{author}{A.~Maity}, \bibinfo{title}{Classical testing in functional
  linear models}, \bibinfo{journal}{Journal of Nonparametric Statistics}
  \bibinfo{volume}{28} (\bibinfo{year}{2016}) \bibinfo{pages}{813--838}.
\bibitem[{Larsen(2013)}]{Larsen:13}
\bibinfo{author}{J.~Larsen}, \bibinfo{title}{Policy institute: Bike-sharing
  programs hit the streets in over 500 cities worldwide},
  \bibinfo{howpublished}{\url{http://www.earth-policy.org/plan_b_updates/
  2013/update112}}, \bibinfo{year}{2013}.
\bibitem[{Lee et~al.(2014)Lee, Noh and Park}]{lee2014model}
\bibinfo{author}{E.~R. Lee}, \bibinfo{author}{H.~Noh}, \bibinfo{author}{B.~U.
  Park}, \bibinfo{title}{Model selection via {B}ayesian information criterion
  for quantile regression models}, \bibinfo{journal}{Journal of the American
  Statistical Association} \bibinfo{volume}{109} (\bibinfo{year}{2014})
  \bibinfo{pages}{216--229}.
\bibitem[{Li et~al.(2015)Li, Staicu and Bondell}]{Li+Staicu+Bondell:15}
\bibinfo{author}{M.~Li}, \bibinfo{author}{A.-M. Staicu}, \bibinfo{author}{H.~D.
  Bondell}, \bibinfo{title}{Incorporating covariates in skewed functional data
  models}, \bibinfo{journal}{Biostatistics} \bibinfo{volume}{16}
  (\bibinfo{year}{2015}) \bibinfo{pages}{413--426}.
\bibitem[{Li et~al.(2007)Li, Liu and Zhu}]{Li2007a}
\bibinfo{author}{Y.~Li}, \bibinfo{author}{Y.~Liu}, \bibinfo{author}{J.~Zhu},
  \bibinfo{title}{{Quantile regression in reproducing kernel Hilbert spaces}},
  \bibinfo{journal}{Journal of the American Statistical Association}
  \bibinfo{volume}{102} (\bibinfo{year}{2007}) \bibinfo{pages}{255--268}.
\bibitem[{Li et~al.(2010)Li, Wang and Carroll}]{Li+:2010}
\bibinfo{author}{Y.~Li}, \bibinfo{author}{N.~Wang}, \bibinfo{author}{R.~J.
  Carroll}, \bibinfo{title}{Generalized functional linear models with
  semiparametric single-index interactions}, \bibinfo{journal}{Journal of the
  American Statistical Association} \bibinfo{volume}{105}
  (\bibinfo{year}{2010}) \bibinfo{pages}{621--633}.
\bibitem[{Li et~al.(2013)Li, Wang and Carroll}]{Li+Wang+Carroll:13}
\bibinfo{author}{Y.~Li}, \bibinfo{author}{N.~Wang}, \bibinfo{author}{R.~J.
  Carroll}, \bibinfo{title}{Selecting the number of principal components in
  functional data}, \bibinfo{journal}{Journal of the American Statistical
  Association} \bibinfo{volume}{108} (\bibinfo{year}{2013})
  \bibinfo{pages}{1284--1294}.
\bibitem[{Morris and Carroll(2006)}]{Morris+Carroll:06}
\bibinfo{author}{J.~S. Morris}, \bibinfo{author}{R.~J. Carroll},
  \bibinfo{title}{Wavelet-based functional mixed models},
  \bibinfo{journal}{Journal of the Royal Statistical Society: Series B
  (Statistical Methodology)} \bibinfo{volume}{68} (\bibinfo{year}{2006})
  \bibinfo{pages}{179--199}.
\bibitem[{Pollard(1991)}]{pollard1991}
\bibinfo{author}{D.~Pollard}, \bibinfo{title}{Asymptotics for least absolute
  deviation regression estimators}, \bibinfo{journal}{Econometric Theory}
  \bibinfo{volume}{7} (\bibinfo{year}{1991}) \bibinfo{pages}{186--199}.
\bibitem[{Ramsay and Silverman(2005)}]{Ramsay+Silverman:05}
\bibinfo{author}{J.~Ramsay}, \bibinfo{author}{B.~Silverman},
  \bibinfo{title}{Functional Data Analysis}, Springer Series in Statistics,
  \bibinfo{publisher}{Springer}, \bibinfo{year}{2005}.
\bibitem[{Redd(2012)}]{Redd:12}
\bibinfo{author}{A.~Redd}, \bibinfo{title}{A comment on the orthogonalization
  of {B}-spline basis functions and their derivatives},
  \bibinfo{journal}{Statistics and Computing} \bibinfo{volume}{22}
  (\bibinfo{year}{2012}) \bibinfo{pages}{251--257}.
\bibitem[{Shi et~al.(2021)Shi, Du, Sun and Zhang}]{Shi2021}
\bibinfo{author}{G.~Shi}, \bibinfo{author}{J.~Du}, \bibinfo{author}{Z.~Sun},
  \bibinfo{author}{Z.~Zhang}, \bibinfo{title}{{Checking the adequacy of
  functional linear quantile regression model}}, \bibinfo{journal}{Journal of
  Statistical Planning and Inference} \bibinfo{volume}{210}
  (\bibinfo{year}{2021}) \bibinfo{pages}{64--75}.
\bibitem[{Staicu et~al.(2012)Staicu, Crainiceanu, Reich and
  Ruppert}]{Staicu+:11}
\bibinfo{author}{A.-M. Staicu}, \bibinfo{author}{C.~M. Crainiceanu},
  \bibinfo{author}{D.~S. Reich}, \bibinfo{author}{D.~Ruppert},
  \bibinfo{title}{Modeling functional data with spatially heterogeneous shape
  characteristics}, \bibinfo{journal}{Biometrics} \bibinfo{volume}{68}
  (\bibinfo{year}{2012}) \bibinfo{pages}{331--343}.
\bibitem[{Staicu et~al.(2015)Staicu, Lahiri and Carroll}]{Staicu+:15}
\bibinfo{author}{A.-M. Staicu}, \bibinfo{author}{S.~N. Lahiri},
  \bibinfo{author}{R.~J. Carroll}, \bibinfo{title}{Significance tests for
  functional data with complex dependence structure}, \bibinfo{journal}{Journal
  of Statistical Planning and Inference} \bibinfo{volume}{156}
  (\bibinfo{year}{2015}) \bibinfo{pages}{1--13}.
\bibitem[{Su et~al.(2017)Su, Di and Hsu}]{Su+:17}
\bibinfo{author}{Y.-R. Su}, \bibinfo{author}{C.-Z. Di},
  \bibinfo{author}{L.~Hsu}, \bibinfo{title}{Hypothesis testing in functional
  linear models}, \bibinfo{journal}{Biometrics} \bibinfo{volume}{73}
  (\bibinfo{year}{2017}) \bibinfo{pages}{551--561}.
\bibitem[{Usset et~al.(2016)Usset, Staicu and Maity}]{Usset+:16}
\bibinfo{author}{J.~Usset}, \bibinfo{author}{A.-M. Staicu},
  \bibinfo{author}{A.~Maity}, \bibinfo{title}{Interaction models for functional
  regression}, \bibinfo{journal}{Computational Statistics \& Data Analysis}
  \bibinfo{volume}{94} (\bibinfo{year}{2016}) \bibinfo{pages}{317--330}.
\bibitem[{Wang et~al.(2012)Wang, Stefanski and Zhu}]{Wang+Stefanski+Zhu:12}
\bibinfo{author}{H.~J. Wang}, \bibinfo{author}{L.~A. Stefanski},
  \bibinfo{author}{Z.~Zhu}, \bibinfo{title}{{Corrected-loss estimation for
  quantile regression with covariate measurement errors}},
  \bibinfo{journal}{Biometrika} \bibinfo{volume}{99} (\bibinfo{year}{2012})
  \bibinfo{pages}{405--421}.
\bibitem[{Wang and Wang(2016)}]{kehuisinica}
\bibinfo{author}{K.~Wang}, \bibinfo{author}{H.~J. Wang},
  \bibinfo{title}{Optimally combined estimation for tail quantile regression},
  \bibinfo{journal}{Statistica Sinica} \bibinfo{volume}{26}
  (\bibinfo{year}{2016}) \bibinfo{pages}{295--311}.
\bibitem[{Wei and Carroll(2009)}]{Wei+Carroll:09}
\bibinfo{author}{Y.~Wei}, \bibinfo{author}{R.~J. Carroll},
  \bibinfo{title}{{Quantile regression with measurement error}},
  \bibinfo{journal}{Journal of the American Statistical Association}
  \bibinfo{volume}{104} (\bibinfo{year}{2009}) \bibinfo{pages}{1129--1143}.
\bibitem[{Wu et~al.(2015)Wu, Ma and Yin}]{Wu+Ma+Yin:14}
\bibinfo{author}{Y.~Wu}, \bibinfo{author}{Y.~Ma}, \bibinfo{author}{G.~Yin},
  \bibinfo{title}{Smoothed and corrected score approach to censored quantile
  regression with measurement errors}, \bibinfo{journal}{Journal of the
  American Statistical Association} \bibinfo{volume}{in press}
  (\bibinfo{year}{2015}).
\bibitem[{Yao et~al.(2005)Yao, M\"{u}ller and Wang}]{Yao+a:05}
\bibinfo{author}{F.~Yao}, \bibinfo{author}{H.-G. M\"{u}ller},
  \bibinfo{author}{J.-L. Wang}, \bibinfo{title}{Functional data analysis for
  sparse longitudinal data}, \bibinfo{journal}{Journal of the American
  Statistical Association} \bibinfo{volume}{100} (\bibinfo{year}{2005})
  \bibinfo{pages}{577--590}.
\bibitem[{Yao et~al.(2017)Yao, Sue-Chee and Wang}]{Yao2017a}
\bibinfo{author}{F.~Yao}, \bibinfo{author}{S.~Sue-Chee},
  \bibinfo{author}{F.~Wang}, \bibinfo{title}{{Regularized partially functional
  quantile regression}}, \bibinfo{journal}{Journal of Multivariate Analysis}
  \bibinfo{volume}{156} (\bibinfo{year}{2017}) \bibinfo{pages}{39--56}.
\bibitem[{Zhang and Chen(2007)}]{Zhang+Chen:07}
\bibinfo{author}{J.-T. Zhang}, \bibinfo{author}{J.~Chen},
  \bibinfo{title}{Statistical inferences for functional data},
  \bibinfo{journal}{The Annals of Statistics} \bibinfo{volume}{35}
  (\bibinfo{year}{2007}) \bibinfo{pages}{1052--1079}.
\bibitem[{Zhao and Xiao(2014)}]{zhao+xiao}
\bibinfo{author}{Z.~Zhao}, \bibinfo{author}{Z.~Xiao}, \bibinfo{title}{Efficient
  regressions via optimally combining quantile information},
  \bibinfo{journal}{Econometric Theory} \bibinfo{volume}{30}
  (\bibinfo{year}{2014}) \bibinfo{pages}{1272--1314}.
\bibitem[{Zhou et~al.(2008)Zhou, Huang and Carroll}]{Zhou+:08}
\bibinfo{author}{L.~Zhou}, \bibinfo{author}{J.~Z. Huang},
  \bibinfo{author}{R.~J. Carroll}, \bibinfo{title}{Joint modelling of paired
  sparse functional data using principal components},
  \bibinfo{journal}{Biometrika} \bibinfo{volume}{95} (\bibinfo{year}{2008})
  \bibinfo{pages}{601--619}.
\bibitem[{Zhu et~al.(2014)Zhu, Yao and Zhang}]{Zhu+:14}
\bibinfo{author}{H.~Zhu}, \bibinfo{author}{F.~Yao}, \bibinfo{author}{H.~H.
  Zhang}, \bibinfo{title}{Structured functional additive regression in
  reproducing kernel hilbert spaces}, \bibinfo{journal}{Journal of the Royal
  Statistical Society: Series B (Statistical Methodology)} \bibinfo{volume}{76}
  (\bibinfo{year}{2014}) \bibinfo{pages}{581--603}.
\bibitem[{Zou and Yuan(2008)}]{zou+yuan}
\bibinfo{author}{H.~Zou}, \bibinfo{author}{M.~Yuan}, \bibinfo{title}{Composite
  quantile regression and the oracle model selection theory},
  \bibinfo{journal}{The Annals of Statistics} \bibinfo{volume}{36}
  (\bibinfo{year}{2008}) \bibinfo{pages}{1108--1126}.

\end{thebibliography}

\end{document}